\documentclass[aps,prl,noeprint,twocolumn,showpacs,amsmath,amssymb]{revtex4-2}
\usepackage{besphysics}
\usepackage{times}
\usepackage{amsmath}
\usepackage{graphicx}
\usepackage{subfigure}
\usepackage{epstopdf}
\usepackage{color}
\usepackage{multirow}
\usepackage{setspace}
\usepackage{overpic}
\usepackage{amssymb}
\usepackage{float}
\usepackage[bookmarksnumbered, pdfstartview=FitH,colorlinks,urlcolor=blue, citecolor=blue,linkcolor=blue]{hyperref}
\usepackage{lineno}
\usepackage{bm}
\usepackage{rotating}
\usepackage[utf8]{inputenc}
\let\oldequation\equation
\let\oldendequation\endequation

\renewenvironment{equation}
{\linenomathNonumbers\oldequation}
{\oldendequation\endlinenomath}

\begin{document}

\title{\bf \boldmath	Observation of a vector charmoniumlike state at 4.7 \texorpdfstring{${\rm GeV}/c^2$}{GeV/c2} and search for \texorpdfstring{$Z_{cs}$}{Zcs} in \texorpdfstring{$e^+e^-\to K^+K^-J/\psi$}{KKJpsi}}
\author{
\small
M.~Ablikim$^{1}$, M.~N.~Achasov$^{4,b}$, P.~Adlarson$^{75}$, O.~Afedulidis$^{3}$, X.~C.~Ai$^{80}$, R.~Aliberti$^{35}$, A.~Amoroso$^{74A,74C}$, Q.~An$^{71,58}$, Y.~Bai$^{57}$, O.~Bakina$^{36}$
% ...
\\
\vspace{0.2cm}
(BESIII Collaboration)\\
\vspace{0.2cm} 
{\it
$^{1}$ Institute of High Energy Physics, Beijing 100049, People's Republic of China\\
% ...
$^{3}$ Bochum Ruhr-University, D-44780 Bochum, Germany\\
$^{4}$ Budker Institute of Nuclear Physics SB RAS (BINP), Novosibirsk 630090, Russia\\
% ...
$^{35}$ Johannes Gutenberg University of Mainz, Johann-Joachim-Becher-Weg 45, D-55099 Mainz, Germany\\
$^{36}$ Joint Institute for Nuclear Research, 141980 Dubna, Moscow region, Russia\\
% ...
$^{57}$ Southeast University, Nanjing 211100, People's Republic of China\\
$^{58}$ State Key Laboratory of Particle Detection and Electronics, Beijing 100049, Hefei 230026, People's Republic of China\\
% ...
$^{71}$ University of Science and Technology of China, Hefei 230026, People's Republic of China\\
% ...
$^{74}$ University of Turin and INFN, (A)University of Turin, I-10125, Turin, Italy; (B)University of Eastern Piedmont, I-15121, Alessandria, Italy; (C)INFN, I-10125, Turin, Italy\\
$^{75}$ Uppsala University, Box 516, SE-75120 Uppsala, Sweden\\
% ...
$^{80}$ Zhengzhou University, Zhengzhou 450001, People's Republic of China\\
\vspace{0.2cm}
% ...
$^{b}$ Also at the Novosibirsk State University, Novosibirsk, 630090, Russia\\
% ...
}
}	

\begin{abstract}
	Using data samples with an integrated luminosity of 5.85~fb$^{-1}$ collected at center-of-mass energies from 4.61 to 4.95 GeV with the BESIII detector operating at the BEPCII storage ring, we measure the cross section for the process $e^+e^-\to K^+K^-J/\psi$. A new resonance with a mass of $M = 4708_{-15}^{+17}\pm21$ MeV/$c^{2}$ and a width of $\Gamma = 126_{-23}^{+27}\pm30$ MeV is observed in the energy-dependent line shape of the $e^+e^-\to K^+K^-J/\psi$ cross section with a significance over $5\sigma$. The $K^{+}\jpsi$ system is also investigated to search for charged charmoniumlike states, but no significant $Z_{cs}^+$ states are observed. Upper limits on the Born cross sections for $\EE\to K^{-} Z_{cs}(3985)^{+}/K^{-} Z_{cs}(4000)^{+} + c.c.$ with $Z_{cs}(3985)^{\pm}/Z_{cs}(4000)^{\pm}\to K^{\pm} \jpsi$ are reported at 90\% confidence levels. The ratio of branching fractions $\frac{\mathcal{B}(Z_{cs}(3985)^{+}\rightarrow K^+ J/\psi)}{\mathcal{B}(Z_{cs}(3985)^{+}\rightarrow (\bar{D}^{0}D_s^{*+} + \bar{D}^{*0}D_s^+))}$ is measured to be less than 0.03 at 90\% confidence level.
\end{abstract}
	
\maketitle
	
\oddsidemargin  -0.2cm
\evensidemargin -0.2cm
	
The charmonium system is an ideal place to study the perturbative and non-perturbative strong interactions of quarks and gluons at the hadronic scale. Below open-charm threshold, the spectrum of $c\bar{c}$ charmonium states is well described by a potential model~\cite{Kwong:1987mj}.  All observed states have been found within expectations, and excellent agreement has been achieved between theories and experiments.  Above open-charm threshold, however, there are still many missing states that have not yet been discovered~\cite{Yuan:2021wpg}, and, surprisingly, several unexpected states, such as the $X(3872)$~~\cite{Belle:2003nnu}, $Y(4260)$~\cite{BaBar:2005hhc} and $Z_c(3900)$~\cite{BESIII:2013ris,Belle:2013yex}, have been observed since 2003. These particles do not match the predictions of the potential models and are widely considered to be good candidates for exotic states~\cite{Swanson:2006st,Chen:2016qju,Guo:2017jvc,Brambilla:2019esw}.
		
Among them, the $Y$ states show strong coupling to hidden-charm final states, such as the experimentally well established $Y(4260)\to\pp\jpsi$~\cite{BaBar:2005hhc,CLEO:2006tct,Belle:2007dxy,BESIII:2016bnd,BESIII:2020oph,BESIII:2022qal} and $Y(4360)/Y(4660)\to\pp\psip$~\cite{Belle:2007umv,Belle:2014wyt,BaBar:2012hpr,BESIII:2017tqk,BESIII:2021njb}. This is in contrast to the vector charmonium states in the same energy region.
In recent electron-positron annihilation experiments running at high energies, structures are observed around 4.66~GeV in both open- and hidden-charm final states~\cite{Belle:2008xmh,Belle:2019qoi,Belle:2020wtd,BESIII:2022quc,BESIII:2022yga,BESIII:2022wjl,BESIII:2023cmv,BESIII:2022kcv}. 
The BESIII experiment has reported evidence for a structure around 4.7 GeV in the study of $e^+e^-\to K_S^0 K_S^0 J/\psi$~\cite{BESIII:2022kcv}. This structure could potentially correspond to the well-known $Y(4660)$ resonance or another excited $Y$ state.
On the other hand, potential models predict the $5S$ and $4D$ charmonium states in this mass region~\cite{Ding:2007rg,Akbar:2020wto,Chaturvedi:2019usm,Sultan:2014oua,Deng:2016stx,Gui:2018rvv,Kanwal:2022ani,A:2021vdw}.  At present, a comprehensive understanding of the $Y$ states above 4.6 GeV is yet to be achieved, due to limited experimental information. More measurements are urgently needed to clarify their nature.
 
In addition, an isospin-1/2 charmoniumlike candidate was recently observed by BESIII in the process $e^+e^- \to K Z_{cs}(3985)$, where the charged $Z_{cs}(3985)^+$, recoiling against a $K^-$, was found decaying to 
$(\bar{D}^0D_s^{*+}+\bar{D}^{*0}D_s^+)$~\cite{BESIII:2020qkh}, and the corresponding neutral $Z_{cs}(3985)^0$, recoiling against a $K_{S}$, was found decaying to 
$(D_s^+D^{*-}+D_s^{*+}D^-)$~\cite{BESIII:2022qzr}. 
The $Z_{cs}(3985)$ is considered to be the strange partner of the $Z_c(3900)$~\cite{Yang:2020nrt,Meng:2020ihj,Yang:2020nrt,Maiani:2021tri,Chen:2022asf}.  
Charge-conjugate modes are implied here and elsewhere unless otherwise specified. 
BESIII also reported a search for $Z_{cs}^{\prime+}\to D_s^{*+}D^{*0}$~\cite{BESIII:2022vxd}, finding an excess of $Z_{cs}^{\prime+}$ candidates with a significance of $2.1\sigma$. 
Meanwhile, LHCb reported tetraquark candidates $Z_{cs}(4000)^+/Z_{cs}(4220)^+\to K^+\jpsi$ in an amplitude analysis of $B^+\to K^+\jpsi \phi$~\cite{LHCb:2021uow}. Although the $Z_{cs}(3985)^+$ and $Z_{cs}(4000)^+$ have comparable masses, their widths are different by nearly an order of magnitude.
There are still ongoing debates on whether these particles are the same state or not~\cite{Shi:2021jyr,Giron:2021sla,Wang:2020rcx,Maiani:2021tri,Wang:2022clw,Karliner:2021qok,Han:2022fup,Meng:2021rdg,Wang:2022ckc}.
Therefore, to enhance our comprehension of these $Z_{cs}$ states and explore the possibility of new $Z_{cs}$ states, it is imperative to conduct further measurements of $\EE\to K Z_{cs}\to K\bar{K}\jpsi$~\cite{Meng:2021rdg,Chen:2013wca,Yang:2020nrt,Wang:2020rcx,ZcsKJpsi1,Wang:2022ckc,ZcsKJpsi2,ZcsKJpsi3}
	
In this Letter, a measurement of the Born cross sections of the process $e^+e^-\to K^+K^-J/\psi$ is presented at  center-of-mass (c.m.)~energies from 4.61 to 4.95~GeV~\cite{BESIII:2020nme}, corresponding to an integrated luminosity of 5.85 ${\rm fb}^{-1}$~\cite{BESIII:2022ulv}. Compared with a previous measurement~\cite{BESIII:2022joj}, new data above 4.6~GeV is analyzed for the first time, which enables us to investigate the $Y$ states above 4.6~GeV with improved precision~\cite{Belle:2007dwu,Belle:2014fgf}. 
To achieve much lower background levels and to improve the statistics, both full reconstruction and partial reconstruction methods are applied.
By investigating the line shape of the $e^+e^-\to K^+K^-J/\psi$ cross section, we report the first observation of the charmoniumlike candidate $Y(4710)$. In addition, a search for the $Z_{cs}^+\to K^{+}J/\psi$ is presented using the same data sample.
	
The BESIII detector~\cite{BESIII:2009fln} records symmetric $e^+e^-$ collisions provided by the BEPCII storage ring~\cite{Yu:2016cof}. 
Simulated data samples produced with a {\sc geant4}-based~\cite{GEANT4:2002zbu} Monte Carlo (MC) toolkit, which includes the geometric description of the BESIII detector and the detector response~\cite{Huang:2022wuo}, are used to determine detection efficiencies and to estimate background contributions. 
Signal events for the process $e^+e^-\to K^+K^-J/\psi$, with the $J/\psi$ decaying into a pair of leptons ($\mu^+\mu^-/e^+e^-$), are generated at each c.m.~energy using a phase space (PHSP) model. 

For $\kk\jpsi$ signal candidates, low-momentum kaons have relatively poor detection efficiency. In order to improve the detection efficiency, only one of the kaons ($K^\pm$) is required to be reconstructed in addition to the pair of leptons.
Charged tracks detected in the multilayer drift chamber (MDC) are required to be within a polar angle $(\theta)$ range of $|\cos\theta|<0.93$, where $\theta$ is defined with respect to the $z$ axis, which is the symmetry axis of the MDC. The distance of closest approach to the interaction point must be less than 10\,cm along the $z$ axis, and less than 1\,cm in the transverse plane. As kaons and leptons are kinematically well separated, two oppositely charged tracks with momentum greater than $0.95\gevc$ in the laboratory frame are assigned as $\ell^\pm$ for data samples with $\sqrt{s}<4.84\gev$.  For $\sqrt{s}\ge4.84\gev$, this value is slightly increased to $1.05\gevc$. The amount of deposited energy in the electromagnetic calorimeter (EMC) is further used to separate muons from electrons. For both muon candidates, the deposited energy in the EMC is required to be less than 0.4$\gev$, while it is required to be greater than 1.0$\gev$ for electrons. 
For the remaining charged tracks, particle identification~(PID), which combines the measurements of the energy deposited in the MDC~(d$E$/d$x$) and the flight time in the time-of-flight system to form likelihoods $\mathcal{L}(h)~(h=K,\pi)$ for each hadron $h$ hypothesis, is used. Tracks are identified as kaons when the kaon hypothesis has a higher likelihood than the pion hypothesis ($\mathcal{L}(K)>\mathcal{L}(\pi)$ and $\mathcal{L}(K)>0$). 
	
For events in which a pair of oppositely charged kaons identified, a four-constraint (4C) kinematic fit imposing energy-momentum conservation is applied to $\EE\to\kk\LL$. 
To remove the radiative Bhabha events, the cosine of the opening angle of the kaon pair is required to be less than 0.98 in the $J/\psi\to\EE$ mode. The $\mu/\pi$ misidentification background in the $J/\psi\to\mu^+\mu^-$ mode is suppressed by requiring at least one of the muon candidates has a penetration depth greater than 30~cm in the muon counter (MUC). 
	
For events in which only one kaon is identified, a one-constraint (1C) kinematic fit is performed under the hypothesis of $\EE\to K^\pm_{\rm{miss}}K^\mp\LL$, where the mass of the missing particle ($K^\pm_\text{miss}$) is constrained to the known $K^+$ mass~\cite{Workman:2022ynf}. 
In the $J/\psi\to e^+e^-$ mode, the dominant background comes from Bhabha events, which are vetoed by requiring $\cos(\theta_{e^+})<0.8$, $\cos(\theta_{e^-})>-0.8$ and $|\cos(\theta_{K^\pm})|<0.8$.
To further reject the radiative Bhabha events ($\gamma\EE$) with $\gamma$ conversion, where the converted electrons are misidentified as kaons, we require $|\cos(\alpha_{K^\pm K^\mp_{\text{miss}}})|<0.95$, $|\cos(\alpha_{K^\pm e^\pm})|<0.95$ and $|\cos(\alpha_{K^\pm e^\mp})|<0.95$, where $\alpha$ is the opening angle between tracks. 
As one kaon is missing, the $\mu/\pi$ misidentification background is higher in the $J/\psi\to\mu^+\mu^-$ mode. Therefore a tighter requirement is imposed by requiring the penetration depth of both muon candidates in the MUC to be greater than 30~cm. 
	
\begin{figure}[htpb]
\centering
\includegraphics[width = 0.40\textwidth]{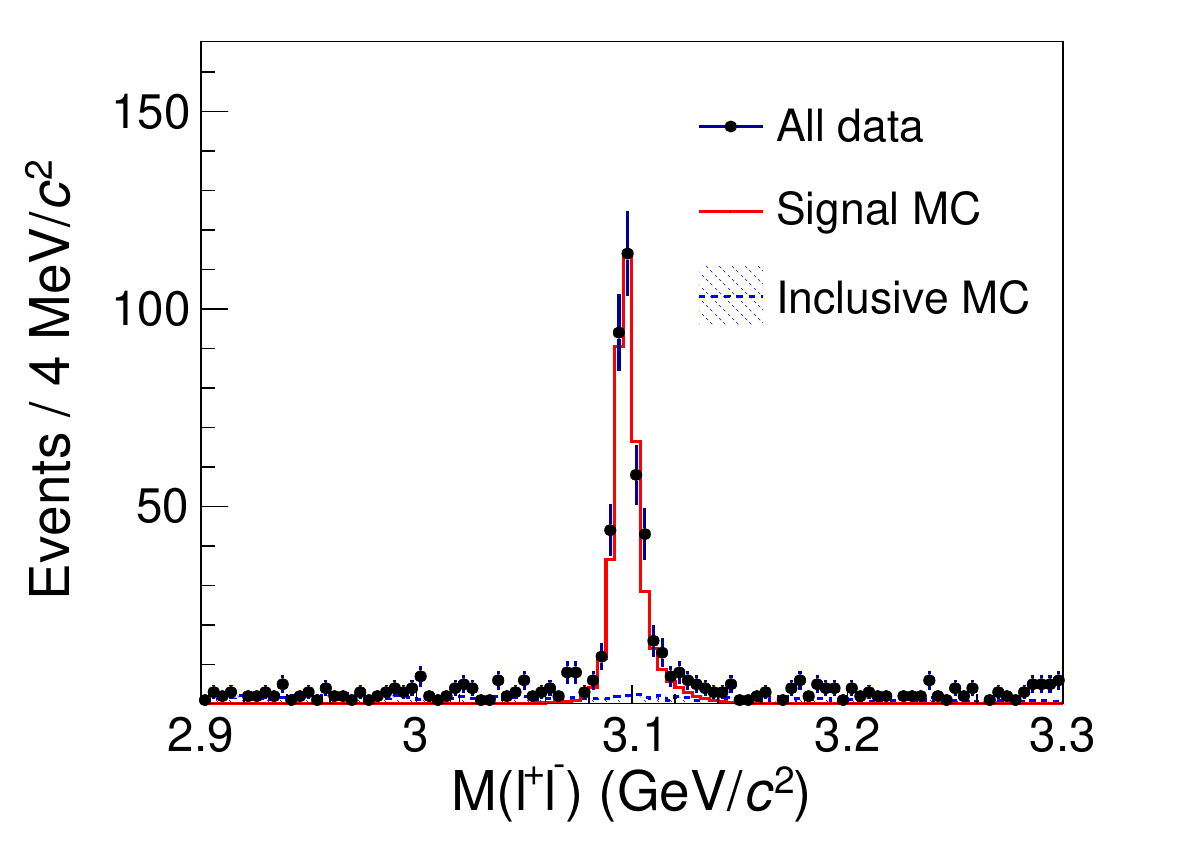}
\caption{The distribution of $M(\LL)$.  Dots with error bars are the selected data at c.m.~energies from 4.61 to 4.95~\rm{GeV}, the red histogram is the PHSP signal MC sample and the blue shaded histogram is the background from the inclusive MC sample.}
\label{fig:mll}
\end{figure}
	
After the above requirements, a clear $\jpsi$ peak is observed in the lepton pair invariant mass distribution, $M(\LL)$, as shown in Fig.~\ref{fig:mll}. A study of the inclusive MC sample~\cite{BESIII:2022kcv} indicates the background level is low and there is no peaking background. The $J/\psi$ signal region is defined as $[3.084, 3.116]\gevcc$, and the sideband regions are defined as $[3.004, 3.068]\gevcc$ and $[3.132, 3.196]\gevcc$, which are four times as wide as the $J/\psi$ signal region. The signal yield ($N^{\rm sig}$) is obtained by sideband subtraction, and the corresponding uncertainty is estimated with the profile likelihood method~\cite{Rolke:2004mj}.

The Born cross section is calculated by
	\begin{equation}
		\sigma^{\rm Born} = \frac{N^{\rm sig}}{\mathcal{L}\, (1+\delta)\, \frac{1}{|1-\Pi|^{2}}\, \epsilon\, \mathcal{B}_{J/\psi}}\;,
		\label{eq:xscalculation}
	\end{equation}
where $\mathcal{L}$ is the integrated luminosity~\cite{BESIII:2022ulv}, $(1+\delta)$ is the ISR correction factor, $\frac{1}{|1-\Pi|^{2}}$ is the correction factor for vacuum polarization~\cite{WorkingGrouponRadiativeCorrections:2010bjp}, $\epsilon$ is the detection efficiency, $\mathcal{B}_{J/\psi}$ is the sum of the branching fractions of $J/\psi\to e^{+} e^{-}$ and $J/\psi\to\mu^{+}\mu^{-}$~\cite{Workman:2022ynf}. The ISR correction factor and detection efficiency are obtained from the signal MC simulations and are corrected by an iterative weighting method~\cite{Sun:2020ehv}. The numbers of signal events, corrected detection efficiencies, and Born cross sections are summarized in the Supplemental Material~\cite{Supplemental}. The dressed cross sections ($\frac{\sigma^{\rm Born}}{|1-\Pi|^2}$) are shown in Fig.~\ref{fig:xsfit} (solid dots with error bars), and a clear structure is found around 4.75 GeV/$c^{2}$. 
	
To determine the parameters of the structures, a maximum likelihood method is used to fit the dressed cross sections obtained in this work and the dressed cross sections with $\sqrt{s}<4.61\gev$~\cite{BESIII:2022joj}. The likelihood is constructed taking into consideration the fluctuations of the numbers of signal and background events. Assuming these $K^{+}K^{-}J/\psi$ signals come from three resonances, the cross section is parameterized as a coherent sum of three relativistic Breit-Wigner functions,
	\begin{equation}
		\sigma^{\rm Dress}(\sqrt{s}) = |B_{1}(\sqrt{s}) + B_{2}(\sqrt{s}) e^{i\phi_{2}} + B_{3}(\sqrt{s}) e^{i\phi_{3}}|\;,
		\label{eq:xsfit}
	\end{equation}
where $B_{j} = \frac{M_{j}} {\sqrt{s}}
\frac{\sqrt{12\pi (\Gamma_{ee}\mathcal{B})_{j} \Gamma_{j}}}{s - M_{j}^{2} + i M_{j}\Gamma_{j}}
\sqrt{\frac{\Phi(\sqrt{s})}{\Phi(M_{j})}}$ with $j = 1$, $2$ or $3$ is the relativistic Breit-Wigner function, $\Phi$ is the three-body PHSP factor, and $\phi_{j}$ with $j = 2$ or $3$ is the relative phase between the Breit-Wigner functions. In the fit, the masses $M_{j}$, the total widths $\Gamma_{j}$, the products of the electronic partial width and the branching fraction to $K^{+}K^{-}J/\psi$ ($(\Gamma_{ee}\mathcal{B})_{j}$), and the relative phases $\phi_{j}$ are free parameters. There are four solutions with the same masses and widths from the fit as shown in the Supplemental Material~\cite{Supplemental}, and one of them is shown in Fig.~\ref{fig:xsfit}.  
The masses and widths of the first and second structures ($Y(4230)$ and $Y(4500)$) are determined to be 
$M_{1} = 4226.0_{-1.4}^{+1.4}$~MeV/$c^{2}$, $\Gamma_{1} = 70.0_{-3.6}^{+3.9}$~MeV, $M_{2} = 4499.4_{-7.6}^{+8.1}$~MeV/$c^{2}$ and $\Gamma_{2} = 124_{-19}^{+22}$~MeV, respectively, which are consistent with that of Ref.~\cite{BESIII:2022joj}. 
The mass and width of the third structure, denoted as $Y(4710)$, are $M_{3} = 4708_{-15}^{+17}$~MeV/$c^{2}$ and $\Gamma_{3} = 126_{-23}^{+27}$~MeV, respectively.
Compared to the total fit result, the peak position of the third structure seems to be shifted due to the strong interference shown in Ref.~\cite{Supplemental}.
The aforementioned uncertainties in the masses and widths are statistical only. 
	
	\begin{figure}[htbp]
		\centering
		\includegraphics[width = 0.40\textwidth]{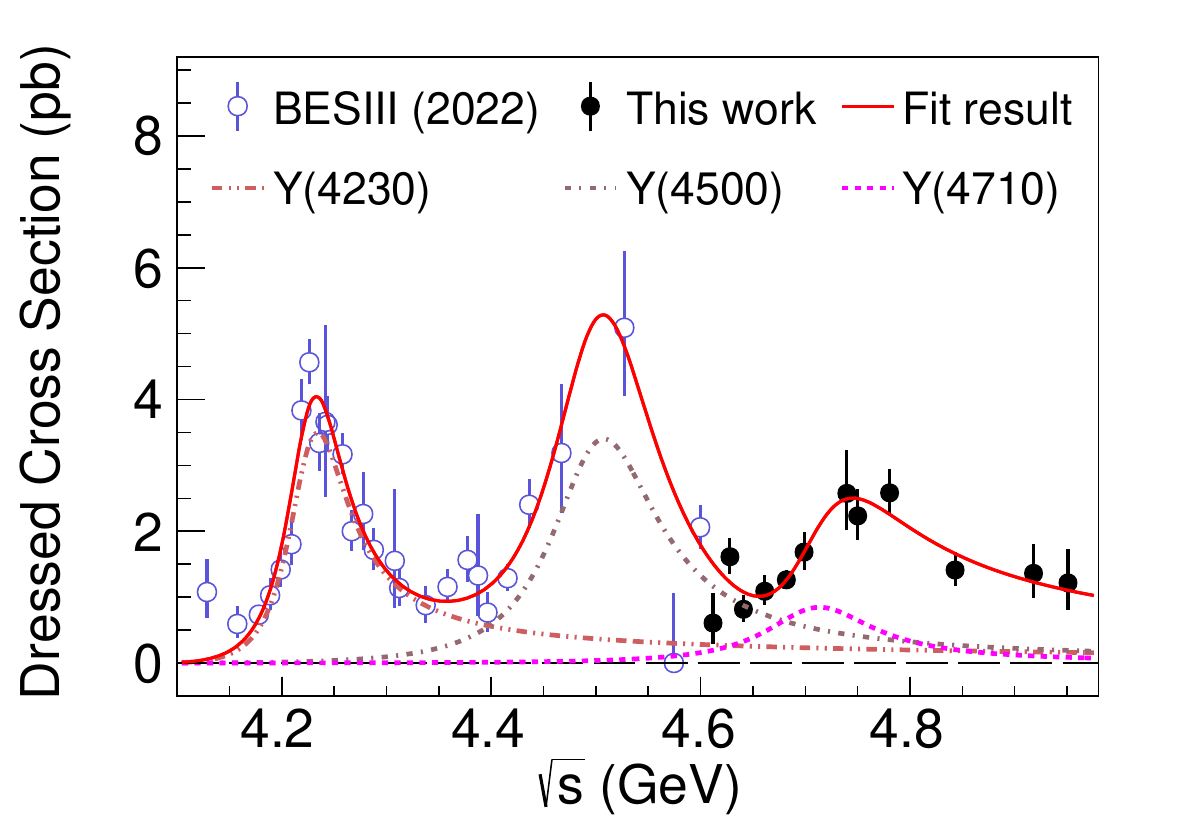}
		\caption{Fit to the dressed cross section of $e^{+}e^{-}\to K^{+}K^{-}J/\psi$ with the coherent sum of three Breit-Wigner functions (solid curve). The dashed, dash-dot-dotted, and dash-dotted curves shows the contributions from the $Y(4710)$, $Y(4230)$, and $Y(4500)$, respectively. The solid dots with error bars are the cross sections from this study, and the open dots with error bars are the cross sections from Ref.~\cite{BESIII:2022joj}. The error bars represent statistical uncertainties only.}
		\label{fig:xsfit}
	\end{figure}
	
Fitting the dressed cross sections with a two-resonance model yields a worse result: the change in the likelihood value from the three-resonance model to the two-resonance model is $|\Delta(-2\,\ln L)| = 43.2$. Taking into account the change in the number of degrees of freedom ($\Delta\text{ndf}=4$), the statistical significance for the three-resonance assumption over the two-resonance assumption  is 5.7$\sigma$. In addition, we also fit the cross sections with the coherent sum of three relativistic Breit-Wigner functions and a three-body PHSP term. This assumption improves the fit quality, but the change of the likelihood value is only $|\Delta(-2\,\ln L)| = 1.2$, which indicates the PHSP term does not make an important contribution.

The systematic uncertainties in the cross section measurement mainly come from the luminosity, tracking efficiency, kinematic fit, MUC response, MC model, radiative correction and branching fraction. The integrated luminosity is measured using Bhabha events with an uncertainty of 0.6\%~\cite{BESIII:2020nme}. The uncertainty of tracking efficiency for the high-momentum leptons is 1.0\% per track. By requiring at least one kaon to be detected, the kaon detection efficiency is very high and the uncertainty is negligible. A track helix parameter
correction method, as discussed in Ref.~\cite{BESIII:2012mpj}, is applied to MC events during the kinematic fit. The difference in efficiencies with and without the correction, 2.1\%, is assigned as the systematic uncertainty from the kinematic fit. The uncertainty from the MUC response is investigated using the $\EE\to\MM$ data sample, and the difference in efficiencies between the data and MC simulation due to the hit depth requirement in the MUC for the muon candidates, 2.0\%, is taken as the systematic uncertainty. Instead of using pure PHSP to model the $\kk\jpsi$ events, we also consider possible $f\to\kk$ in MC simulation based on a partial wave analysis~\cite{Supplemental}, {\it e.g.},~$f_0(980)$, $f_0(1500)$. The efficiency difference between this model and a three-body PHSP one is 5.9\% (8.7\%) for the data with $\sqrt{s}\leq 4.70\gev$ ($\sqrt{s}> 4.70\gev$). 
To estimate the systematic uncertainty from the ISR correction factor,
we replace the description of the default dressed cross section line shape with the coherent sum of three relativistic Breit-Wigner functions and a PHSP function, or change the parameterization of the Breit-Wigner function. The maximum difference due to the line shape, 0.6\%, is assigned as the systematic uncertainty due to the ISR correction factor.

The uncertainty from the branching fraction of $\jpsi\to\LL$ (0.4\%) is taken from the Particle Data Group~\cite{Workman:2022ynf}. 
Assuming all the sources are independent, the total systematic uncertainty is calculated by adding them in quadrature, resulting in 6.9\% (9.4\%) for the cross section measurement at $\sqrt{s}\leq 4.70\gev$ ($\sqrt{s}> 4.70\gev$).
	
The systematic uncertainties in the resonance parameters mainly come from the absolute c.m.~energy measurement, the c.m.~energy spread, 
the parameterization of the fit function, and the systematic uncertainty on the cross section measurement. The systematic uncertainty of the c.m.~energy is common for all the energies and will propagate to the mass measurement directly. The uncertainty from the c.m.~energy spread is estimated by convolving the fit formula with a Gaussian function, whose width is set as the energy dependent beam spread~\cite{Abakumova:2011rp}.  
To estimate the uncertainty from the parameterization of the Breit-Wigner function, the $\Gamma_{j}$ in the denominator of the Breit-Wigner function is replaced with a mass dependent width $\Gamma_{j}\frac{\Phi(\sqrt{s})}{\Phi(M_{j})}$. The uncertainties from the cross section measurement are divided into two parts. The first one is uncommon uncertainties of the measured cross sections among the different c.m.~energies, which mainly come from the MC model. The corresponding uncertainty is estimated by including the uncommon uncertainties in the dressed cross section fit, and the differences on the parameters are taken as the corresponding uncertainties. The second part, including all the other uncertainties of the measured cross sections, is common for all the energies, and only affects the parameter $(\Gamma_{ee}\mathcal{B})$. 

Intermediate states decaying into $K^+J/\psi$, denoted as $Z_{cs}^+$, are of great interest due to their exotic nature. The distribution of the maximum of the invariant masses of the $K^+J/\psi$ and $K^-J/\psi$ combinations, $M_{\rm max}(K^\pm J/\psi)$, is useful for identifying these exotic states~\cite{BESIII:2013ris}. A simultaneous fit to the $M_{\rm max}(K^\pm J/\psi)$ spectra from datasets within the energy region $4.63\gev\leq\sqrt{s}\leq 4.92\gev$ is performed. The data sets at $\sqrt{s}=4.61\gev$ and $4.95\gev$ are not included in the fit due to their relatively low statistics. In the fit, following the model in Ref.~\cite{BESIII:2022qzr}, the $Z_{cs}^+$ component is modeled by the product of an $S$-wave Breit-Wigner shape with a mass-dependent width:
	\begin{equation}
		\mathcal{F}(M) \propto\left|\frac{\sqrt{q \cdot p}}{M^2-m_0^2+i m_0 \Gamma(M)}\right|^2
	\end{equation}
where $\Gamma(M)=\Gamma_0 \cdot\left(p / p^*\right) \cdot\left(m_0 / M\right)$, 
$M$ is the reconstructed mass, $m_{0}$ is the resonance mass, $\Gamma_{0}$ is the width, 
$q$ is the bachelor $K^{-}$ momentum in the initial $\ee$ system,
and $p$ is the $K^{+}$ momentum in the rest frame of the $K^{+}J/\psi $ system. 
To account for detector resolution and efficiency, we use an efficiency-weighted $\mathcal{F}$ convolved with a resolution function based on MC simulation in the fits.
Inspecting the Dalitz plots shown in the Supplemental Material~\cite{Supplemental},
we add shapes from PHSP MC simulation for the fits at $\sqrt{s}\leq4.70\gev$ as no intermediate state is evident. For the fits at $\sqrt{s}>4.70\gev$, we add shapes for $f$ states according to partial wave analysis~\cite{Supplemental}. Both shapes are derived from a kernel estimation~\cite{Cranmer:2000du} thus no free parameters are introduced in these shapes. The $m_0$ and $\Gamma_0$ of the $Z_{cs}$ as well as the yields of all components are free. The fit results are shown in Fig.~\ref{fig:fit Zcs}, with a small excess of $Z_{cs}$ over other components. The fitted mass and width are $4.044\pm0.006\,\mathrm{GeV}/c^{2}$ and $0.036\pm0.016\,\mathrm{GeV}$, respectively. The uncertainties are statistical only. By comparing fits with and without $Z_{cs}$ components, which gives $|\Delta(-2\ln L)|=23.8$ and $\Delta\text{ndf}=12$, the statistical significance is determined to be $2.3\sigma$. Despite no significant $Z_{cs}$, upper limits on the production of the $Z_{cs}(3985)^{+}$ and $Z_{cs}(4000)^{+}$ are of interest to further understand their properties and search for them in future experiments. The upper limits at 90\% confidence level for $\sigma^{\rm Born}(e^+e^-\rightarrow K^-Z_{cs}(3985)^{+})\times\mathcal{B}(Z_{cs}(3985)^+\rightarrow K^+ J/\psi)$ are $\mathcal{O}(1)$ pb and the upper limits for $\sigma^{\rm Born}(e^+e^-\rightarrow K^-Z_{cs}(4000)^{+})\times\mathcal{B}(Z_{cs}(4000)^+\rightarrow K^+ J/\psi)$ are $\mathcal{O}(3)$ pb. These upper limits include systematic uncertainties and are summarized in the Supplemental Material~\cite{Supplemental}.
	
	\begin{figure}[htbp]
		\centering
		\includegraphics[width = 0.40\textwidth]{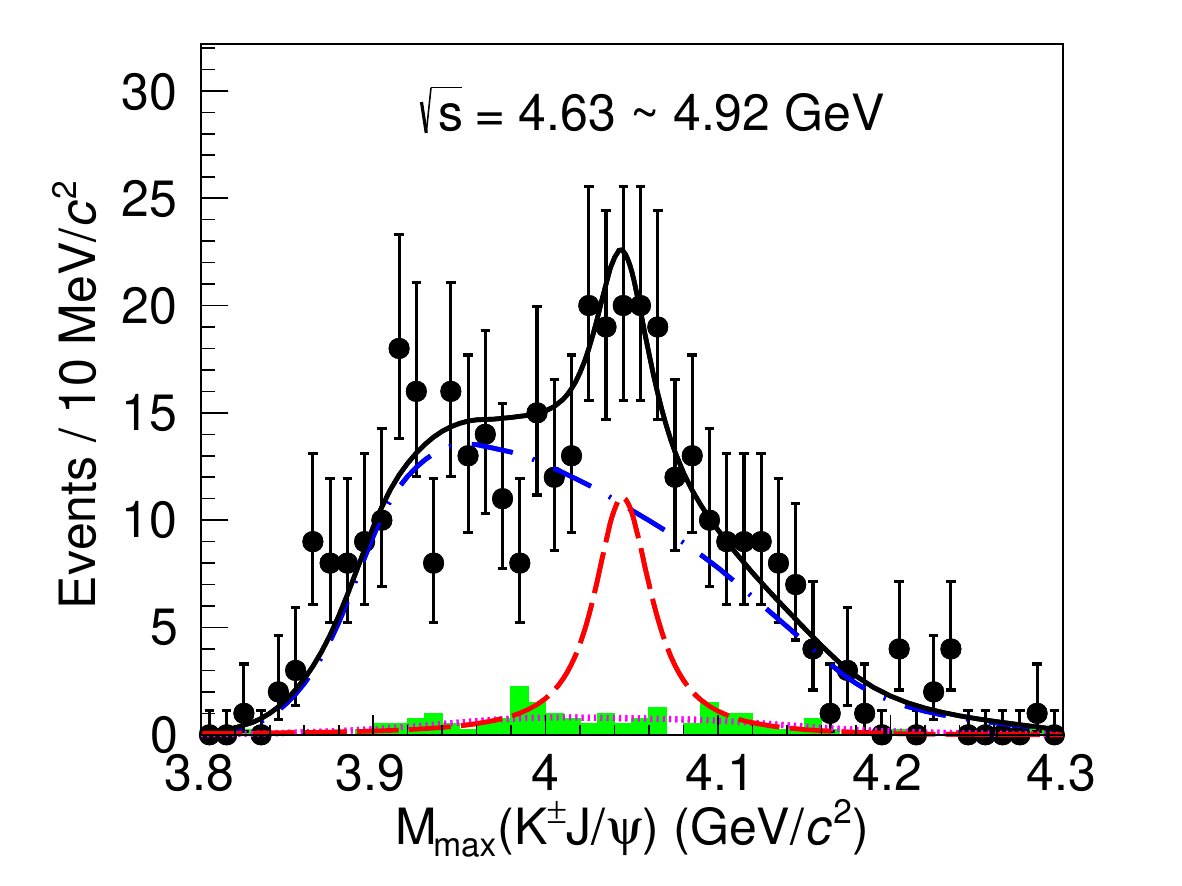}
		\caption{The $M_\mathrm{max}(K^{\pm}J/\psi)$ spectrum with the fit results overlaid. The red dashed line is the signal component of $Z_{cs}$, the blue dash-dotted line is PHSP (or $f$ states) and the pink dotted line is the combinatorial background. The sideband events are scaled to the fitted size of combinatorial background, which are shown in green.}
		\label{fig:fit Zcs}
	\end{figure}

The upper limit on the ratio of branching fractions 
	\begin{equation}
		R_{B}\equiv\frac{\mathcal{B}(Z_{cs}(3985)^{+}\rightarrow K^+ J/\psi)}{\mathcal{B}(Z_{cs}(3985)^{+}\rightarrow (\bar{D}^{0}D_s^{*+} + \bar{D}^{*0}D_s^+))}
	\end{equation}
is determined at $\sqrt{s}=4.68$ GeV since the $Z_{cs}(3985)^{+}$ is the most significant at this c.m.~energy.  
We extract the distribution of $\sigma^{\rm Born}\times\mathcal{B}(Z_{cs}(3985)^{+}\rightarrow K^+ J/\psi)$ from the smeared likelihood values as shown in Ref.~\cite{Supplemental}, which is denoted as $u(\sigma^{\rm Born}\times\mathcal{B}(Z_{cs}(3985)^{+}\rightarrow K^+ J/\psi))$. We model the distribution of $\sigma^{\rm Born}\times\mathcal{B}(Z_{cs}(3985)^{+}\rightarrow \left(\bar{D}^0D_s^{*+}+ \bar{D}^{* 0}D_s^{+}\right))$ by a Gaussian function $G(\sigma^{\rm Born}\times\mathcal{B}(Z_{cs}(3985)^{+}\rightarrow \left(\bar{D}^0D_s^{*+}+ \bar{D}^{* 0}D_s^{+}\right)))$ of which the mean and width are set to the reported center value and uncertainty~\cite{BESIII:2020qkh}. Then the upper limit $R_{B}^{\rm UL}$ of $R_{B}$ at the 90\% confidence level is derived from the convolution of these two distributions,
    \vspace{-0.6cm}

    \begin{align}
        g(R_{B})\equiv\int u(zR_{B})\cdot G(z)\mathrm{d}z,\\
		\frac{\int_{0}^{R_{B}^{\rm UL}} g(R_{B}) \mathrm{d}R_{B}} {\int_{0}^{+\infty} g(R_{B})\mathrm{d}R_{B}} = 0.9,
    \end{align}
where $g(R_{B})$ is the convolved function and $R_{B}^{\rm UL}$ is determined to be 0.03 by numerical integration.

In summary, the cross sections of $e^+e^-\to K^+K^-J/\psi$ at c.m.~energies between $4.61$ and $4.95~\rm{GeV}$ are measured.  
Fitting the cross sections from this work and Ref.~\cite{BESIII:2022joj} with three resonances ($Y(4230)$, $Y(4500)$, and $Y(4710)$), we obtain the mass and width of the $Y(4710)$ to be $M(4710) = 4708_{-15}^{+17}\pm21$~MeV/$c^{2}$ and $\Gamma(4710) = 126_{-23}^{+27}\pm30$~MeV, where the first uncertainties are statistical and the second systematic.
A new resonance structure $Y(4710)$ is observed with a statistical significance over $5\sigma$, which is one of the heaviest vector charmoniumlike states. Our new results confirm that the structure previously reported as evidence in Ref.~\cite{BESIII:2022kcv} is indeed the $Y(4710)$ resonance.
On the other hand, it is also possible that the $Y(4710)$ is an excited charmonium state (e.g. $5S$ state) predicted by the potential model~\cite{Ding:2007rg,Gui:2018rvv}.
This observation brings new insights into the charmonium(like) states above the open-charm threshold. 

We also investigate the $Z_{cs}$ states in the $K\jpsi$ system, but no significant structure is observed. Thus the upper limits of the product of the Born cross sections $\sigma^B[\EE\to K^- Z_{cs}(3985)^+/K^- Z_{cs}(4000)^+]$ and the branching fraction of $Z_{cs}(3985)^+/Z_{cs}(4000)^+\to K^+ \jpsi$ are determined at 90\% confidence level. The ratio of branching fractions $\frac{\mathcal{B}(Z_{cs}(3985)^{+}\rightarrow K^+ J/\psi)}{\mathcal{B}(Z_{cs}(3985)^{+}\rightarrow (\bar{D}^{0}D_s^{*+} + \bar{D}^{*0}D_s^+))}$ is determined to be less than 0.03 at 90\% confidence level.
The suppression of the decay $Z_{cs}(3985)^+\rightarrow K^+ J/\psi$ disfavors the QCD sum rule calculation under the molecular state assumption in Ref.~\cite{ZcsKJpsi1}. 
It supports the $Z_{cs}(3985)^+$ and $Z_{cs}(4000)^+$ as two different states~\cite{Meng:2021rdg}. Our measurements provide important inputs for the understanding of the nature of the $Z_{cs}(3985)$.
To further improve studies of the potential $Z_{cs}$ state, more statistics are necessary to conduct a partial wave analysis.

The BESIII Collaboration thanks the staff of BEPCII and the IHEP computing center for their strong support. This work is supported in part by National Key R\&D Program of China under Contracts Nos. 2020YFA0406300, 2020YFA0406400; National Natural Science Foundation of China (NSFC) under Contracts Nos. 11635010, 11735014, 11835012, 11935015, 11935016, 11935018, 11961141012, 12022510, 12025502, 12035009, 12035013, 12061131003, 12192260, 12192261, 12192262, 12192263, 12192264, 12192265, 12221005, 12225509, 12235017; the Chinese Academy of Sciences (CAS) Large-Scale Scientific Facility Program; the CAS Center for Excellence in Particle Physics (CCEPP); Joint Large-Scale Scientific Facility Funds of the NSFC and CAS under Contract No. U1832207; CAS Key Research Program of Frontier Sciences under Contracts Nos. QYZDJ-SSW-SLH003, QYZDJ-SSW-SLH040; 100 Talents Program of CAS; The Institute of Nuclear and Particle Physics (INPAC) and Shanghai Key Laboratory for Particle Physics and Cosmology; ERC under Contract No. 758462; European Union's Horizon 2020 research and innovation programme under Marie Sklodowska-Curie grant agreement under Contract No. 894790; German Research Foundation DFG under Contracts Nos. 455635585, Collaborative Research Center CRC 1044, FOR5327, GRK 2149; Istituto Nazionale di Fisica Nucleare, Italy; Ministry of Development of Turkey under Contract No. DPT2006K-120470; National Research Foundation of Korea under Contract No. NRF-2022R1A2C1092335; National Science and Technology fund of Mongolia; National Science Research and Innovation Fund (NSRF) via the Program Management Unit for Human Resources \& Institutional Development, Research and Innovation of Thailand under Contract No. B16F640076; Polish National Science Centre under Contract No. 2019/35/O/ST2/02907; The Swedish Research Council; U. S. Department of Energy under Contract No. DE-FG02-05ER41374.
		
	\bibliographystyle{apsrev4-2}
	\bibliography{bibitem}
	
\onecolumngrid
\newpage

%
%
%
%
%
%
%
{\large\bf
   Supplemental Material for ``Observation of a vector charmonium-like state at 4.7 \texorpdfstring{${\rm GeV}/c^2$}{GeV/c2} and search for \texorpdfstring{$Z_{cs}$}{Zcs} states in \texorpdfstring{$e^+e^-\to K^+K^-J/\psi$}{KKJpsi}''
 }
	
%
%
	
	
	\maketitle
	
	\oddsidemargin  -0.2cm
	\evensidemargin -0.2cm
	
	\section{\texorpdfstring{$M(\LL)$}{M(ll)} at each center-of-mass energy}
	
	Figure~\ref{fig:mll} shows the $M(\LL)$ at each c.m.~energy ($\sqrt{s}=4.61-4.95$ GeV)
	
	\begin{figure}[H]
		\centering
		\includegraphics[width=0.32\linewidth]{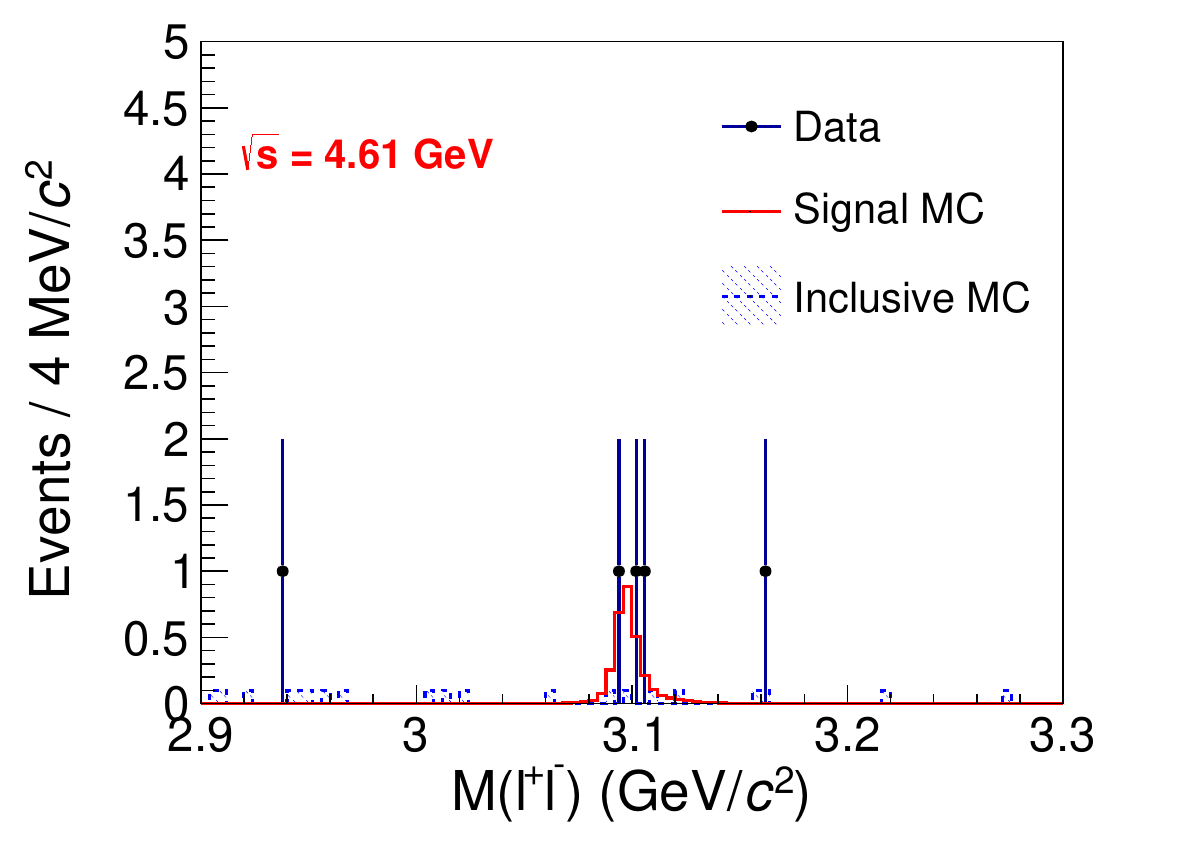}
		\includegraphics[width=0.32\linewidth]{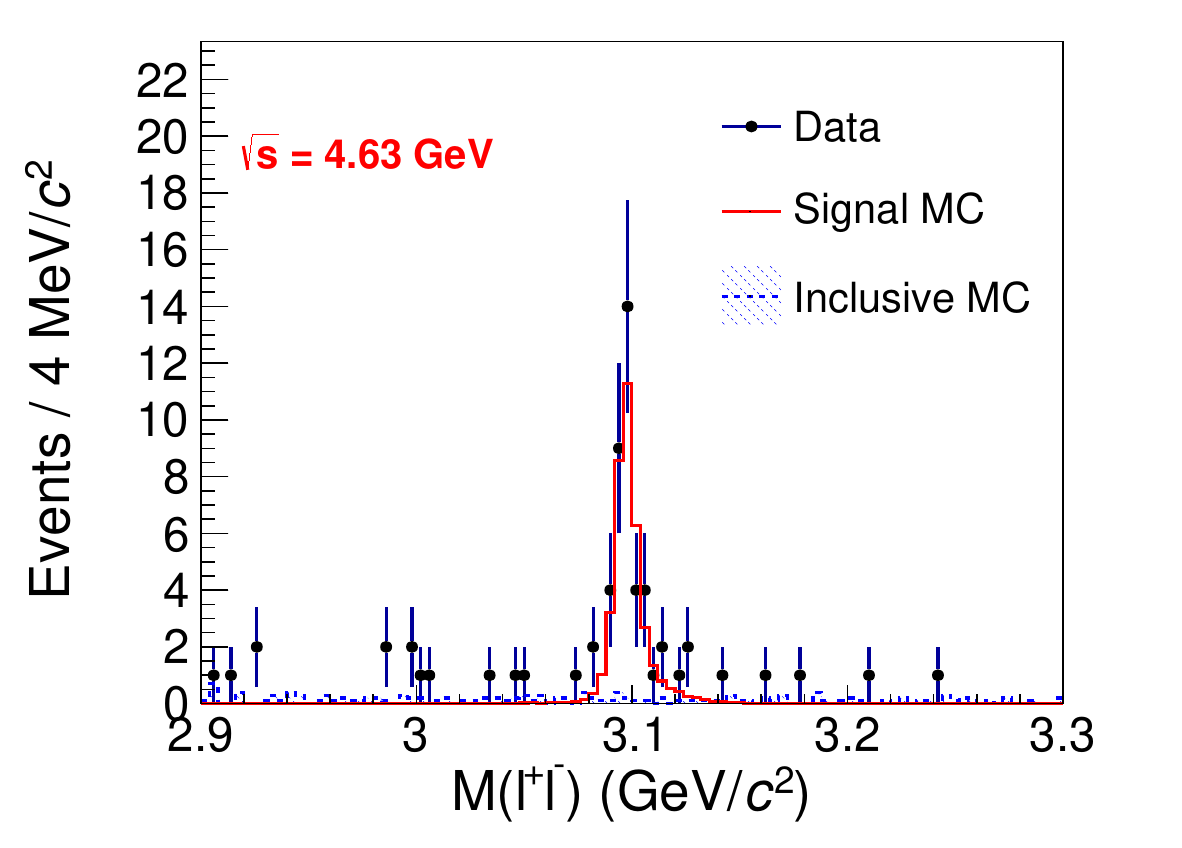}
		\includegraphics[width=0.32\linewidth]{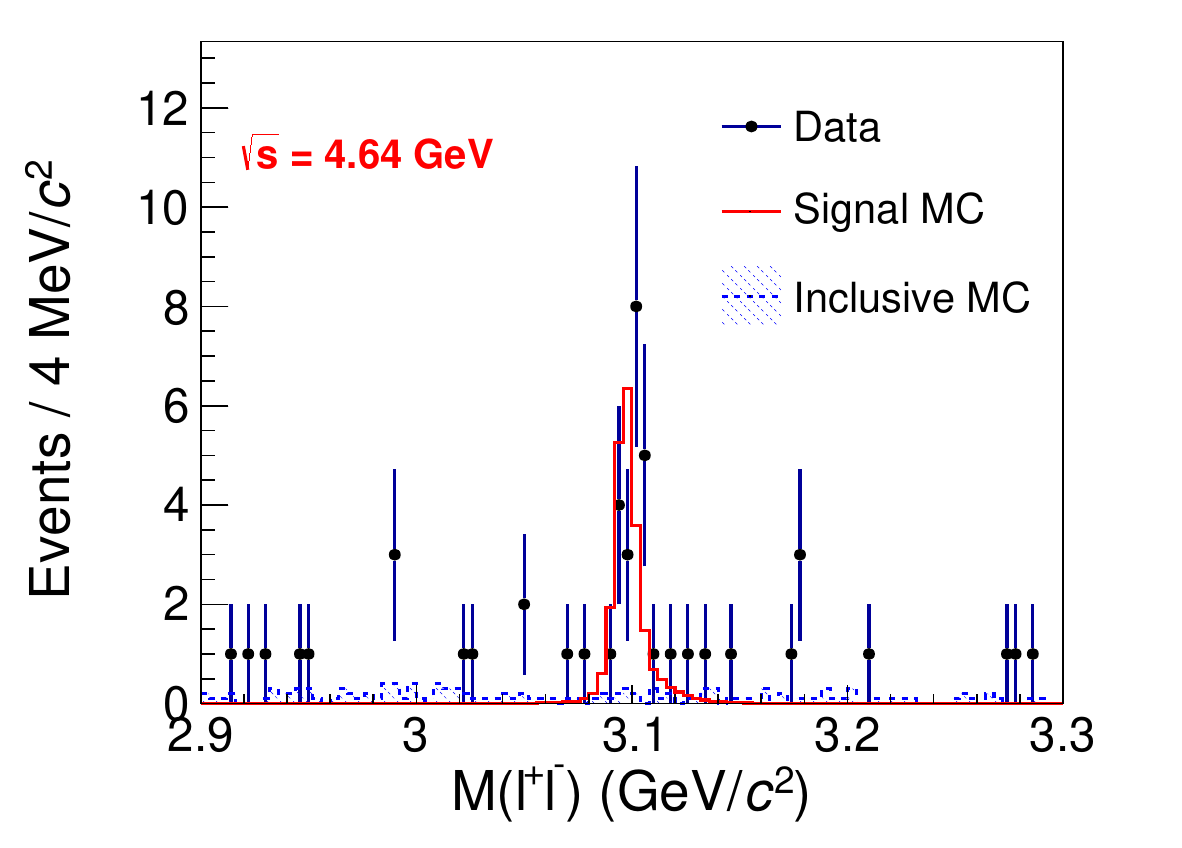}
		\vfill
		\includegraphics[width=0.32\linewidth]{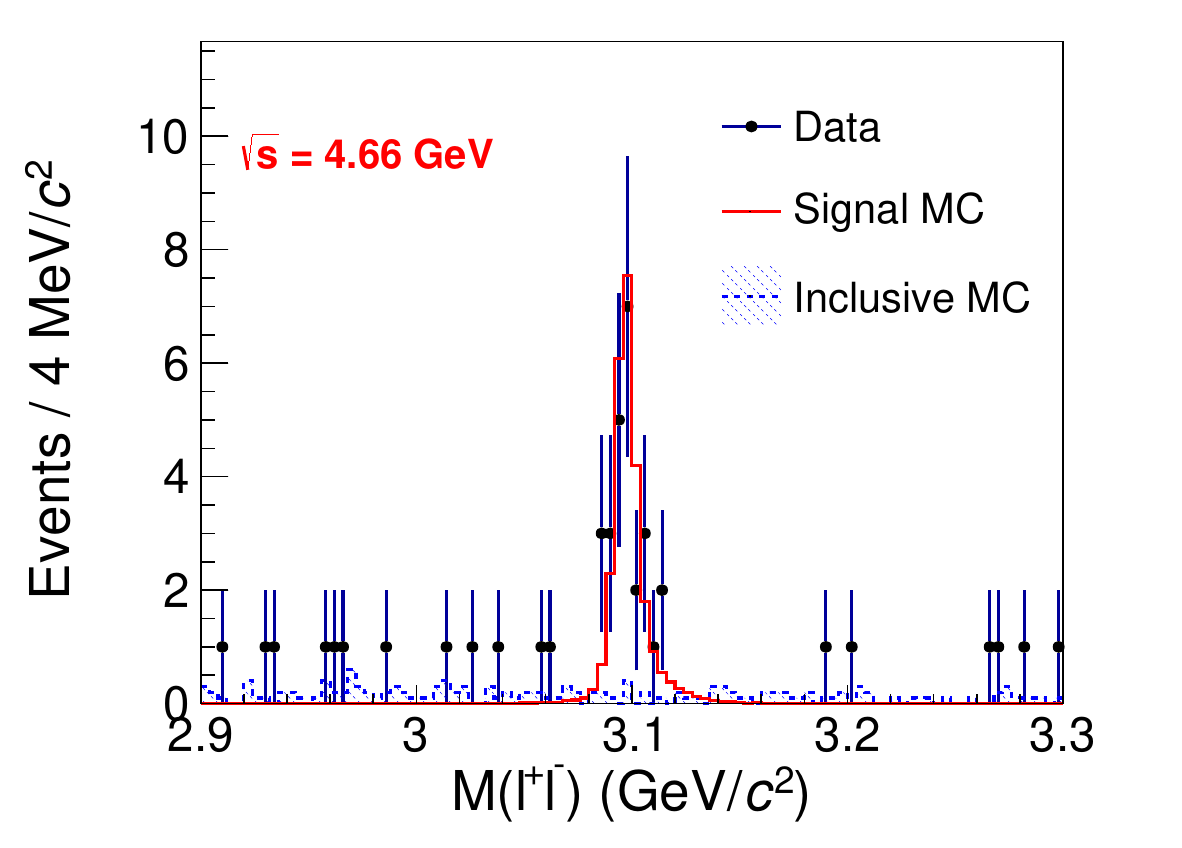}
		\includegraphics[width=0.32\linewidth]{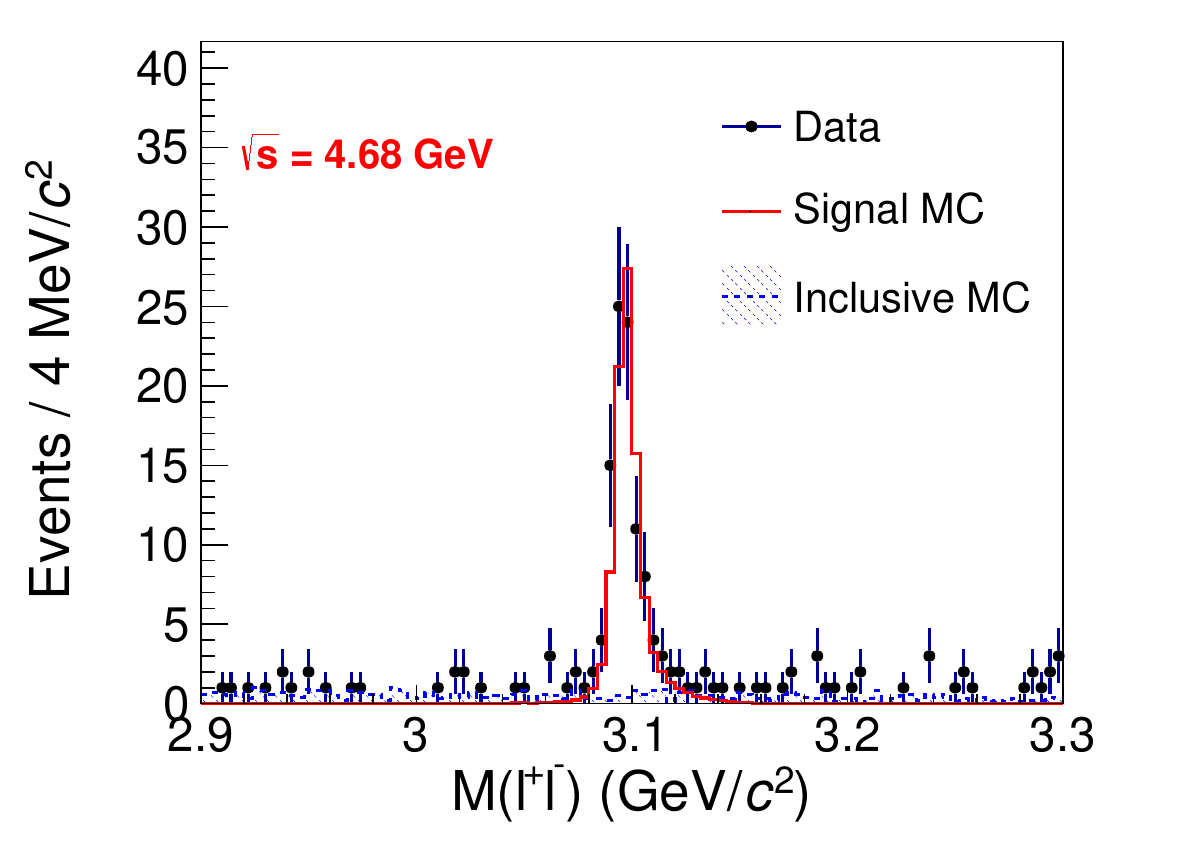}
		\includegraphics[width=0.32\linewidth]{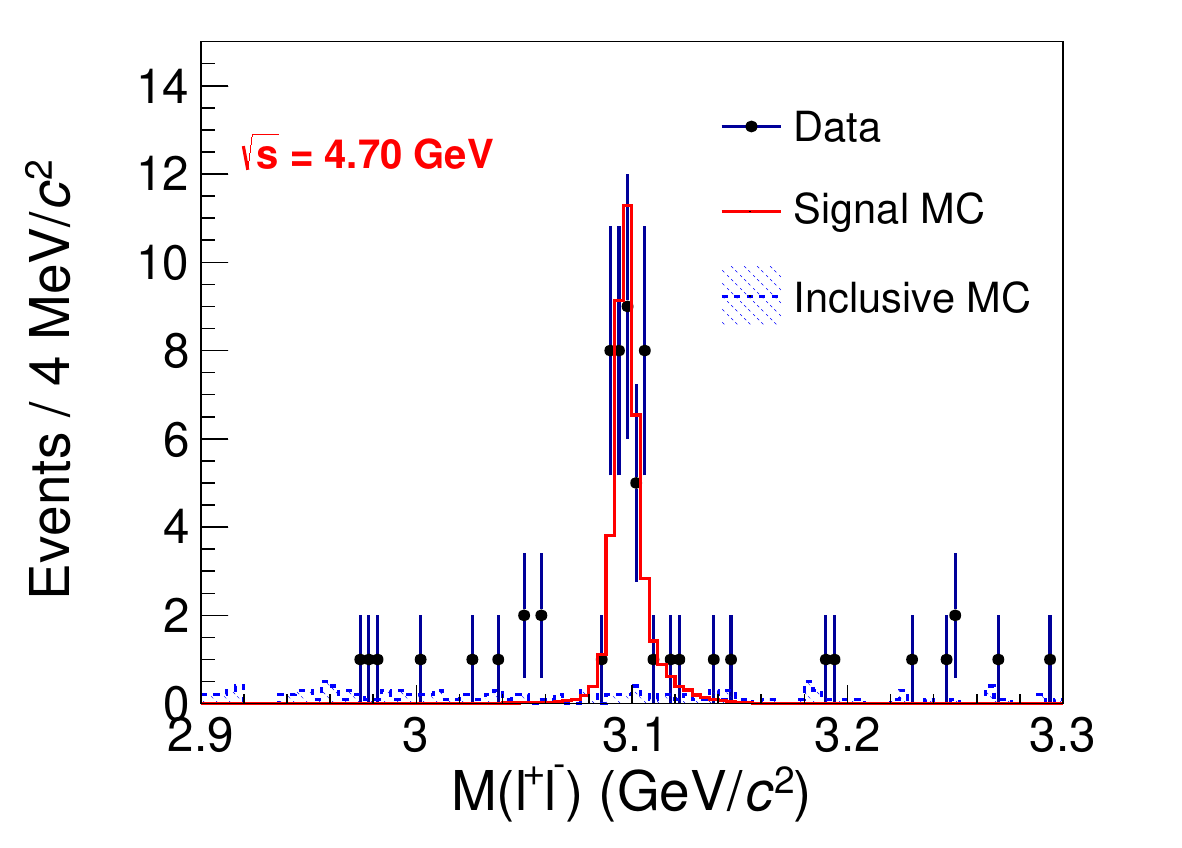}
		\vfill
		\includegraphics[width=0.32\linewidth]{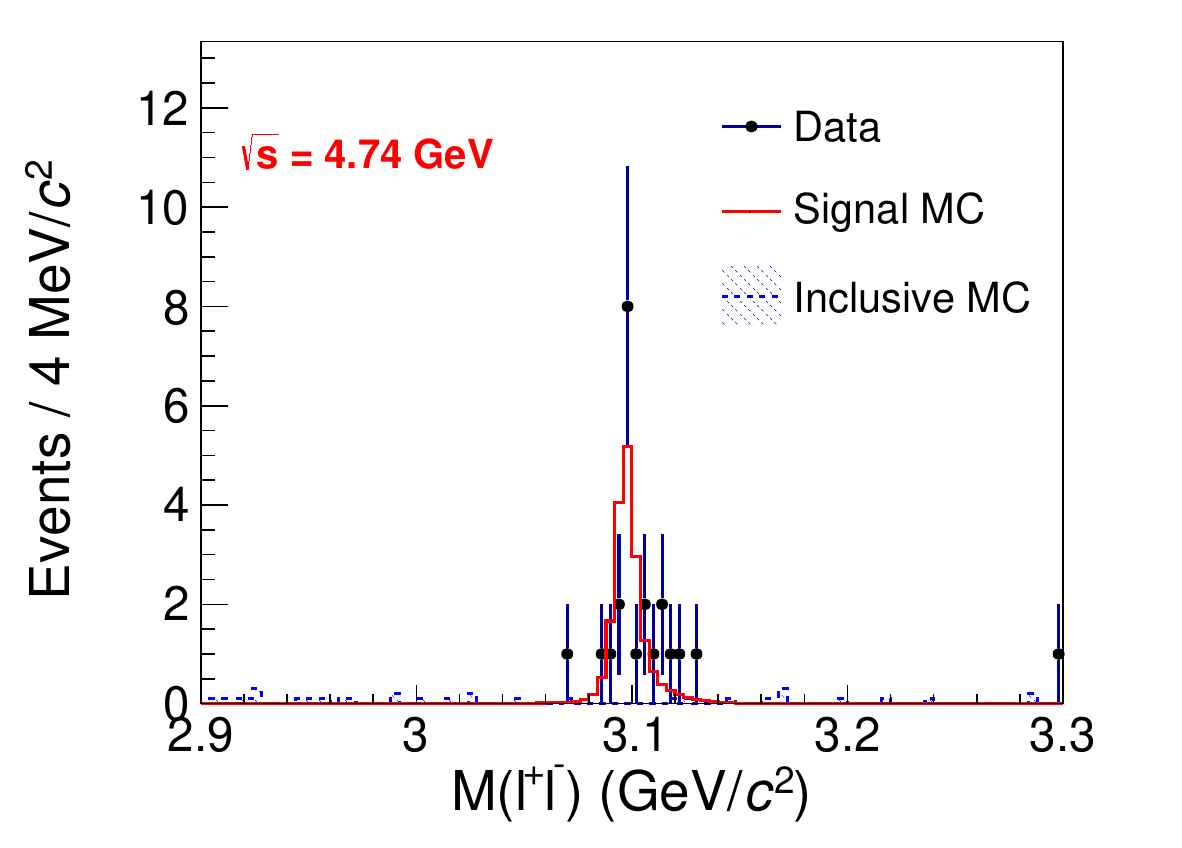}
		\includegraphics[width=0.32\linewidth]{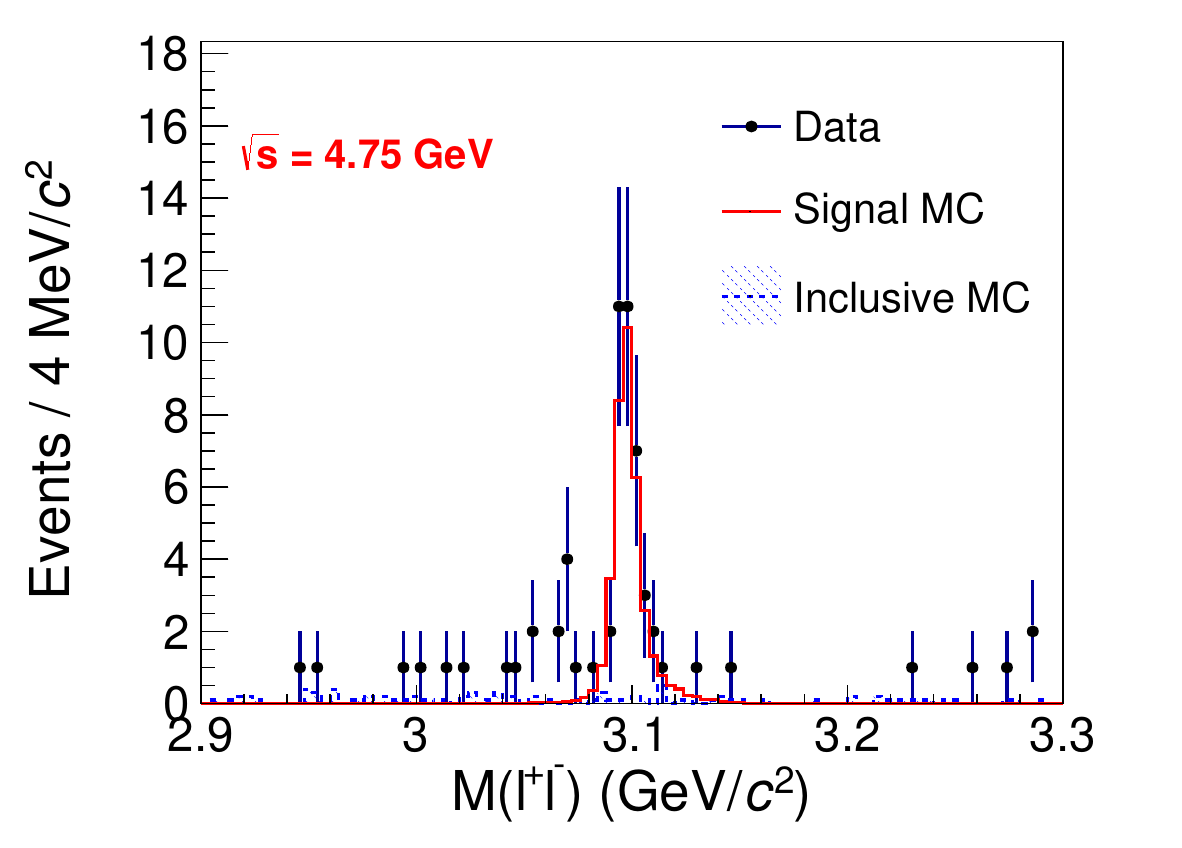}
		\includegraphics[width=0.32\linewidth]{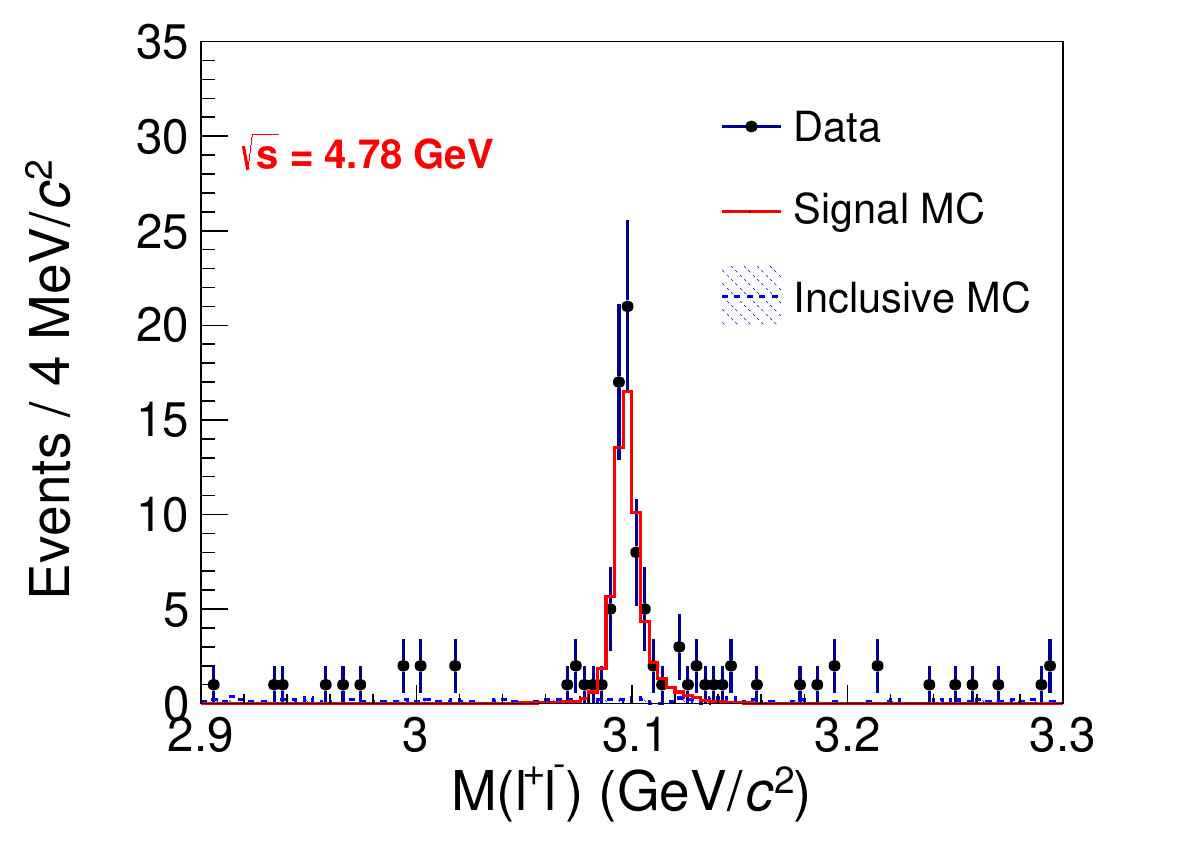}
		\vfill
		\includegraphics[width=0.32\linewidth]{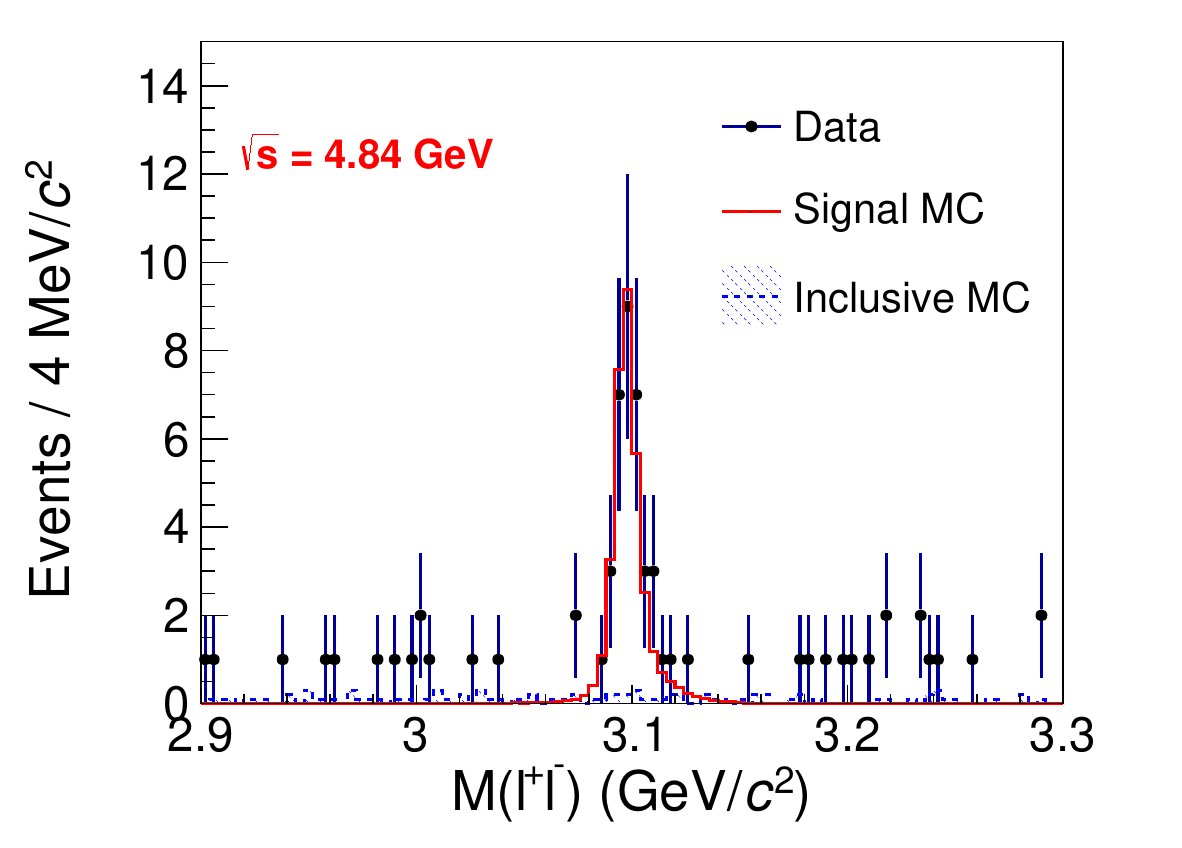}
		\includegraphics[width=0.32\linewidth]{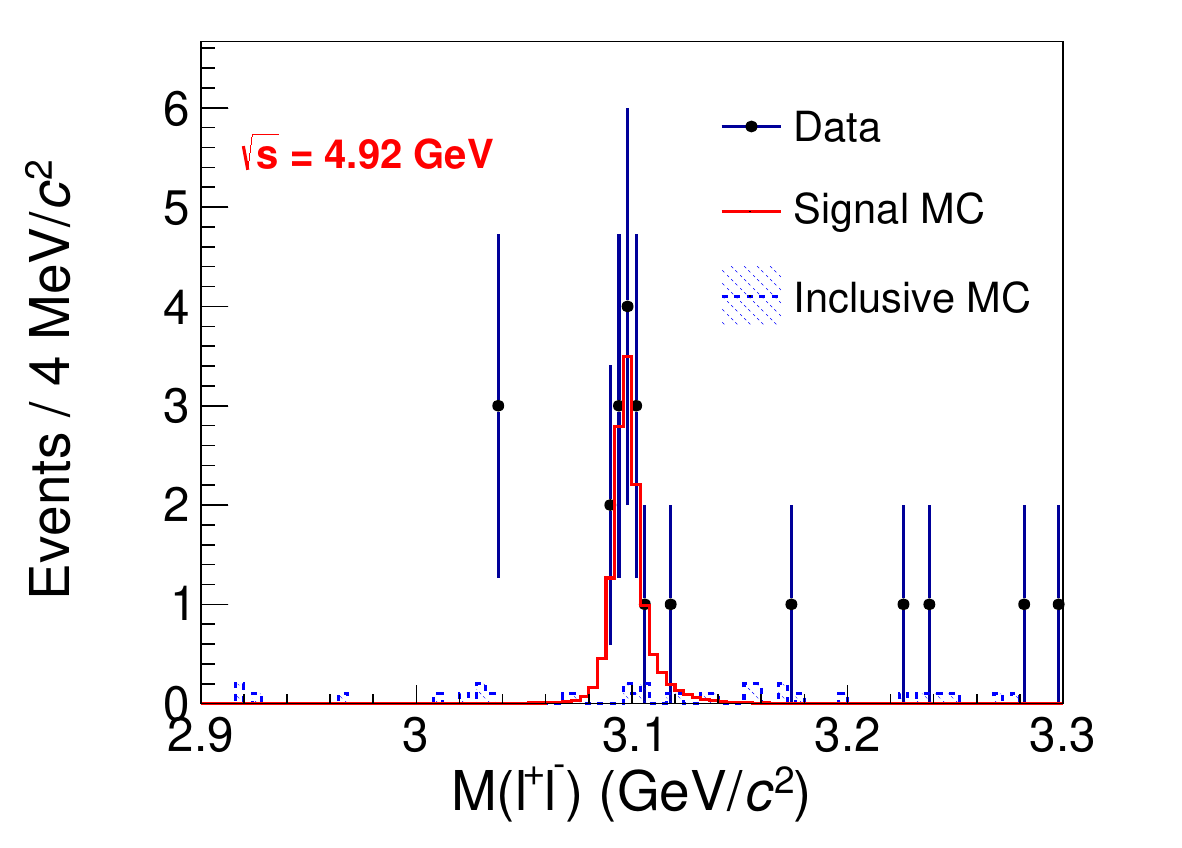}
		\includegraphics[width=0.32\linewidth]{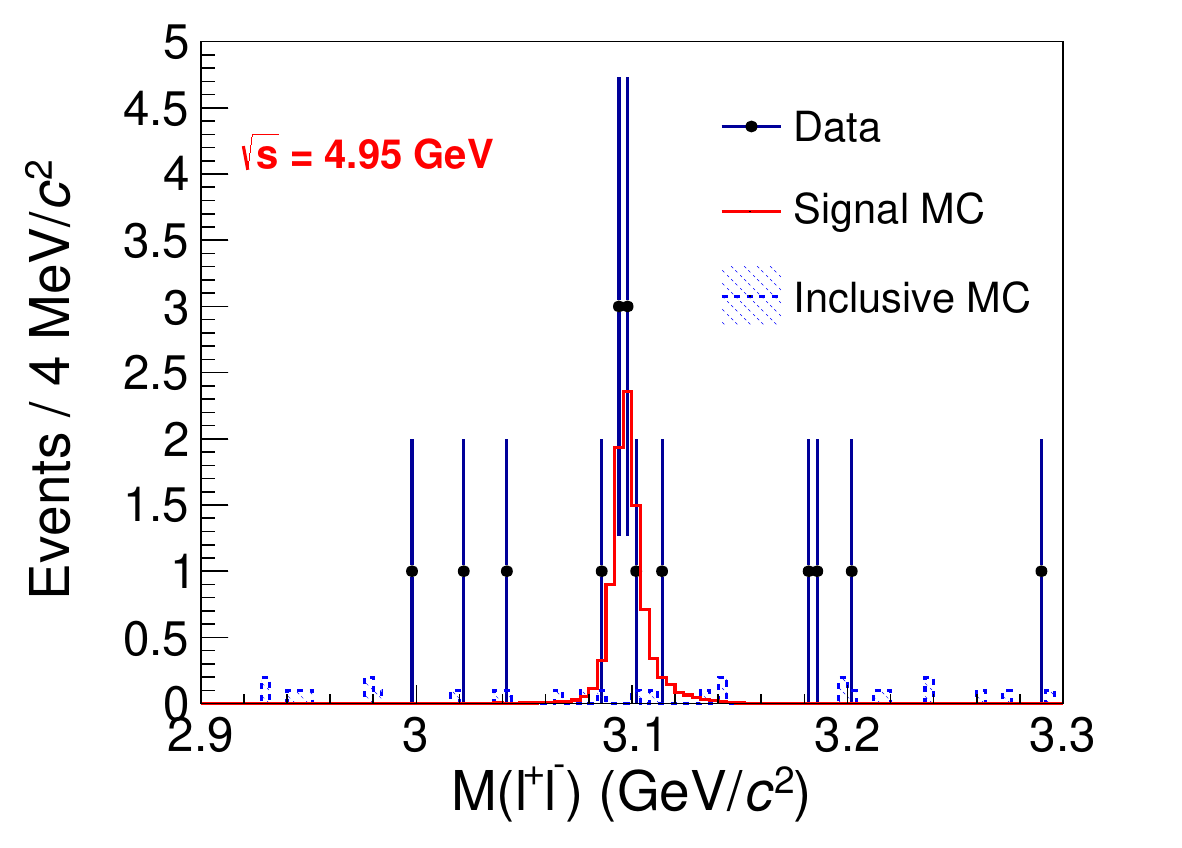}
		
		\caption{The $M(\LL)$ distribution at each c.m.~energy. Dots with error bars are data, the red histograms are signal MC and the blue shaded histograms stand for the background from the inclusive MC samples.}
		\label{fig:mll}
	\end{figure}
	
\begin{figure}[H]
		\centering
		\includegraphics[width=0.32\linewidth]{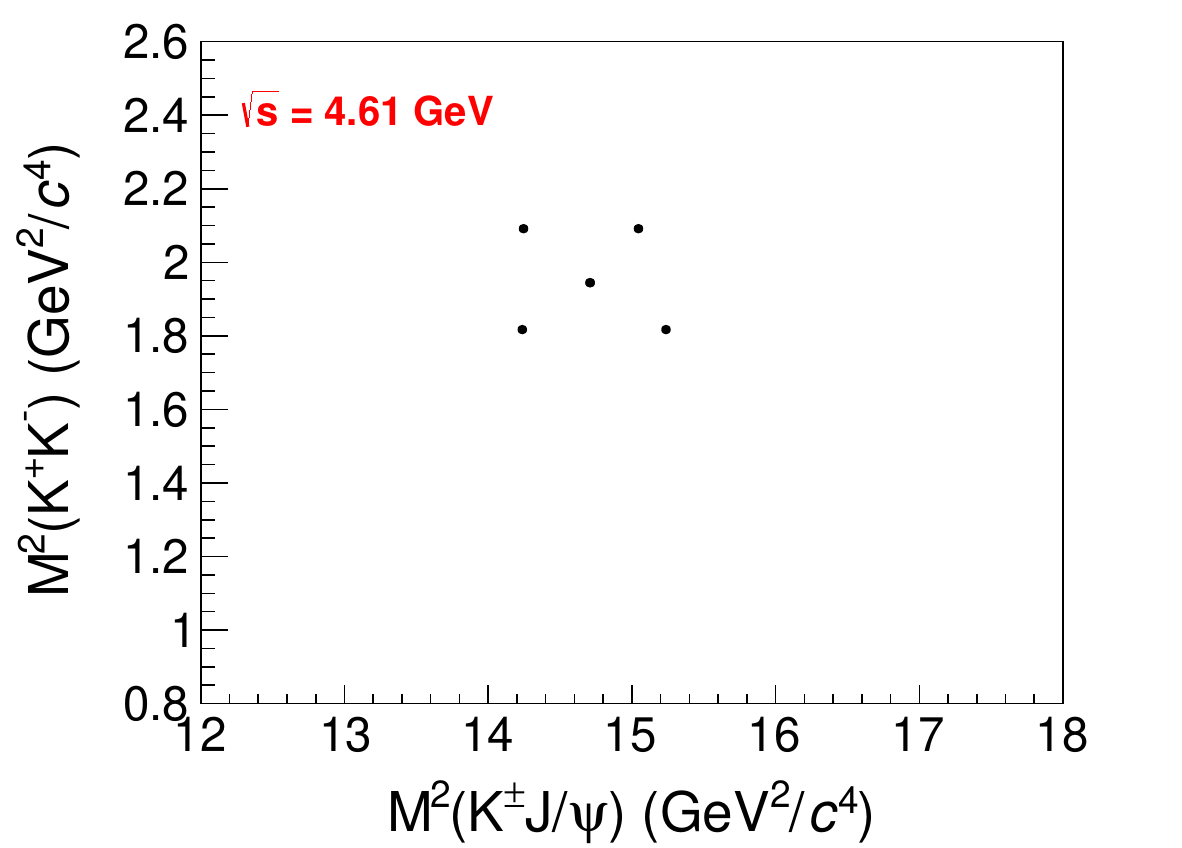}
		\includegraphics[width=0.32\linewidth]{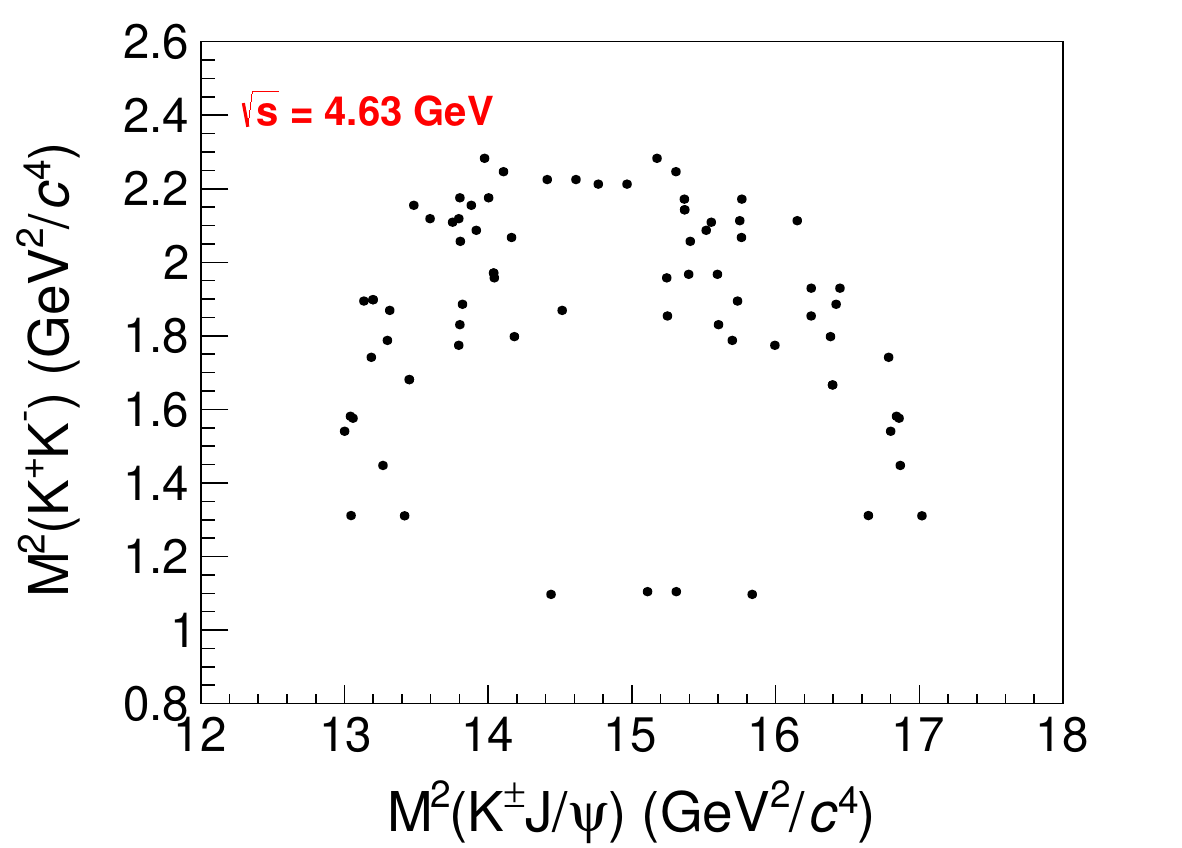}
		\includegraphics[width=0.32\linewidth]{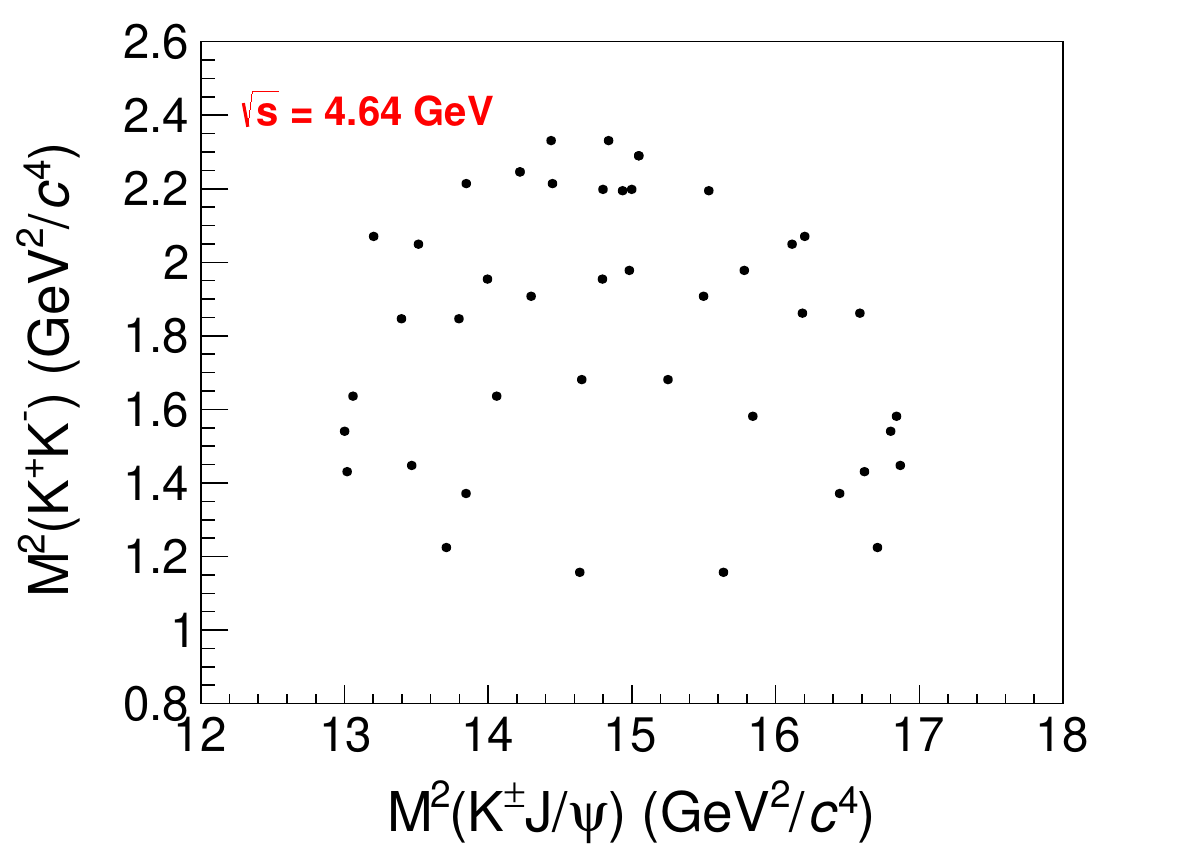}
		
		\includegraphics[width=0.32\linewidth]{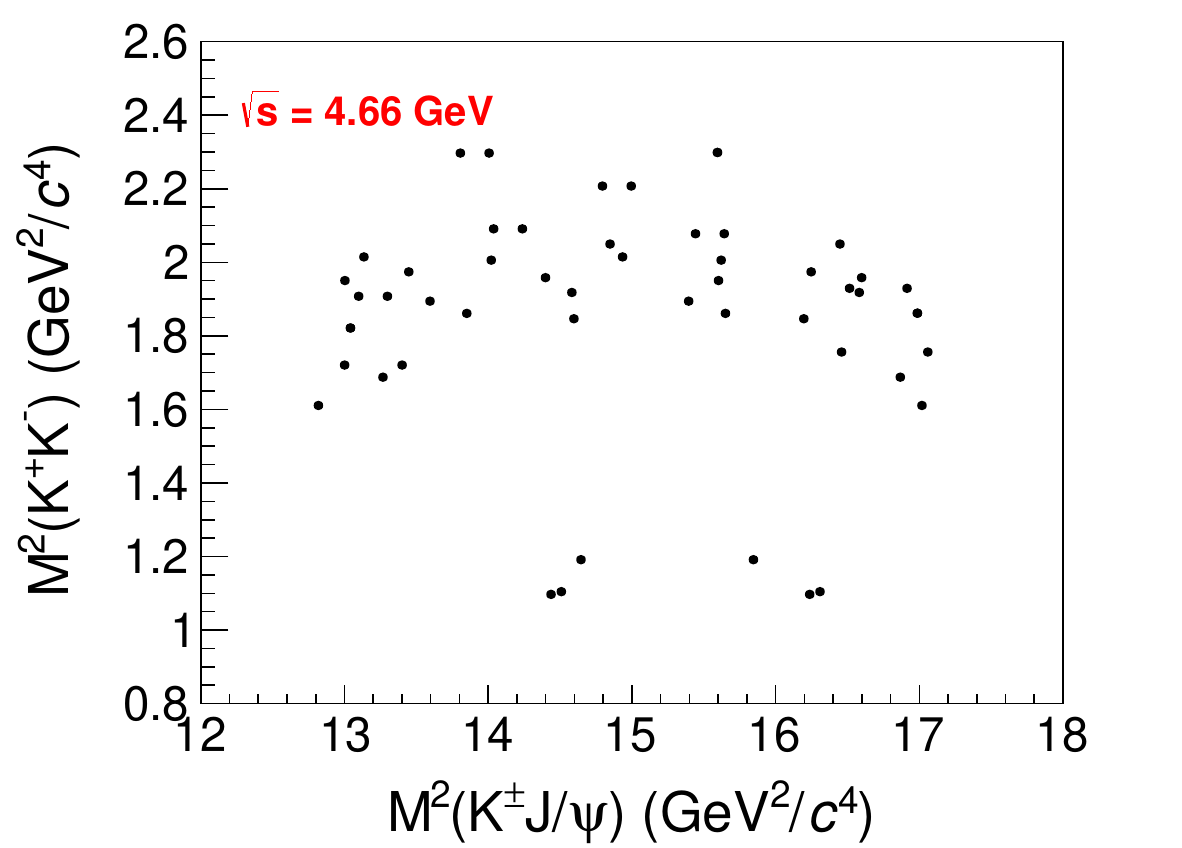}
		\includegraphics[width=0.32\linewidth]{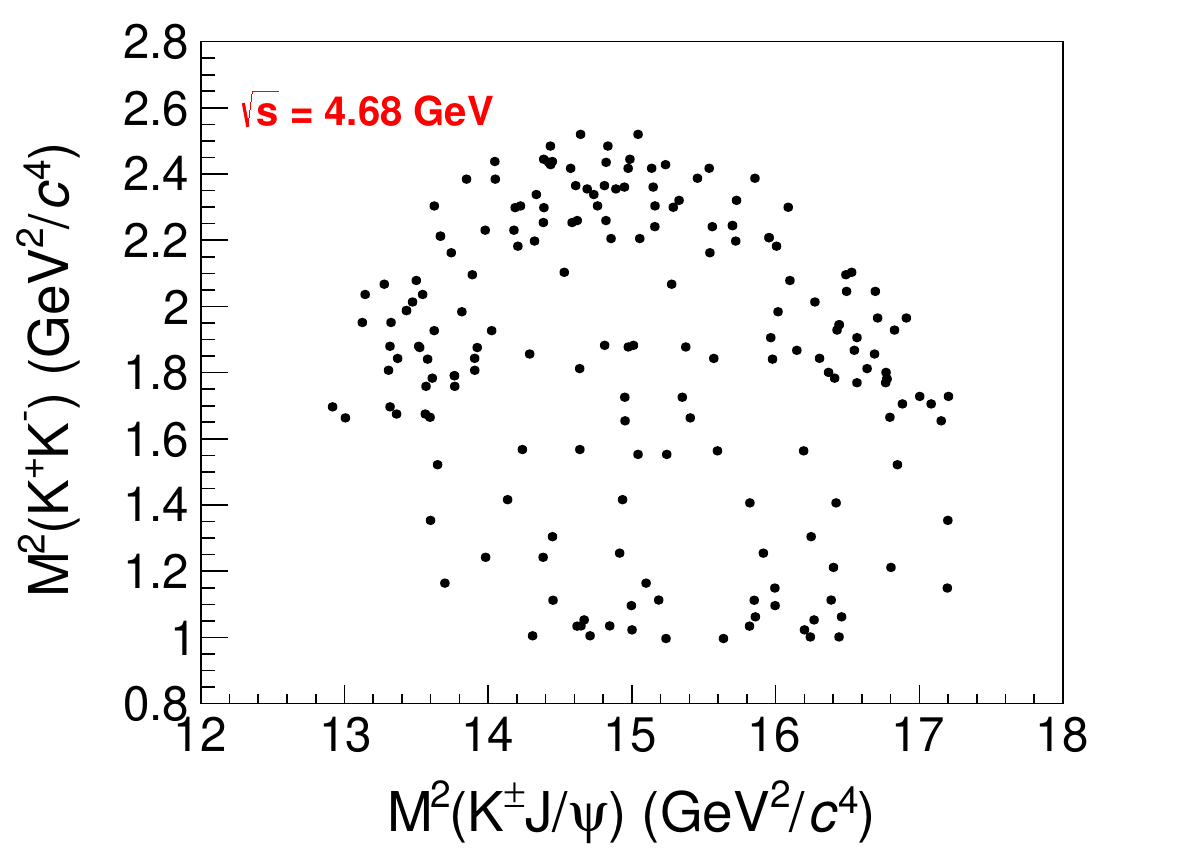}
		\includegraphics[width=0.32\linewidth]{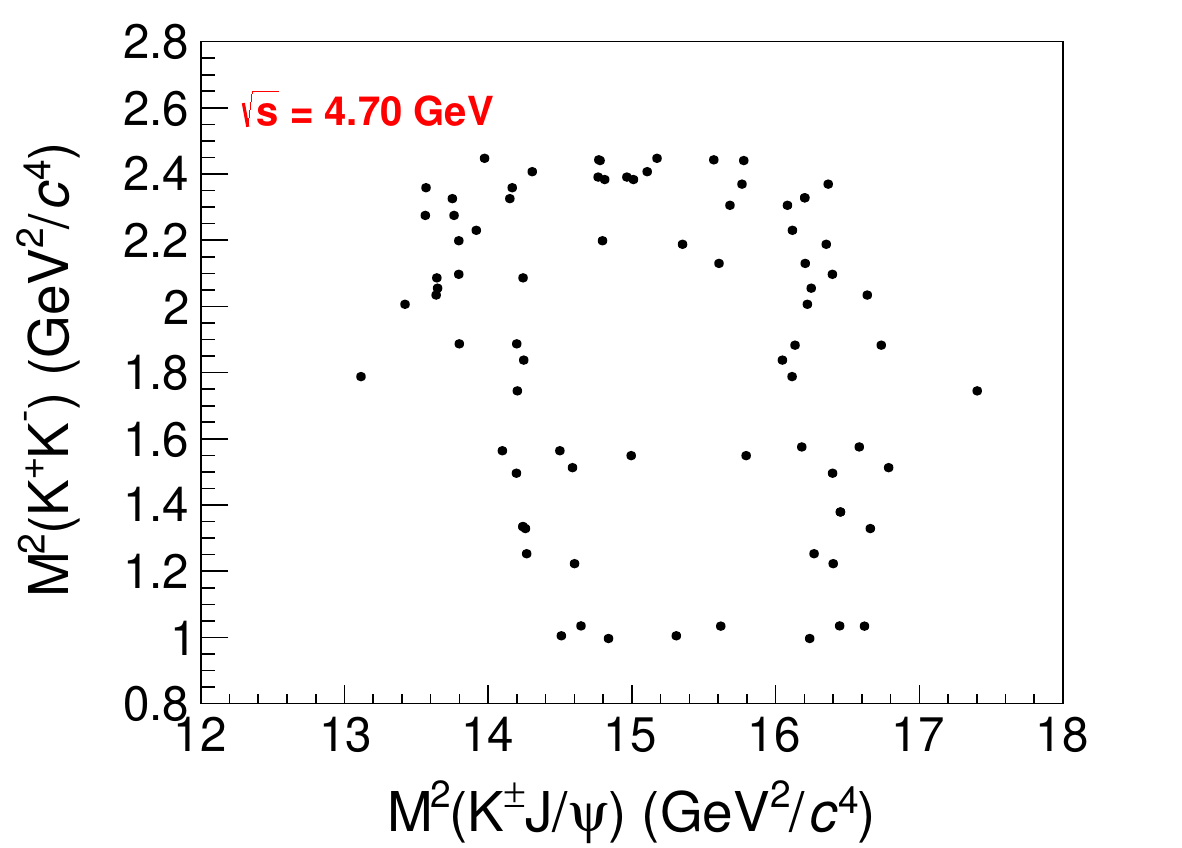}
		
		\includegraphics[width=0.32\linewidth]{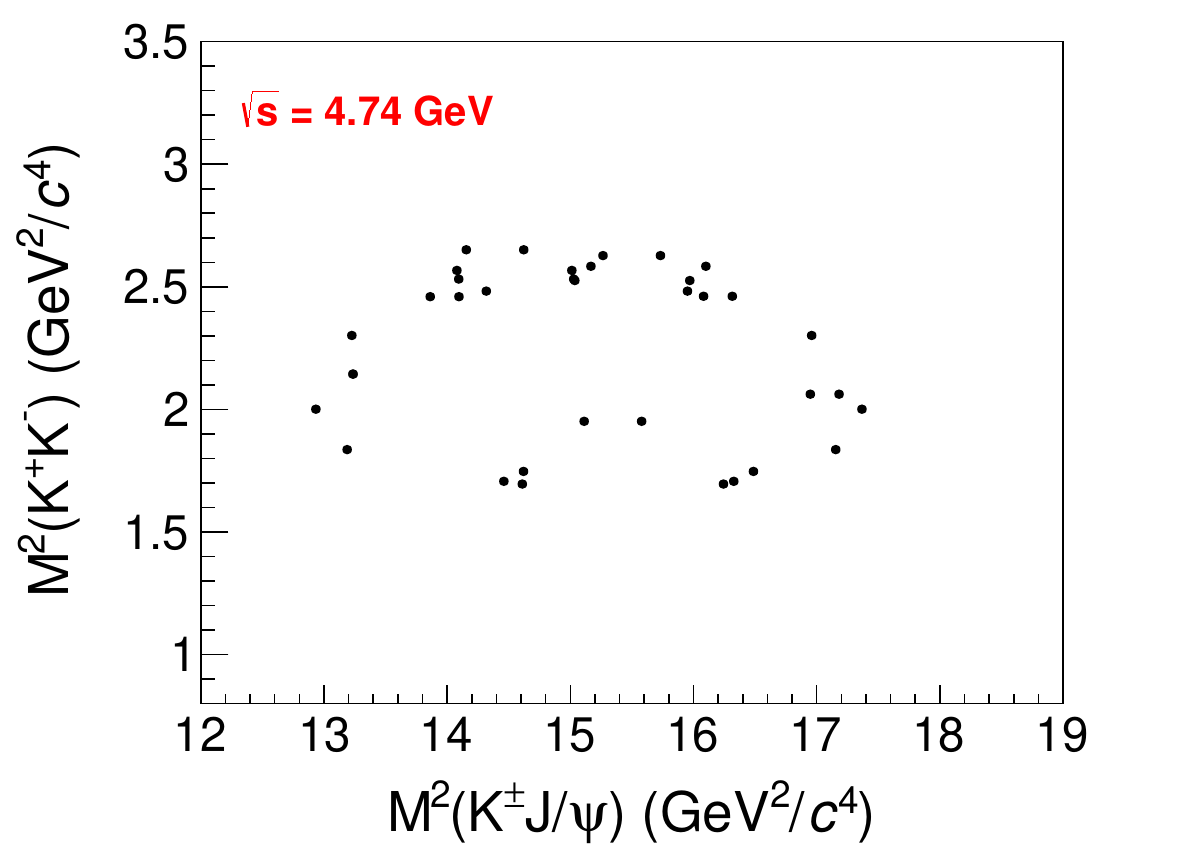}
		\includegraphics[width=0.32\linewidth]{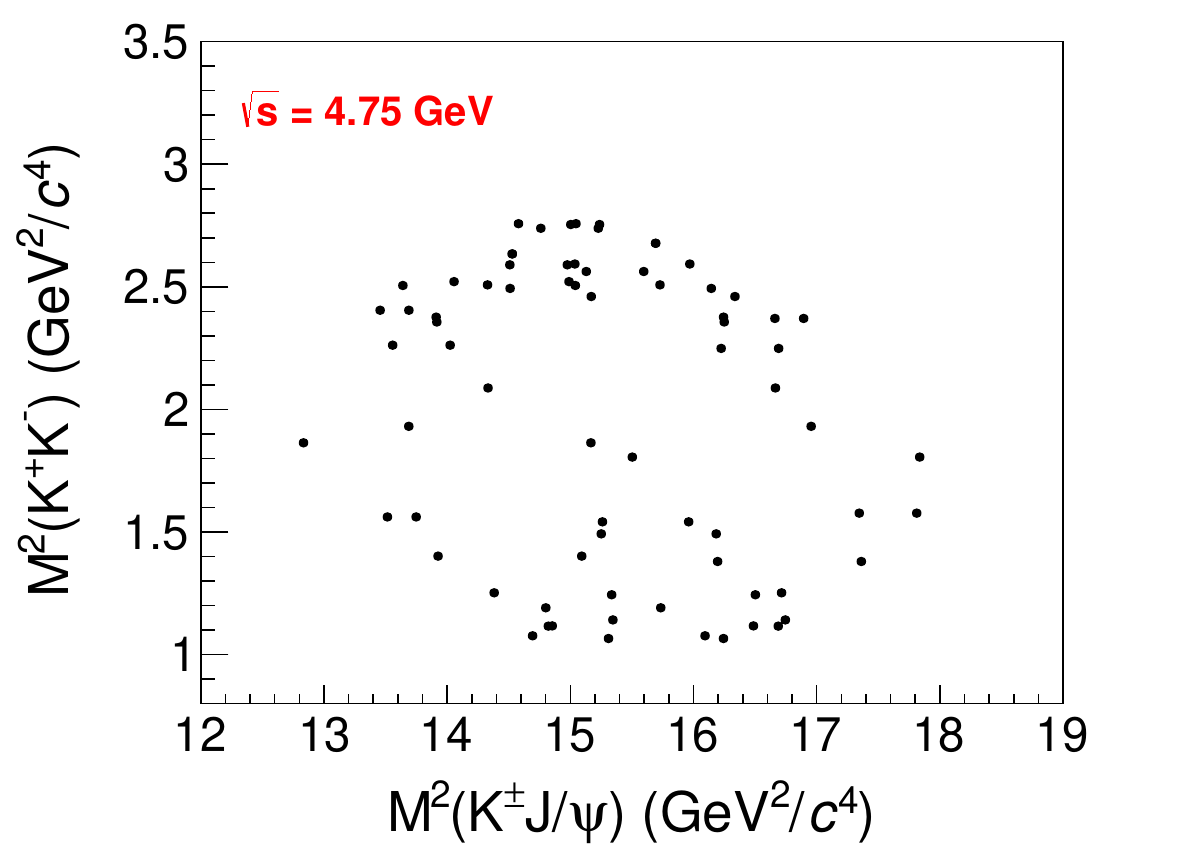}
		\includegraphics[width=0.32\linewidth]{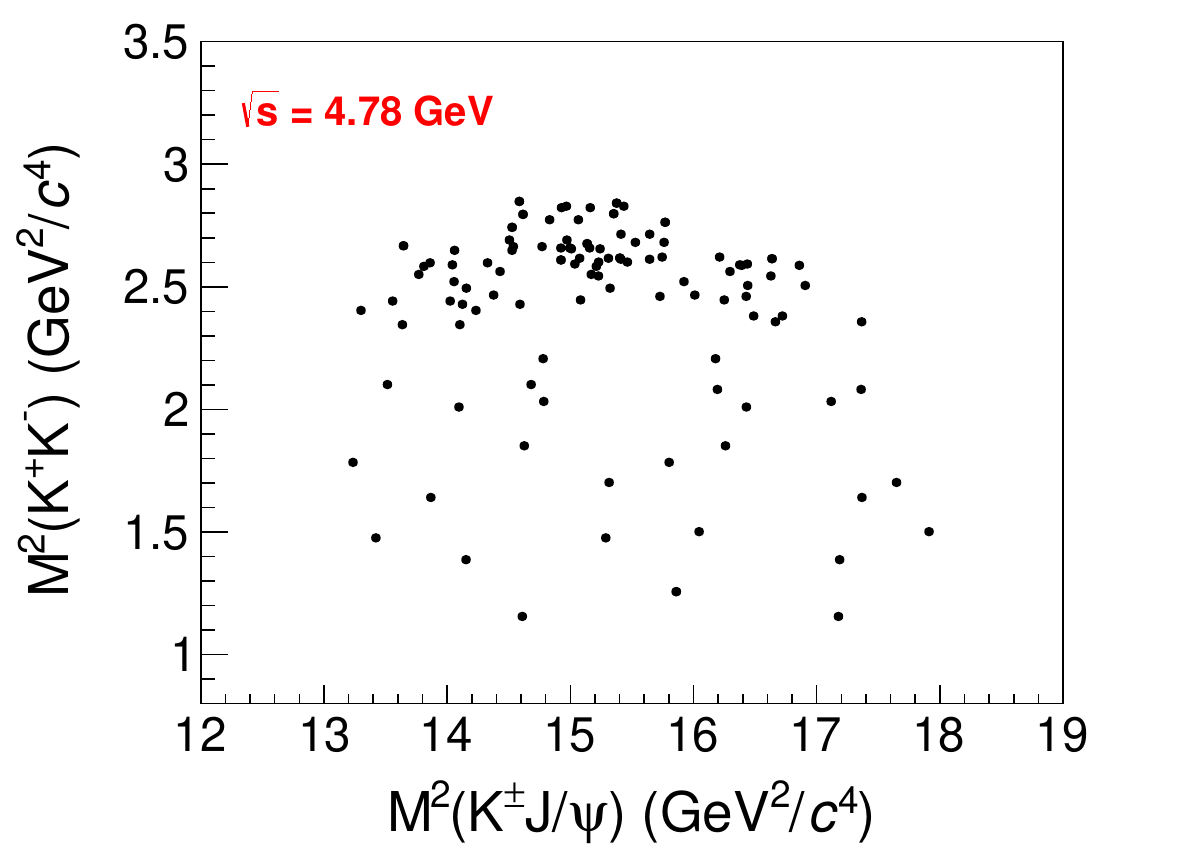} 
		
		\includegraphics[width=0.32\linewidth]{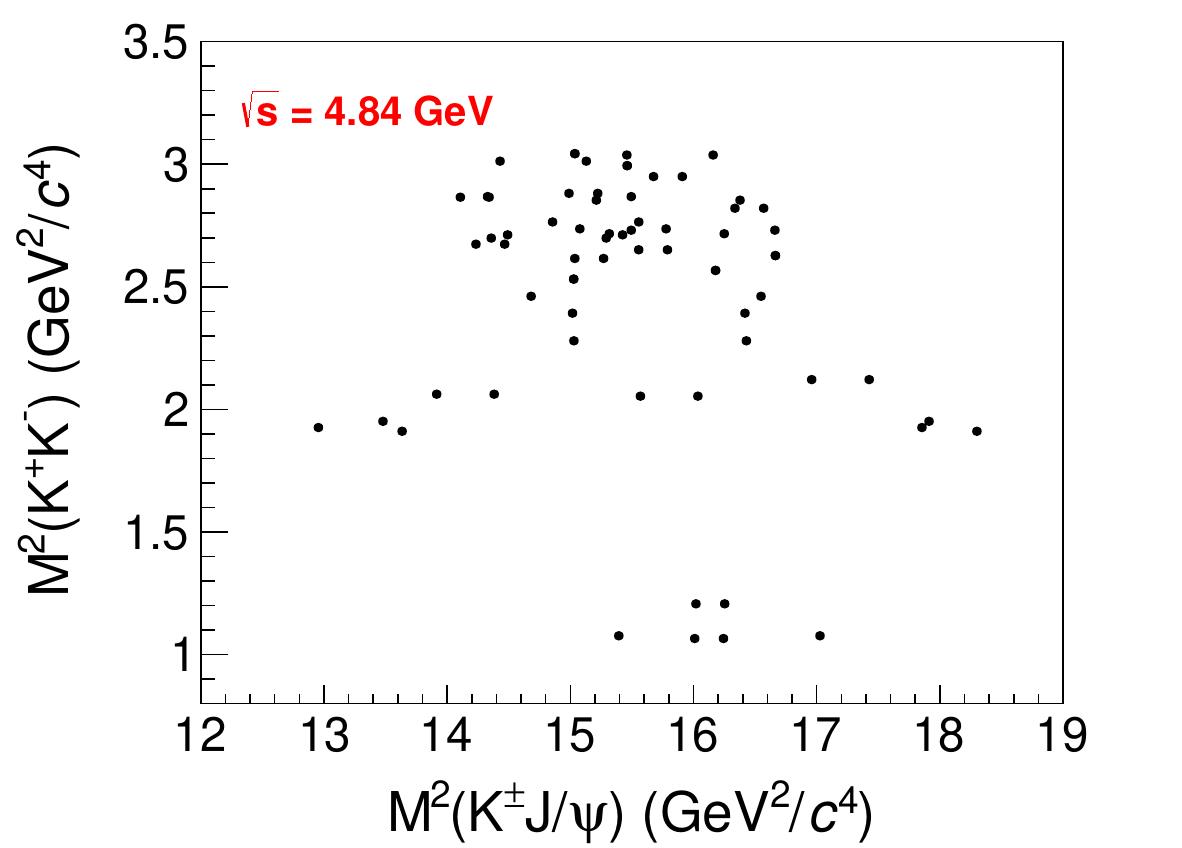}
		\includegraphics[width=0.32\linewidth]{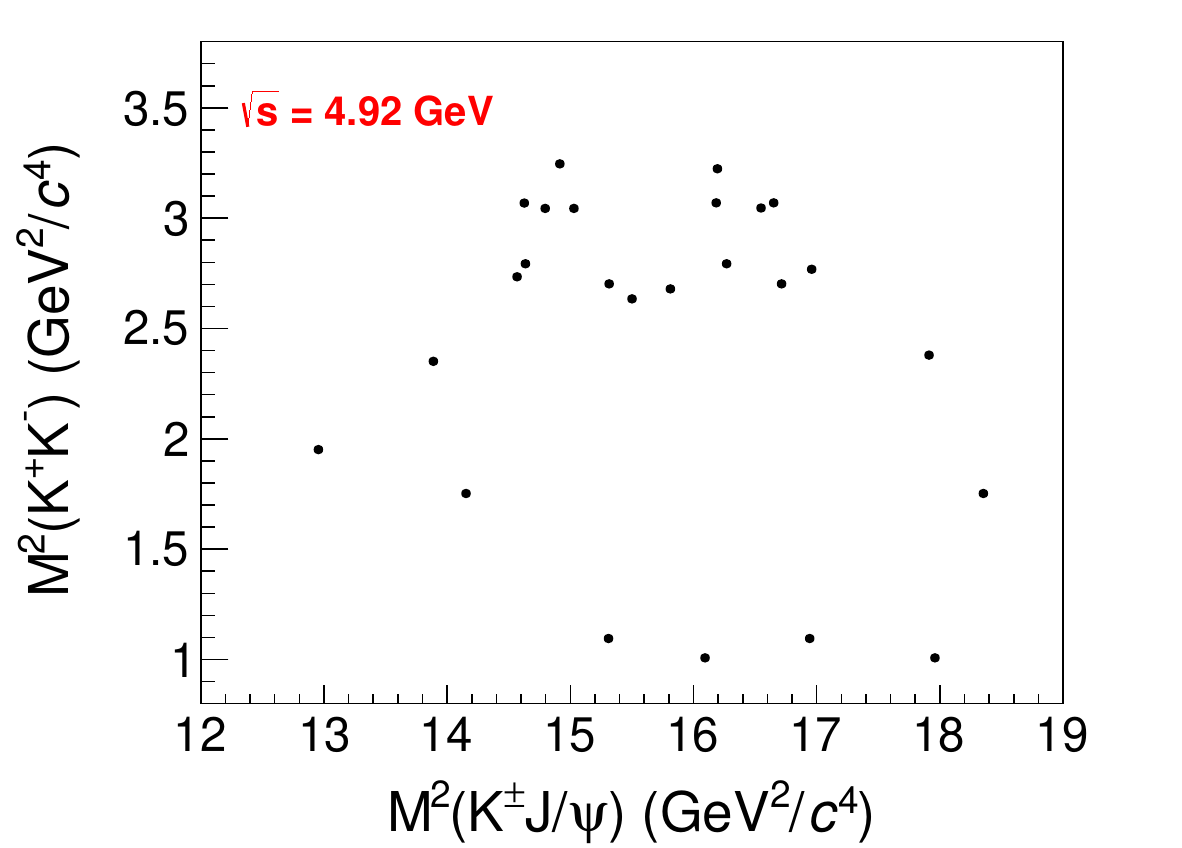}
		\includegraphics[width=0.32\linewidth]{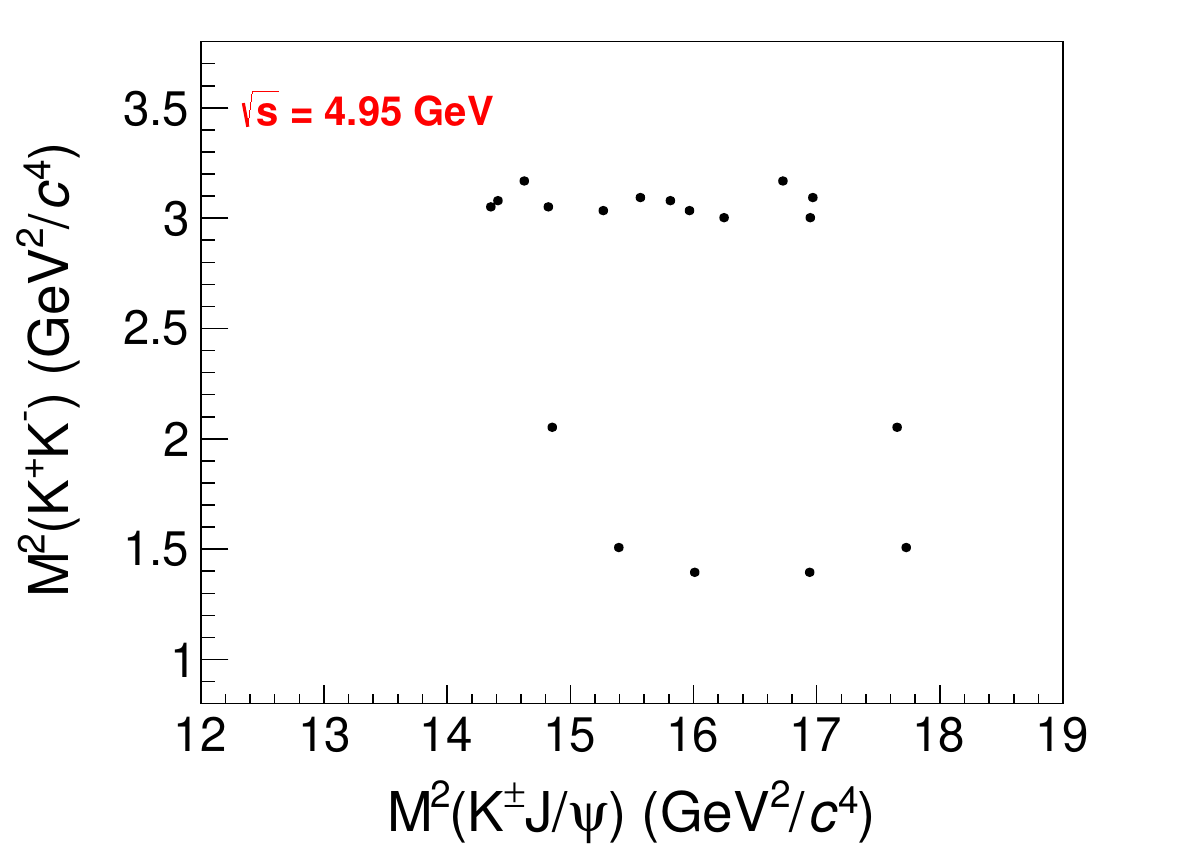}
		
		\caption{The Dalitz plots of the data samples at each c.m.~energy. \label{fig:dalitz}}
	\end{figure}

	Figure~\ref{fig:dalitz} shows the Daltiz plots at each c.m.~energy,  evident $f$ states can be seen above $\sqrt{s}=4.70$ GeV.
	
	\section{cross sections}
	The c.m.~energies ($\sqrt{s}$), integrated luminosities ($\mathcal{L}$), 
	numbers of events in the signal region ($N^{\rm obs}$) and in the sideband regions ($N^{\rm side}$), signal yields ($N^{\rm sig}$), event selection efficiencies ($\epsilon$), ISR correction factors ($(1+\delta)$), vacuum polarization factors ($|1+\Pi|^{2}$), Born cross sections ($\sigma^{\rm Born}$), and Born cross section ratios of $e^{+}e^{-}\to\ks\ks\jpsi$ to $e^{+}e^{-}\to K^{+}K^{-}\jpsi$ ($\frac{\sigma^{\rm Born}(\ks\ks\jpsi)}{\sigma^{\rm Born}(K^{+}K^{-}\jpsi)}$) are shown in Table~\ref{tab:bxs}.
	
	\begin{table}[htbp]
		\centering\addtolength{\tabcolsep}{0.1cm}
		\caption{The c.m.~energies ($\sqrt{s}$), integrated luminosities ($\mathcal{L}$), 
			numbers of events in the signal region ($N^{\rm obs}$) and in the sideband regions ($N^{\rm side}$), signal yields ($N^{\rm sig}$), event selection efficiencies ($\epsilon$), ISR correction factors ($(1+\delta)$), vacuum polarization factors ($\frac{1}{|1-\Pi|^{2}}$), Born cross sections ($\sigma^{\rm Born}$), and Born cross section ratios $\frac{\sigma^{\rm Born}(\ks\ks\jpsi)}{\sigma^{\rm Born}(K^{+}K^{-}\jpsi)}$.
			The first uncertainties of $\sigma^{\rm Born}$ and $\frac{\sigma^{\rm Born}(\ks\ks\jpsi)}{\sigma^{\rm Born}(K^{+}K^{-}\jpsi)}$ are statistical, and the second ones systematic. The uncertainties of $N^{\rm sig}$ are only statistical.}
		\begin{tabular}{cccccccccc}
			\hline
			\hline
			$\sqrt{s}$~(GeV) &$\mathcal{L}$~(pb$^{-1}$) &$N^{\rm obs}$ &$N^{\rm side}$ &$N^{\rm sig}$  &$\epsilon$ &$(1+\delta)$ &$\frac{1}{|1-\Pi|^{2}}$ &$\sigma^{\rm Born}$~(pb) &$\frac{\sigma^{\rm Born}(\ks\ks\jpsi)}{\sigma^{\rm Born}(K^{+}K^{-}\jpsi)}$ \\
			\hline
			4.61   &103.65   &3   &1    &$2.75 _{-1.45}^{+2.09 }$  &0.322 &1.137     &1.055    &$0.57_{-0.30}^{+0.44}\pm0.04$  &$0.379_{-0.354}^{+0.674}\pm0.026$\\
			4.63   &521.53   &38  &7    &$36.25_{-5.88}^{+6.54 }$  &0.306 &1.181     &1.054    &$1.53_{-0.25}^{+0.28}\pm0.10$  &$0.116_{-0.084}^{+0.162}\pm0.008$\\
			4.64   &551.65   &22  &10   &$19.50_{-4.45}^{+5.09 }$  &0.304 &1.200     &1.054    &$0.77_{-0.18}^{+0.20}\pm0.05$  &$0.832_{-0.306}^{+0.525}\pm0.056$\\
			4.66   &529.43   &26  &6    &$24.50_{-4.82}^{+5.47 }$  &0.308 &1.158     &1.054    &$1.03_{-0.20}^{+0.23}\pm0.07$  &$0.654_{-0.266}^{+0.351}\pm0.044$ \\
			4.68   &1667.39  &94  &26   &$87.50_{-9.46}^{+10.12}$  &0.334 &1.043     &1.054    &$1.20_{-0.13}^{+0.14}\pm0.08$  &$0.772_{-0.158}^{+0.172}\pm0.052$\\
			4.70   &536.54   &40  &10   &$37.50_{-6.06}^{+6.71 }$  &0.365 &0.953     &1.055    &$1.60_{-0.26}^{+0.29}\pm0.11$  &$0.635_{-0.175}^{+0.247}\pm0.043$\\
			4.74   &163.87   &18  &0    &$18.00_{-3.92}^{+4.59 }$  &0.403 &0.886     &1.055    &$2.44_{-0.53}^{+0.62}\pm0.23$  &$0.111_{-0.099}^{+0.184}\pm0.007$\\
			4.75   &366.55   &37  &9    &$34.75_{-5.81}^{+6.47 }$  &0.398 &0.892     &1.055    &$2.12_{-0.35}^{+0.39}\pm0.20$  &$0.562_{-0.184}^{+0.248}\pm0.038$\\
			4.78   &511.47   &60  &12   &$57.00_{-7.47}^{+8.13 }$  &0.388 &0.931     &1.055    &$2.45_{-0.32}^{+0.35}\pm0.23$  &$0.399_{-0.133}^{+0.172}\pm0.027$\\
			4.84   &525.16   &34  &7    &$32.25_{-5.55}^{+6.21 }$  &0.358 &1.015     &1.056    &$1.34_{-0.23}^{+0.26}\pm0.12$  &$0.474_{-0.172}^{+0.240}\pm0.032$\\
			4.92   &207.82   &13  &4    &$12.00_{-3.33}^{+3.98 }$  &0.328 &1.082     &1.056    &$1.29_{-0.36}^{+0.43}\pm0.12$  &$0.384_{-0.223}^{+0.348}\pm0.026$ \\
			4.95   &159.28   &9   &4    &$8.00 _{-2.73}^{+3.38 }$  &0.312 &1.103     &1.056    &$1.15_{-0.39}^{+0.49}\pm0.11$  &$0.180_{-0.169}^{+0.332}\pm0.012$ \\
			\hline
			\hline
		\end{tabular}
		\label{tab:bxs}
	\end{table}
	
	In the maximum likelihood fit to the dressed cross sections of $e^{+}e^{-}\to K^{+}K^{-}J/\psi$,  assuming the obtained signal events obey Poisson ($N^{\rm sig} \leq 10$) or asymmetric Gaussian ($N^{\rm sig} > 10$). The Poisson is defined as
	\begin{equation}
		P_{\rm Poisson} = (N^{\rm fit} + f\cdot N^{\rm side})^{N^{\rm obs}}\cdot\frac{e^{-(N^{\rm fit} + f\cdot N^{\rm side})}}{N^{\rm obs}\,!},
	\end{equation}
	while the asymmetric Gaussian is defined as
	\begin{equation}
		P_{\rm Gaussian} = \left\{
		\begin{array}{lr}
			\frac{1}{\sqrt{2\pi}\cdot(\sigma_{l}+\sigma_{h})}\cdot e^{-\frac{(N^{\rm fit}-N^{\rm sig})^{2}}{2\sigma_{l}^{2}}}, &N^{\rm fit}<N^{
				\rm sig}  \\
			\frac{1}{\sqrt{2\pi}\cdot(\sigma_{l}+\sigma_{h})}\cdot e^{-\frac{(N^{\rm fit}-N^{\rm sig})^{2}}{2\sigma_{h}^{2}}}, &N^{\rm fit} \geq N^{\rm sig}
		\end{array}
		\right.
	\end{equation}
	where $f$ is the normalization factor of signal to sideband region (0.5), $N^{\rm fit}$ is the expected number of signal events, and $\sigma_{l}$ and $\sigma_{h}$ are the low and high uncertainties of the number of signal events. The likelihood is structured as
	\begin{equation}
		L = \prod \limits_{i} P_{i}, \\
	\end{equation}
	where $P_{i}$ is $P_{\rm Poisson}$ if $N^{\rm sig} \leq10$ or $P_{\rm Gaussian}$ if $N^{\rm sig} > 10$.
	
\begin{figure}[htbp]
		\centering
		\includegraphics[width = 0.360\textwidth]{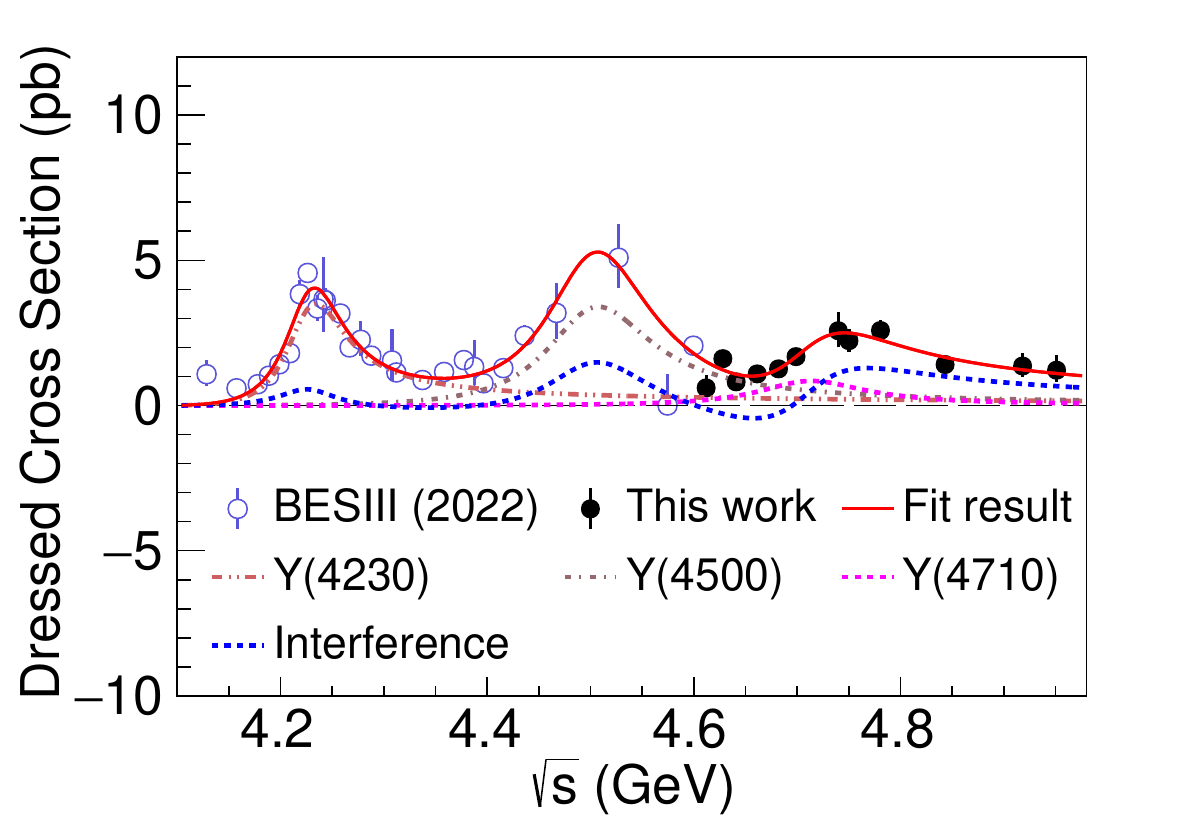}
		\includegraphics[width = 0.360\textwidth]{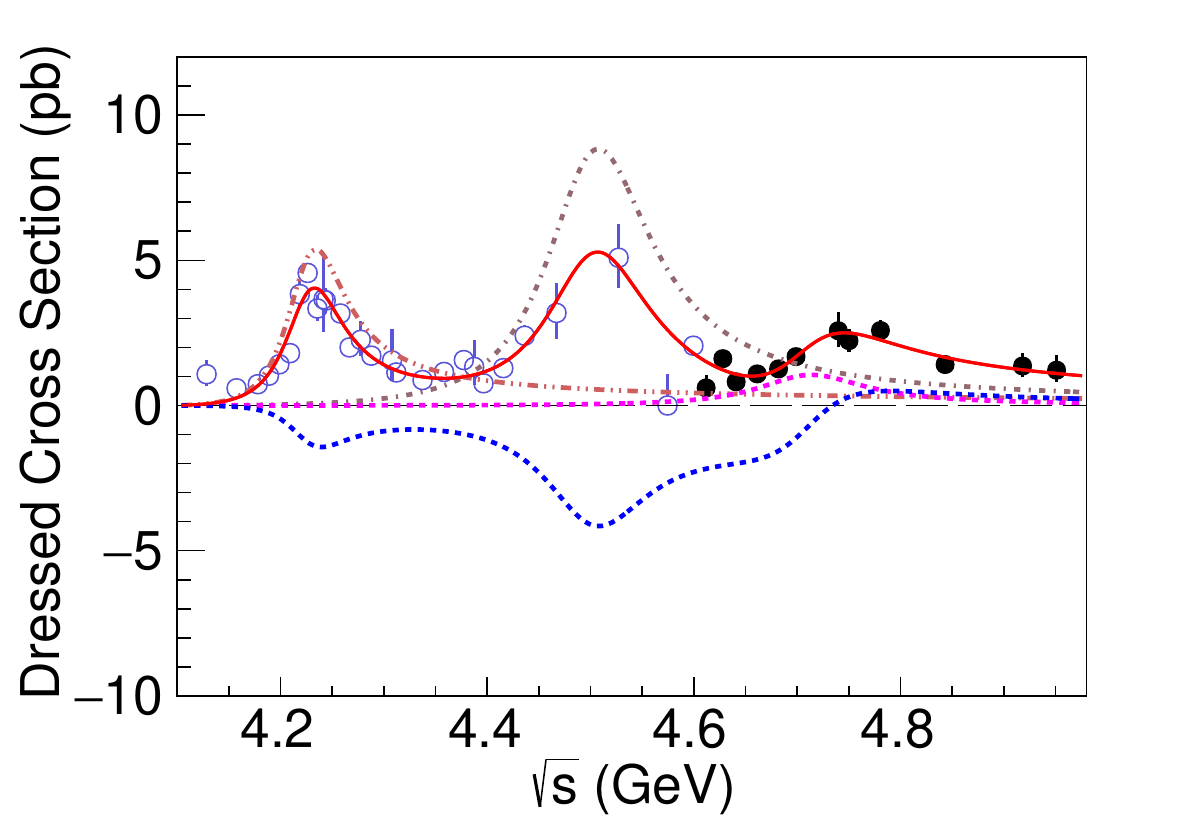}
		\includegraphics[width = 0.360\textwidth]{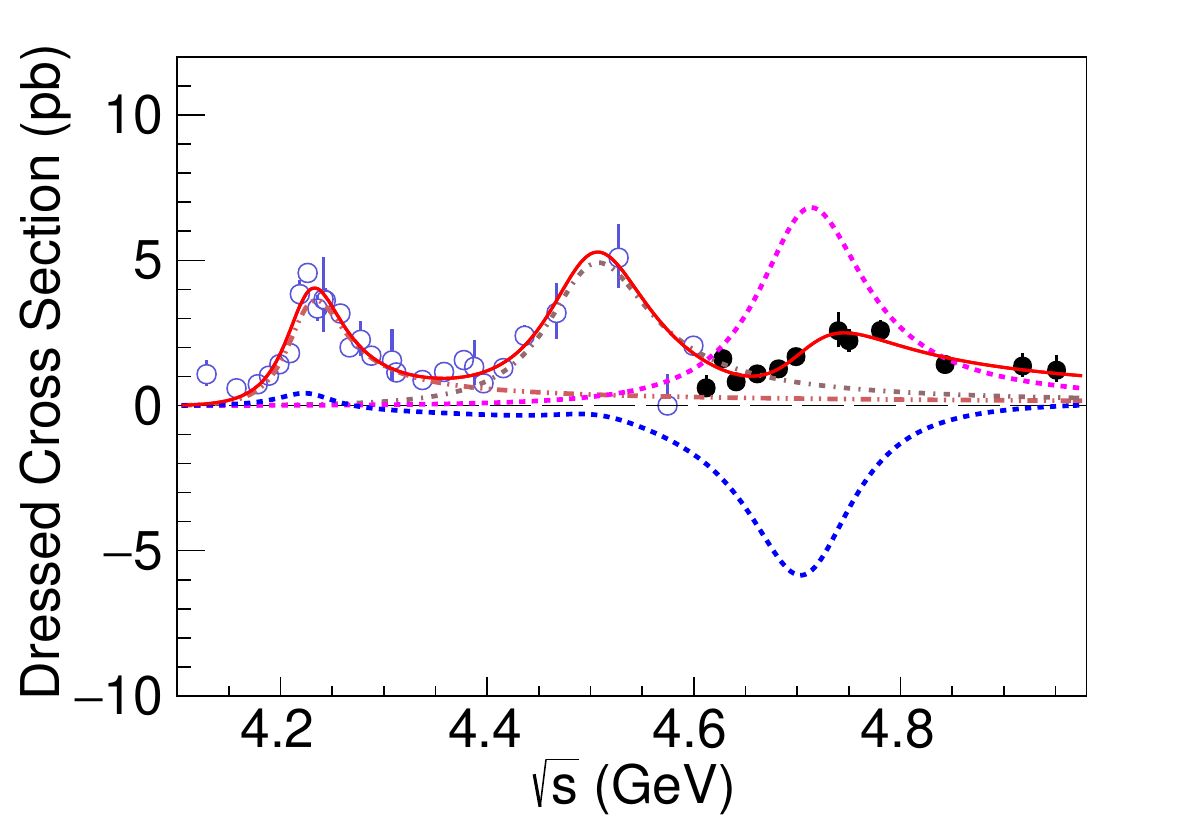}
		\includegraphics[width = 0.360\textwidth]{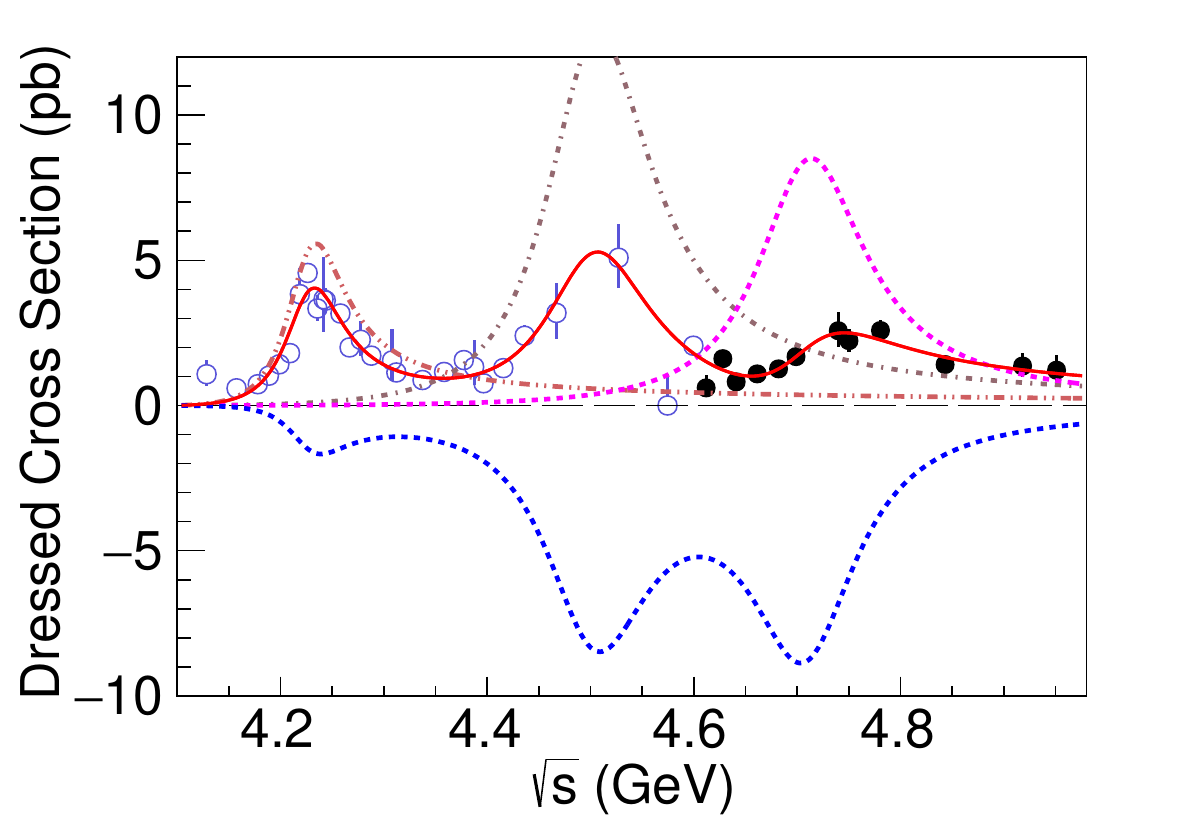}
		\caption{Four solutions of the fits to the dressed cross sections of $e^{+}e^{-}\to K^{+} K^{-} J/\psi$ with the coherent sum of three Breit-Wigner functions (solid curve). The dash (dash-dot-dot or dash-dot) curve shows the contribution from the three structures $Y(4710)$ ($Y(4230)$ or $Y(4500)$). The solid dots with error bars are the cross sections from this study, and the dash dots with error bars are the cross sections from Ref.~\cite{BESIII:2022joj}. The error bars are statistical uncertainty only.}
		\label{fig:fitxs_3bw_nobelle}
	\end{figure}

	Figure~\ref{fig:fitxs_3bw_nobelle} shows the four solutions of the fits to the dressed cross sections of $e^{+}e^{-}\to K^{+} K^{-} J/\psi$ with the coherent sum of three Breit-Wigner functions, and Table~\ref{tab:gamee} shows the quantities of the four solutions.
	The systematic uncertainties of the fit parameters are listed in Table~\ref{tab:errfitdcs}.
	
	\begin{table}
		\centering\addtolength{\tabcolsep}{0.2cm}
		\caption{The four solutions of $(\Gamma_{ee}\mathcal{B})$ and $\phi$ for the third Breit-Wigner function, which represents the resonance $Y(4710)$. The uncertainties are statistical only.}
		\begin{tabular}{c|cccc}
			\hline
			\hline
			$(\Gamma_{ee}\mathcal{B})_{3}$~(eV) &$0.16\pm0.04$ &$0.20\pm0.06$  &$1.29\pm0.23$  &$1.61\pm0.36$ \\
			$\phi_{3}$~(rad)                    &$0.25\pm0.43$ &$-1.60\pm0.45$ &$-0.92\pm0.19$ &$-2.76\pm0.16$ \\
			\hline
			\hline
		\end{tabular}
		\label{tab:gamee}
	\end{table}
	
	\begin{table}[htbp]
		\centering\addtolength{\tabcolsep}{0.2cm}
		\caption{The systematic uncertainties in the measurement of resonance parameters, including that due to the c.m. energy ($\sqrt{s}$),
			the parameterization of the fit function (Fitting), the c.m.~energy spread (ES), and the uncommon ($\sigma^{\rm Dress}_{1}$) and common ($\sigma^{\rm Dress}_{2}$) systematic uncertainties from the cross section measurement. The symbol ``--'' represents the uncertainty, which can be neglected.}
		\begin{tabular}{ccccccc}
			\hline
			\hline
			Parameter     &$\sqrt{s}$  &Fitting &ES &$\sigma^{\rm Dress}_{1}$ &$\sigma^{\rm Dress}_{2}$ &Sum \\
			\hline
			$M_{3}$~(MeV/$c^{2}$) &0.80  &20.91 &0.11 &3.03 &-- &21.1 \\
			$\Gamma_{3}$ (MeV)    &--    &29.67 &0.01 &1.09 &-- &29.7 \\
			\hline
			\multirow{4}{*}{$(\Gamma_{ee}\mathcal{B})_{3}$ (eV)}
			&--      &0.09 &--   &--   &0.01 &0.09 \\
			&--      &0.01 &--   &--   &0.01 &0.01 \\
			&--      &0.26 &--   &0.03 &0.04 &0.26 \\
			&--      &0.06 &--   &0.03 &0.05 &0.08 \\
			\hline
			\multirow{4}{*}{$\phi_{3}$ (rad)}
			&--    &0.40 &--   &0.08 &-- &0.41 \\
			&--    &0.69 &--   &0.08 &-- &0.69 \\
			&--    &0.42 &0.01 &0.02 &-- &0.42 \\
			&--    &0.70 &--   &0.02 &-- &0.70 \\
			\hline
			\hline
		\end{tabular}
		\label{tab:errfitdcs}
	\end{table}

 \par
\begin{figure}[htbp]
  \centering
  \includegraphics[width = 0.36\textwidth]{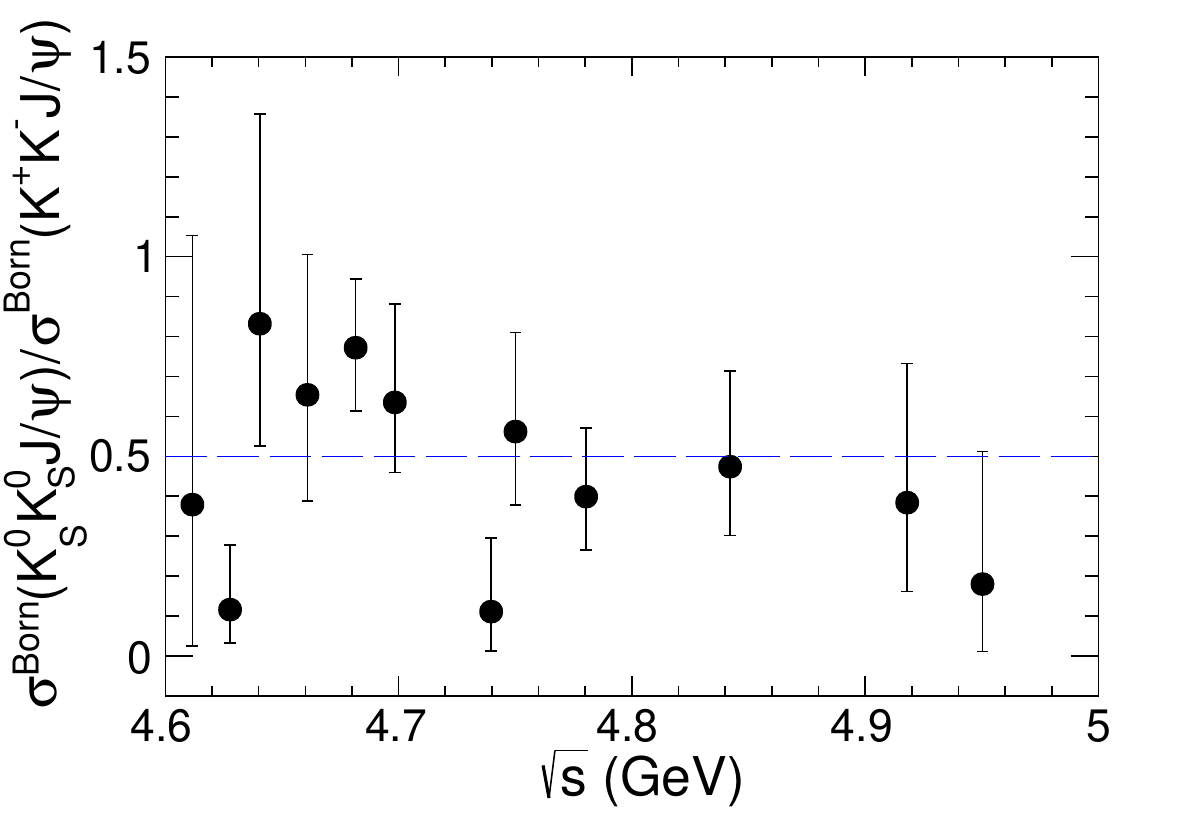}
  \caption{Ratio of Born cross section
           $\sigma^{\rm Born}(e^{+}e^{-} \to \ks \ks \jpsi)$ to $\sigma^{\rm Born}(e^{+}e^{-} \to K^{+}K^{-} \jpsi)$, where the error bars are statistical only.}
  \label{fig:ratio}
\end{figure}

 The Born cross section ratios $\frac{\sigma^{\rm Born}(\ks\ks\jpsi)}{\sigma^{\rm Born}(K^{+}K^{-}\jpsi)}$, shown in Table~\ref{tab:bxs} and Fig.~\ref{fig:ratio}, are determined by the ratio likelihood simulations, where the Born cross sections of $e^{+}e^{-}\to\ks\ks\jpsi$ ($\sigma^{\rm Born}(\ks\ks\jpsi)$) are obtained from Ref.~\cite{BESIII:2022kcv}. The systematic uncertainties of $\frac{\sigma^{\rm Born}(\ks\ks\jpsi)}{\sigma^{\rm Born}(K^{+}K^{-}\jpsi)}$ are estimated by taking into account correlations among uncertainties (the kinematic fit, ISR correction, $\ks$ reconstruction, and MUC depth are unrelated).

 \par The average Born cross ratio $\frac{\sigma^{\rm Born}(\ks\ks\jpsi)}{\sigma^{\rm Born}(K^{+}K^{-}\jpsi)}$ over $\sqrt{s} = 4.61-4.95$ GeV is determined to be $0.512_{-0.060}^{+0.074}\pm0.035$ based on combined ratio likelihood simulations, where the first uncertainties are statistical, while the second one systematic. The common items of the systematic uncertainties have been canceled. The P-value for the ratio being greater then 0.5 is 0.621 which indicates a $0.31\sigma$ significance isospin-violation effect in $e^{+}e^{-}\to K \bar{K} \jpsi$.

	\clearpage
	\section{Systematic Uncertainties on the \texorpdfstring{$Z_{cs}$}{Zcs(4050)} measurement}\label{sec:syszcs}
	
	Systematic uncertainties for the Born cross sections of $Z_{cs}$ states include the detection efficiencies, vacuum polarization  and ISR factor, which are estimated in the cross section measurement of $e^+e^-\rightarrow K^+K^-J/\psi$. These uncertainties are multiplicative. 
	
	\begin{table}[htbp]
		\centering
		\caption{Systematic uncertainties on the $Z_{cs}$ Born cross sections. The first five uncertainties are additive systematic uncertainties on the yields of $Z_{cs}$ while the last two uncertainties are multiplicative.}
		\begin{tabular}{cc}
			\hline\hline
			Source                &Systematic uncertainty on \emph{yields} or efficiencies\\
			\hline
			Detector resolution & 0.1 \\
			Efficiency curves & Negligible \\
			Signal model & 0.2 \\
			Backgrounds & 0.1 \\
			$f$ states & 0.3 \\
			\hline
			$Z_{cs}$ resonance parameters & See the main texts\\
			Detection efficiency & 15\% \\
			\hline\hline
		\end{tabular}
		\label{tab:Zcs syst}
	\end{table}

	In addition, we take into account the systematic uncertainties from the detector resolution, efficiency curves, signal models of $Z_{cs}$, backgrounds and $f$ states, $Z_{cs}$ resonance parameters, which are summarized in Table~\ref{tab:Zcs syst}. The difference of detector resolution between data and MC is estimated to be 3.2 MeV$/c^{2}$ by studying the control sample of $e^+e^-\rightarrow K^+ D^{*0} D_s^{*-}$. We smear the resolution function and redo the $M_{\rm max}(K^\pm J/\psi)$ fit and the change is taken as the systematic uncertainty. Through the similar procedure to estimate systematic uncertainties, we vary the efficiency curves within $\pm 1\sigma$ uncertainties, change the signal model under different $J^P$ assumptions, vary the kernel width parameter of the background shapes and add shapes of $f$ state at $\sqrt{s}>4.70$~GeV. We also generate signal MC samples for each signal model, and we take the largest difference between the resultant efficiencies and the nominal efficiencies as the systematic uncertainty. We vary the resonance parameters within $\pm 1\sigma$ regions and take the largest changes of the yields as the uncertainties.
	
	These additive systematic uncertainties on the fitted yields are then converted into Born cross sections. Then the the root of quadratic sum of converted uncertainties and multiplicative uncertainties, which will be discussed later, is assigned as the $\sigma$ of a Gaussian function. Then the Gaussian function will be used to convolve with the distributions of the nominal $-\ln L$ values.
	
	Multiplicative uncertainties are related to the detection efficiencies, vacuum polarization and ISR factor, which part of which has been estimated in the measurement of the line shape. Except for the uncertainty of MC model which is irrelevant to detection efficiencies of $Z_{cs}$, all uncertainties in the measurement of line shape will be considered in the convolution of systematic uncertainties. Different spin-parity assumptions will affect the angular distribution of final states thus affect the detection efficiencies. We take the largest relative change among all energies to the nominal uncertainty in different spin-parity assumptions, 15.0\%, as the systematic uncertainty.
	
\begin{figure}
		\centering
		\includegraphics[width=0.3\textwidth]{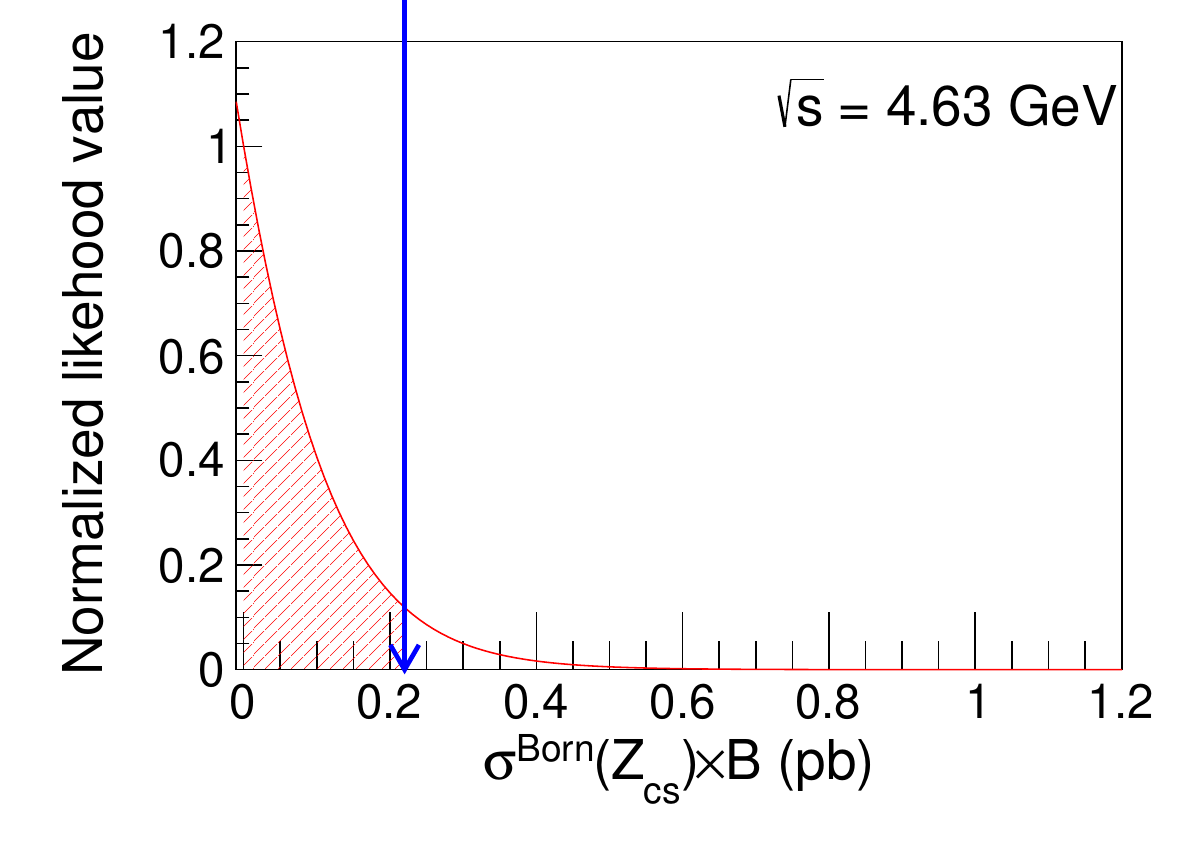}
		\includegraphics[width=0.3\textwidth]{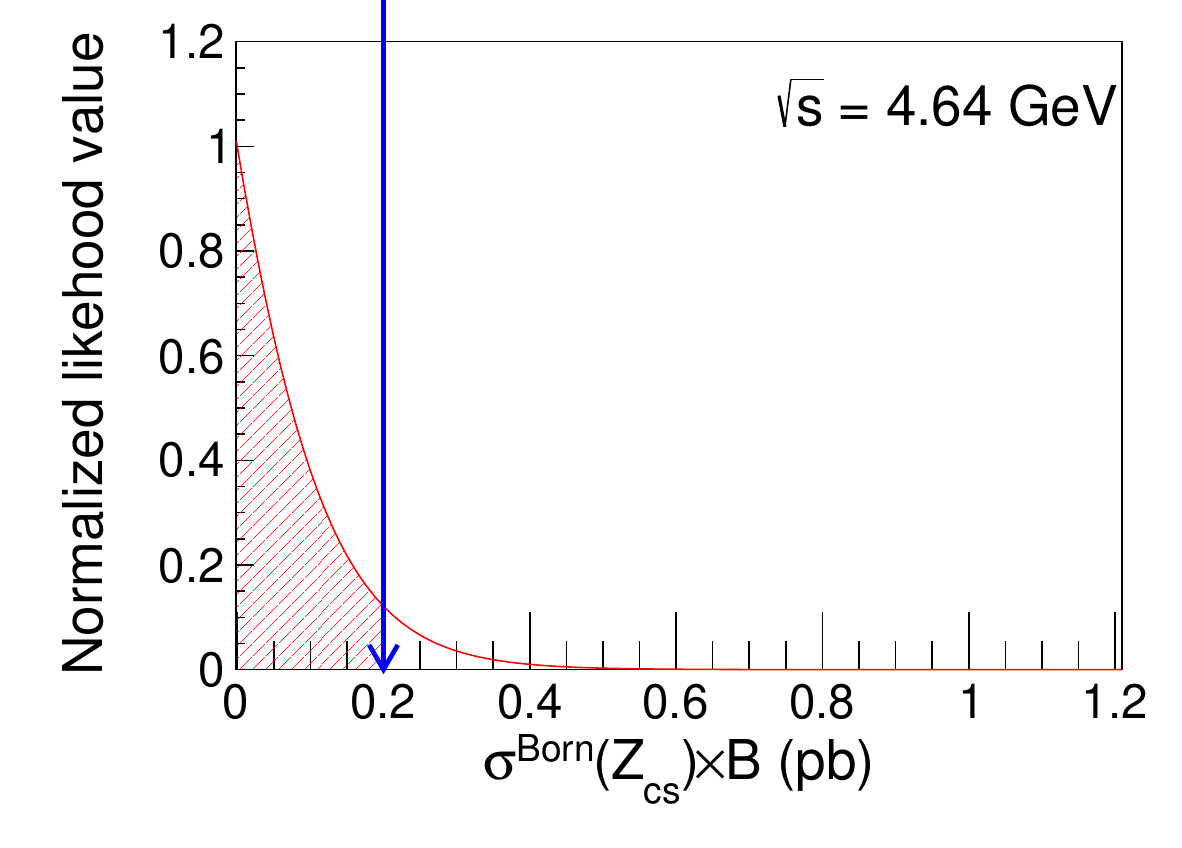}
		\includegraphics[width=0.3\textwidth]{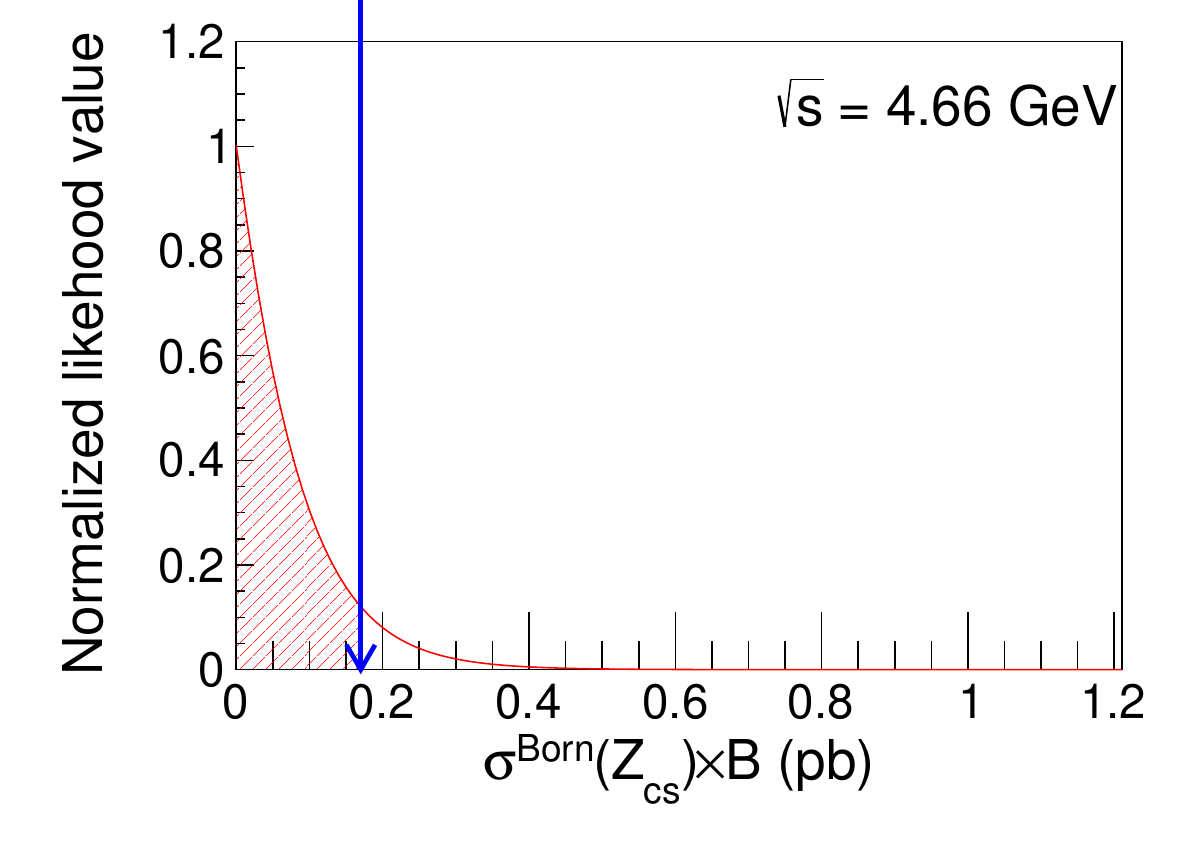}
		\includegraphics[width=0.3\textwidth]{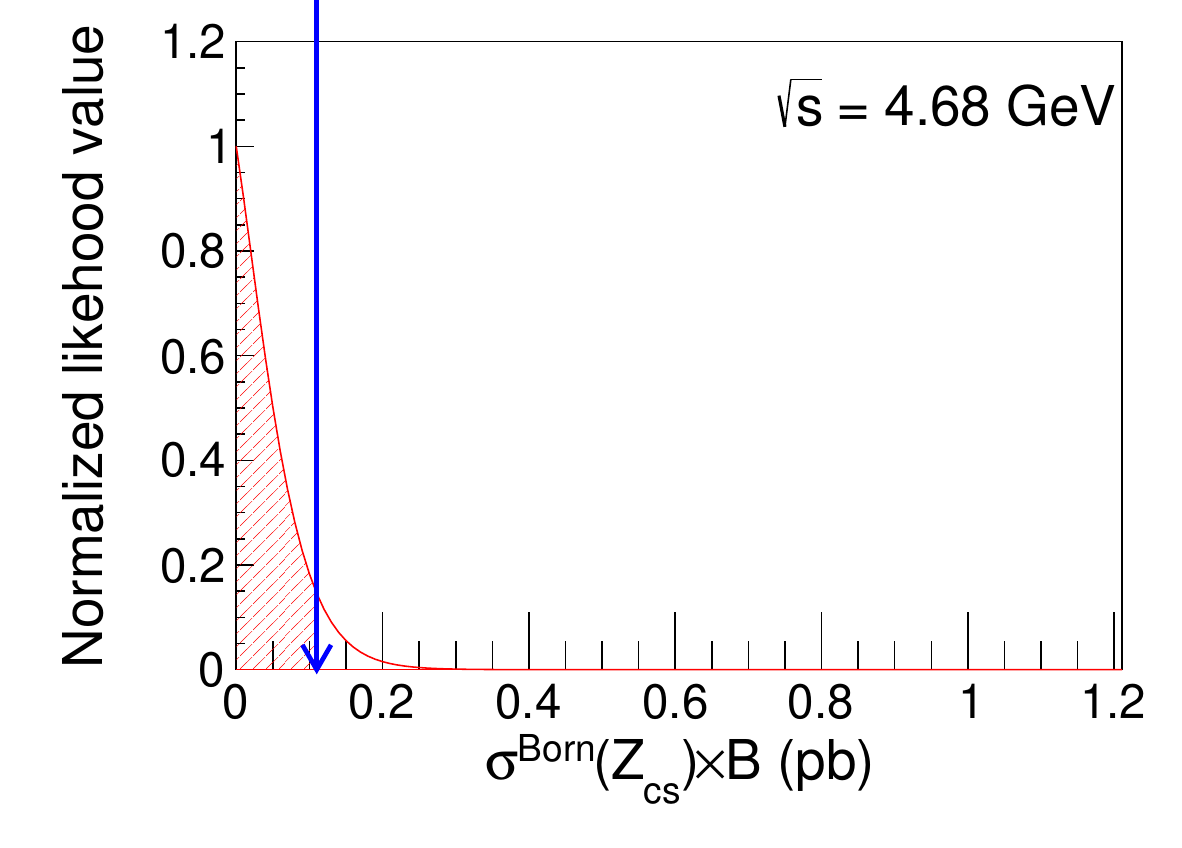}
		\includegraphics[width=0.3\textwidth]{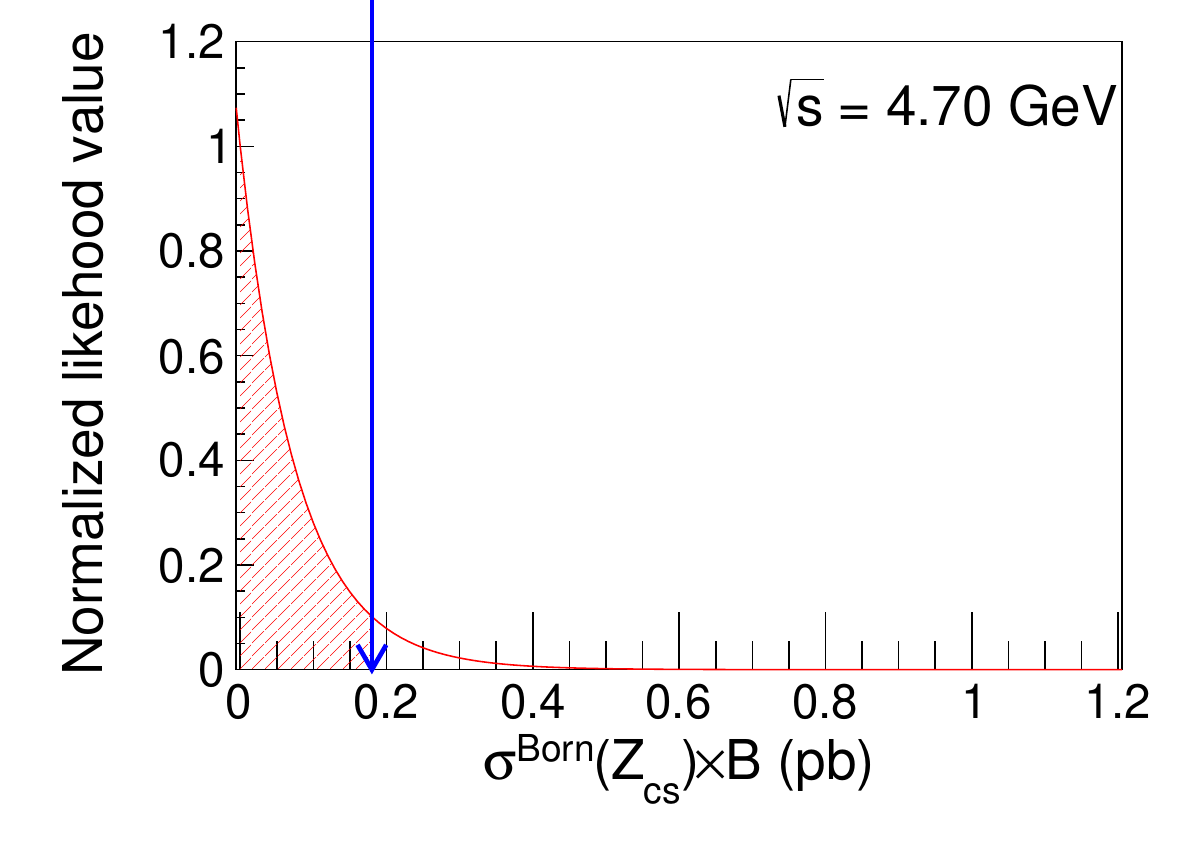}
		\includegraphics[width=0.3\textwidth]{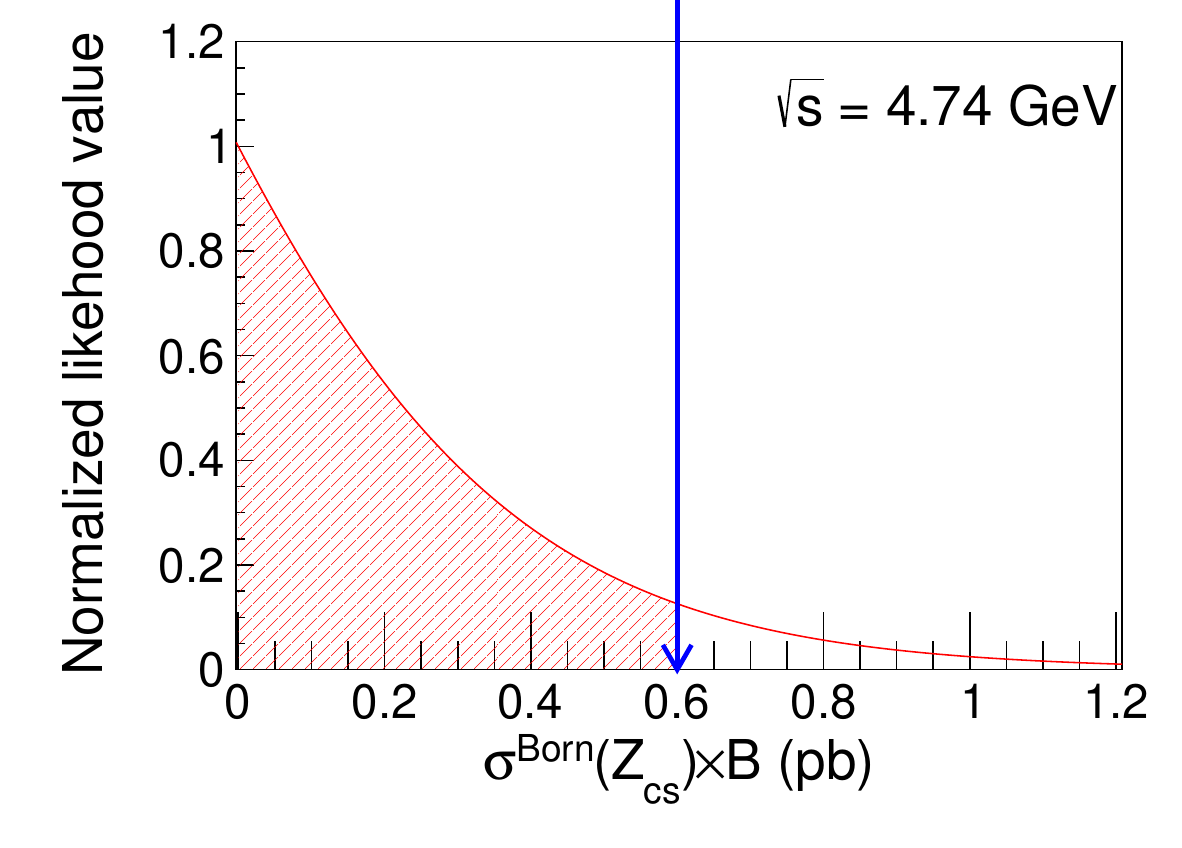}
		\includegraphics[width=0.3\textwidth]{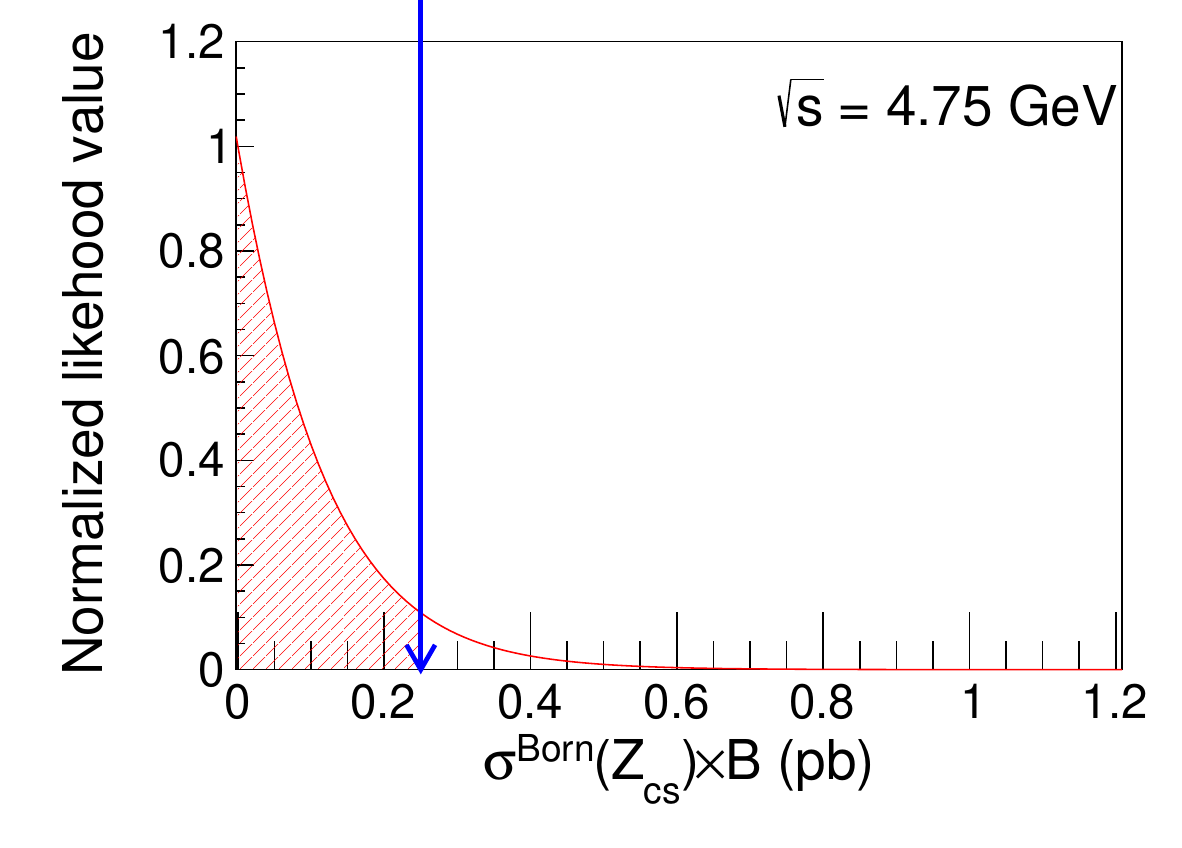}
		\includegraphics[width=0.3\textwidth]{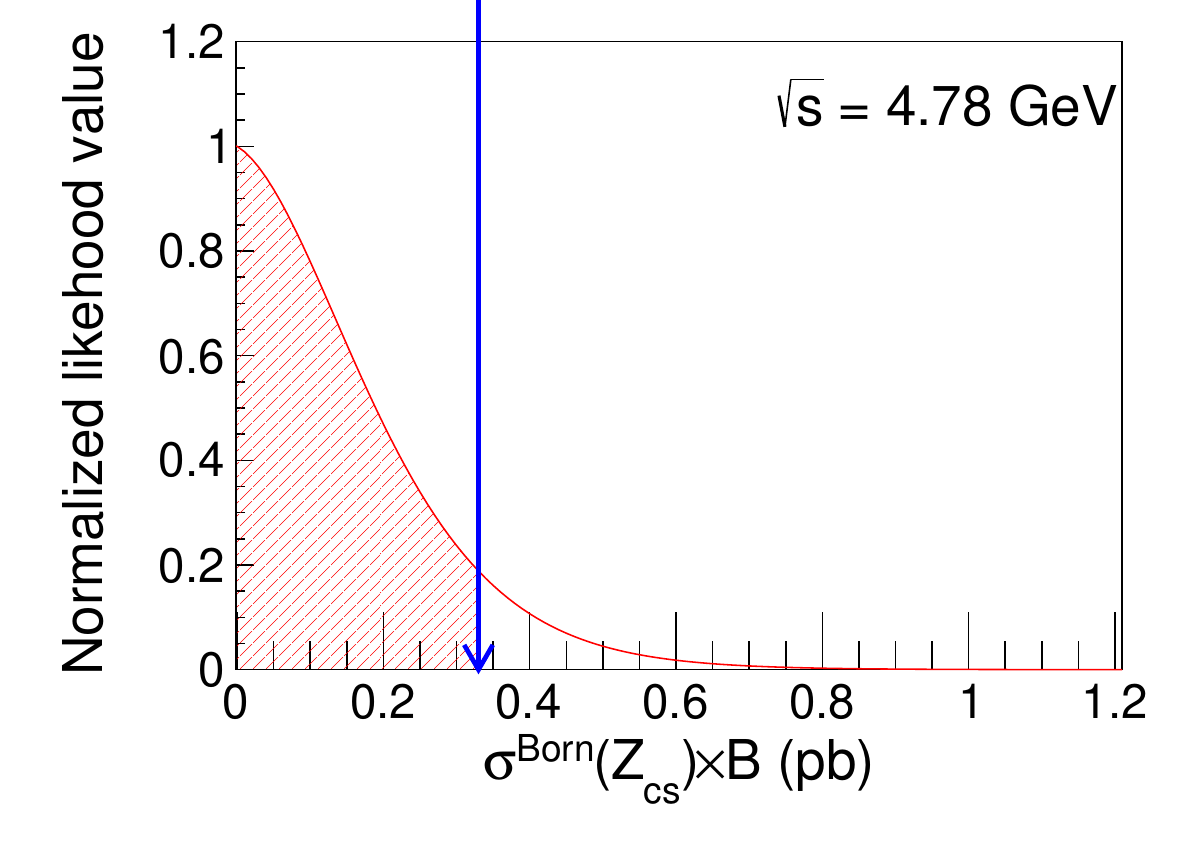}
		\includegraphics[width=0.3\textwidth]{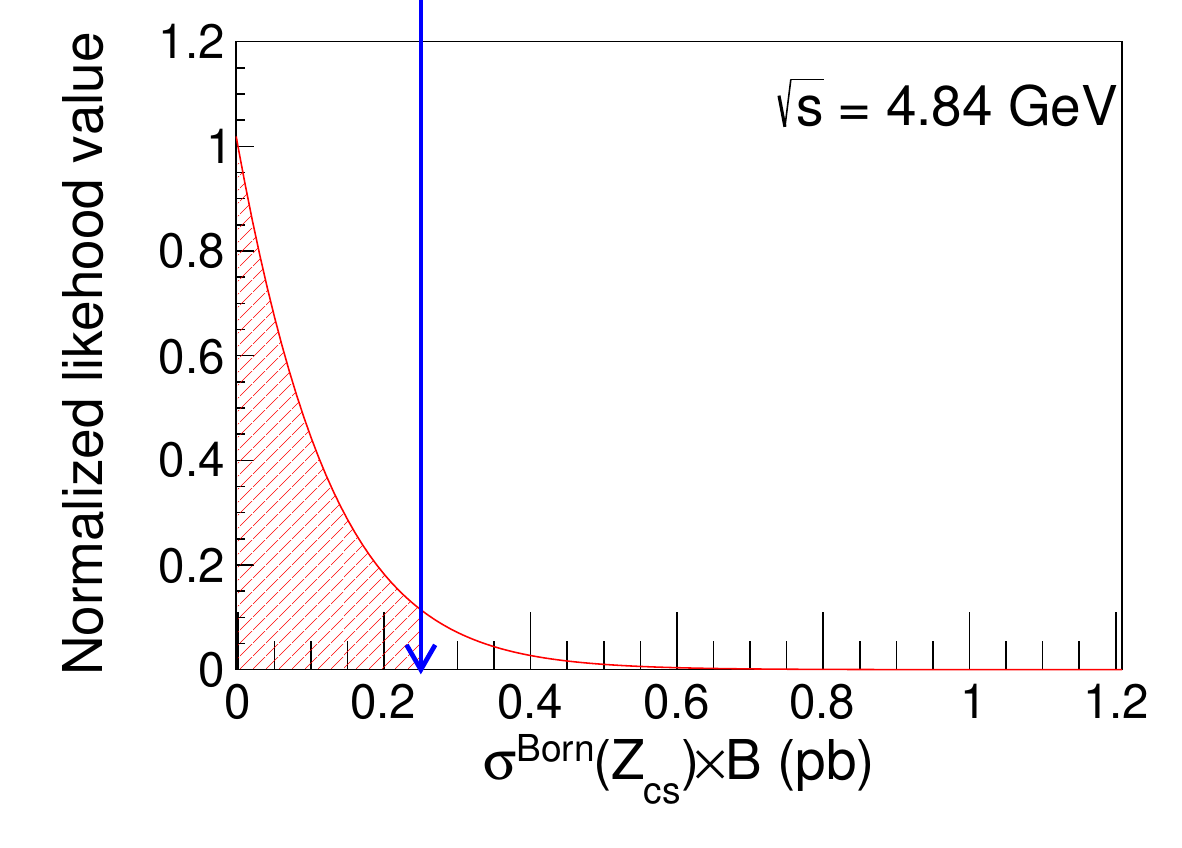}
		\includegraphics[width=0.3\textwidth]{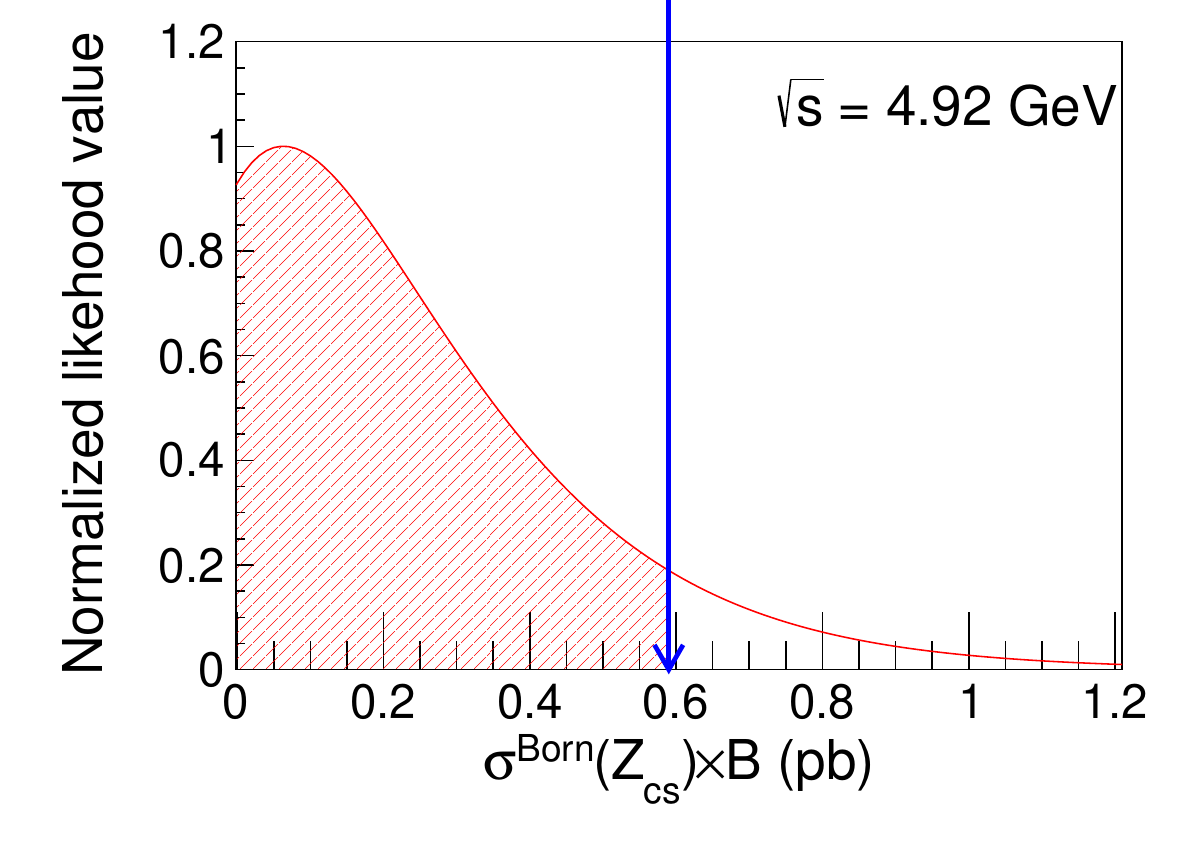}
		\caption{The scans for the upper limits of Born cross sections of $\ee\to K^- Z_{cs}(3985)^{+} + c.c.$. The blue arrows indicate the upper limits at the 90\% confidence level by integrating the red regions consisting of smeared likelihood values with systematic uncertainties considered.}
		\label{fig:UL Zcs3985}
	\end{figure}

\begin{figure}
		\centering
		\includegraphics[width=0.3\textwidth]{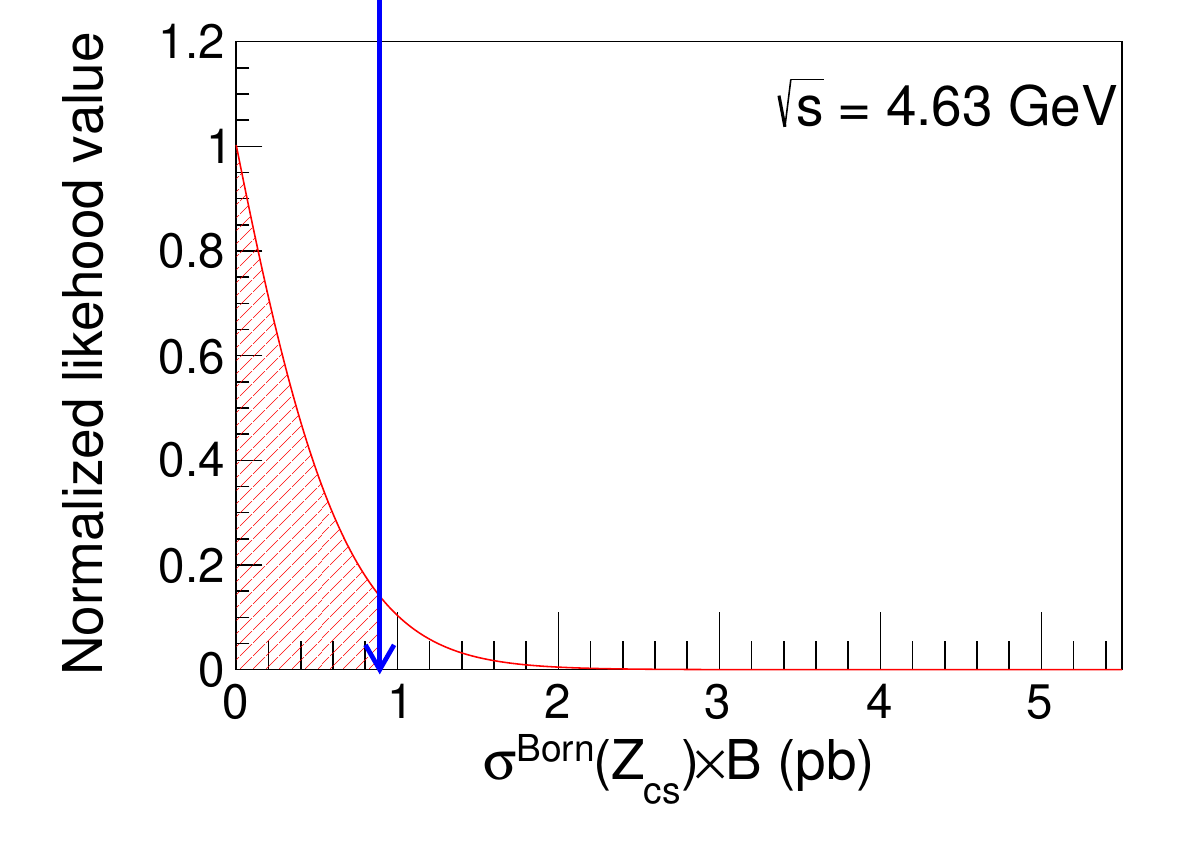}
		\includegraphics[width=0.3\textwidth]{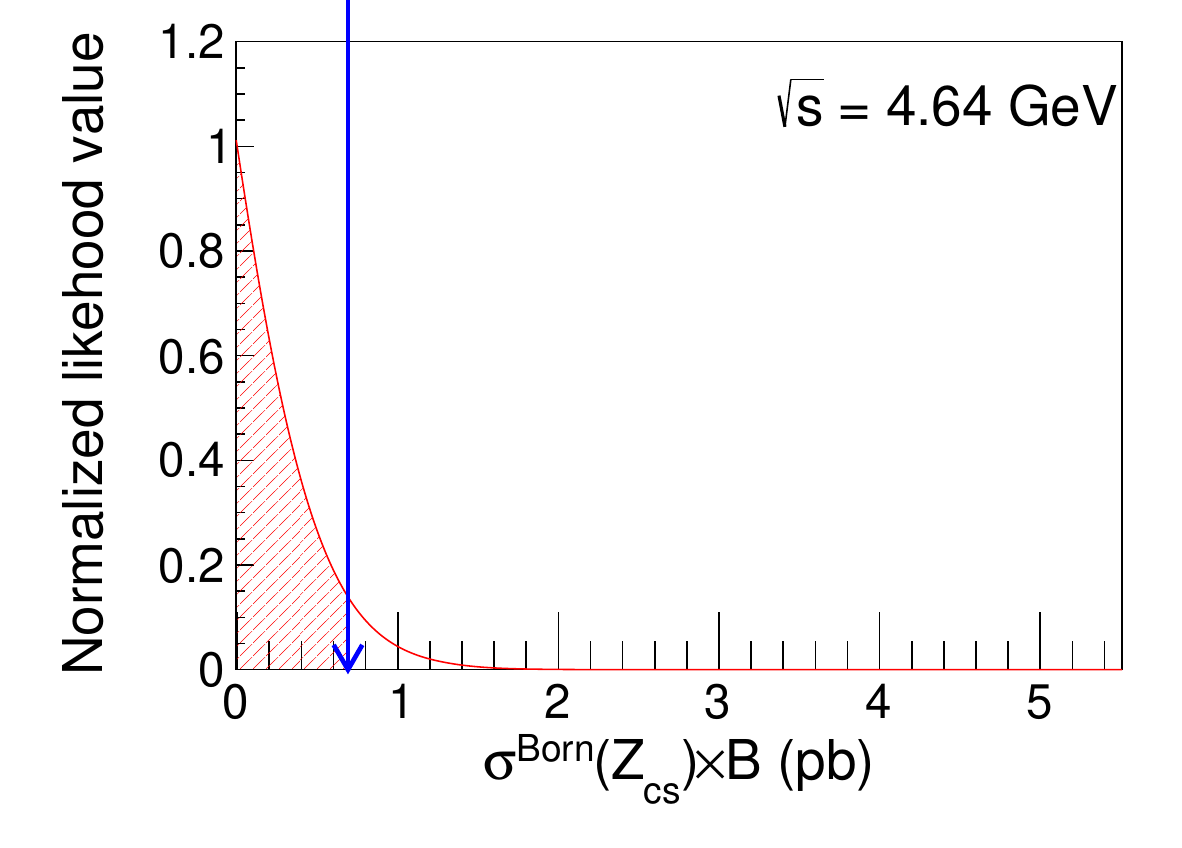}
		\includegraphics[width=0.3\textwidth]{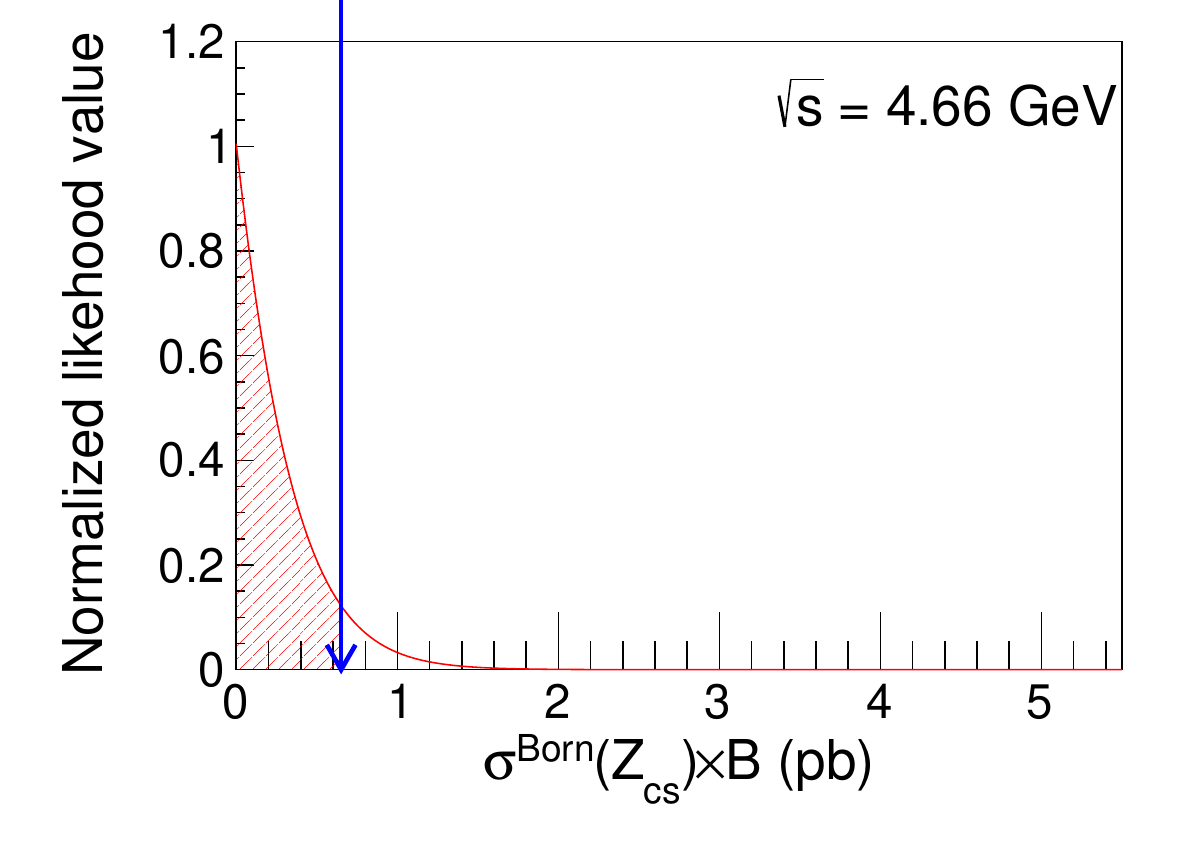}
		\includegraphics[width=0.3\textwidth]{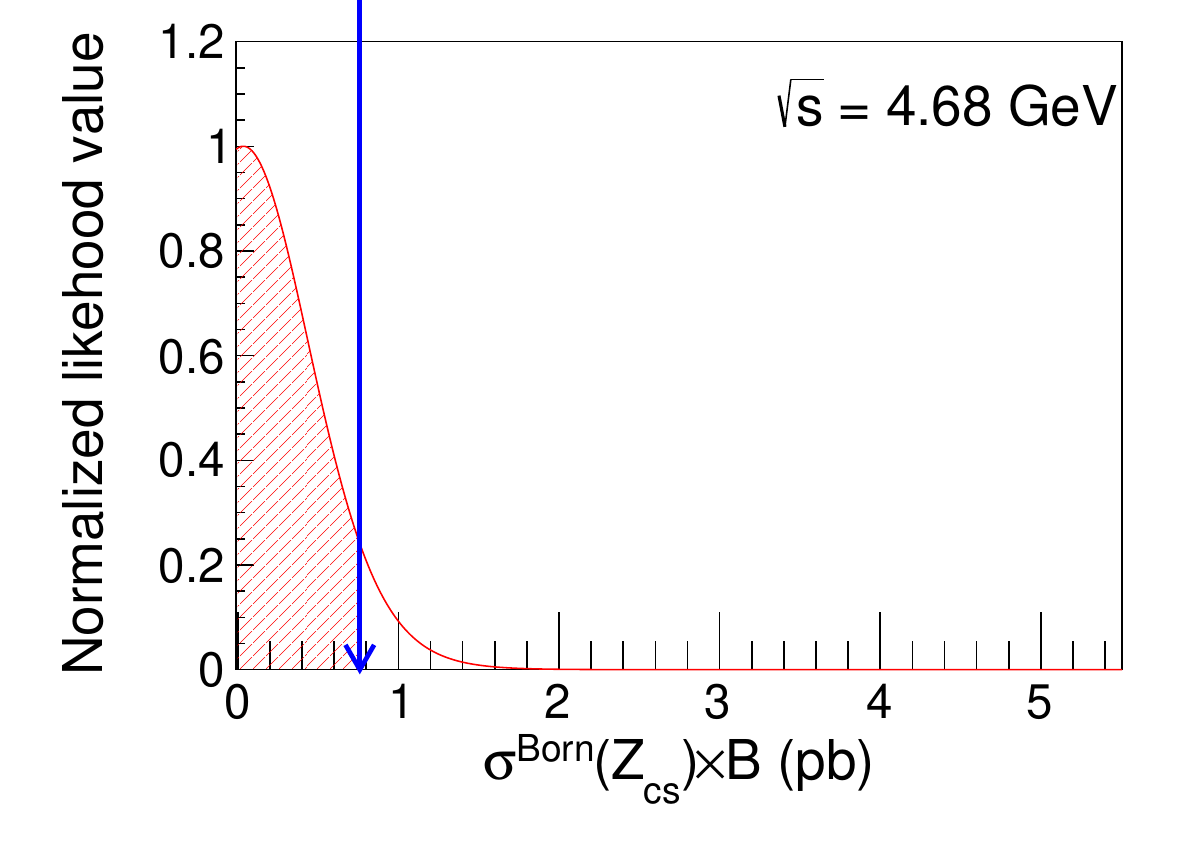}
		\includegraphics[width=0.3\textwidth]{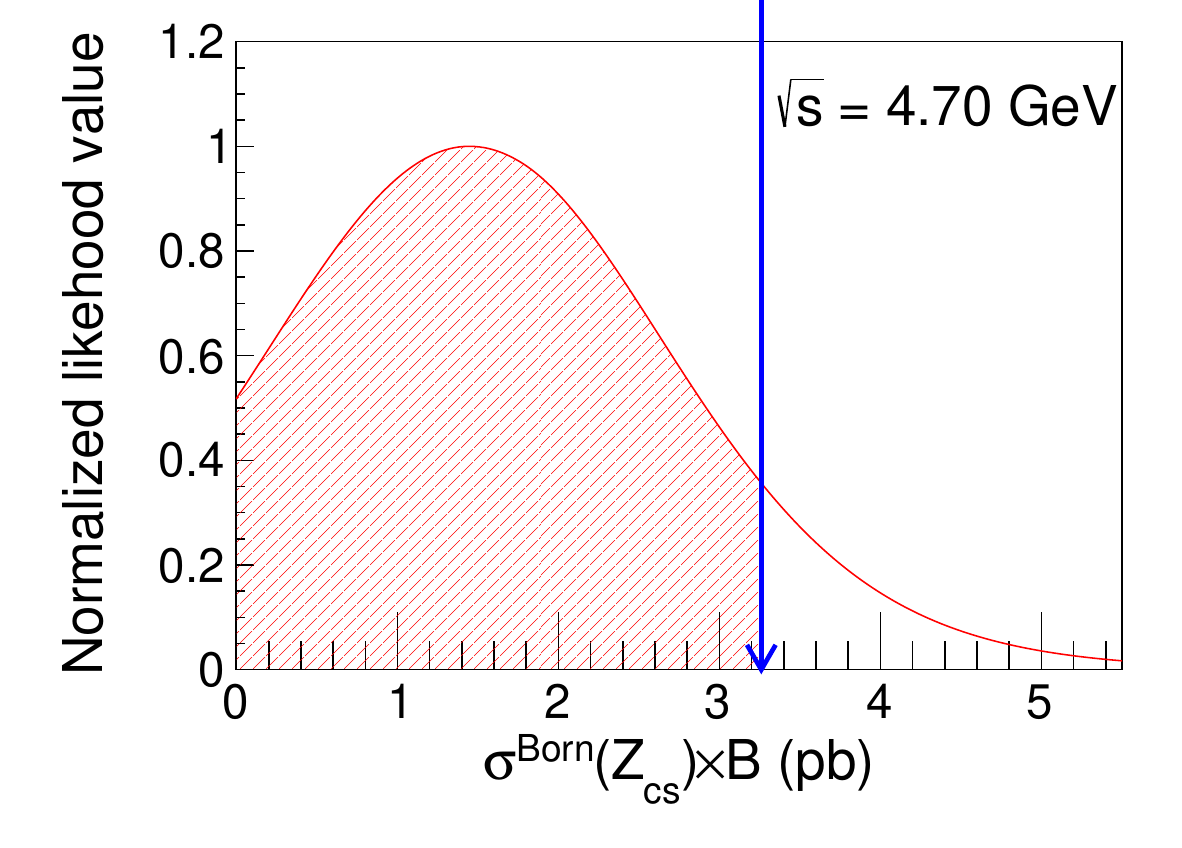}
		\includegraphics[width=0.3\textwidth]{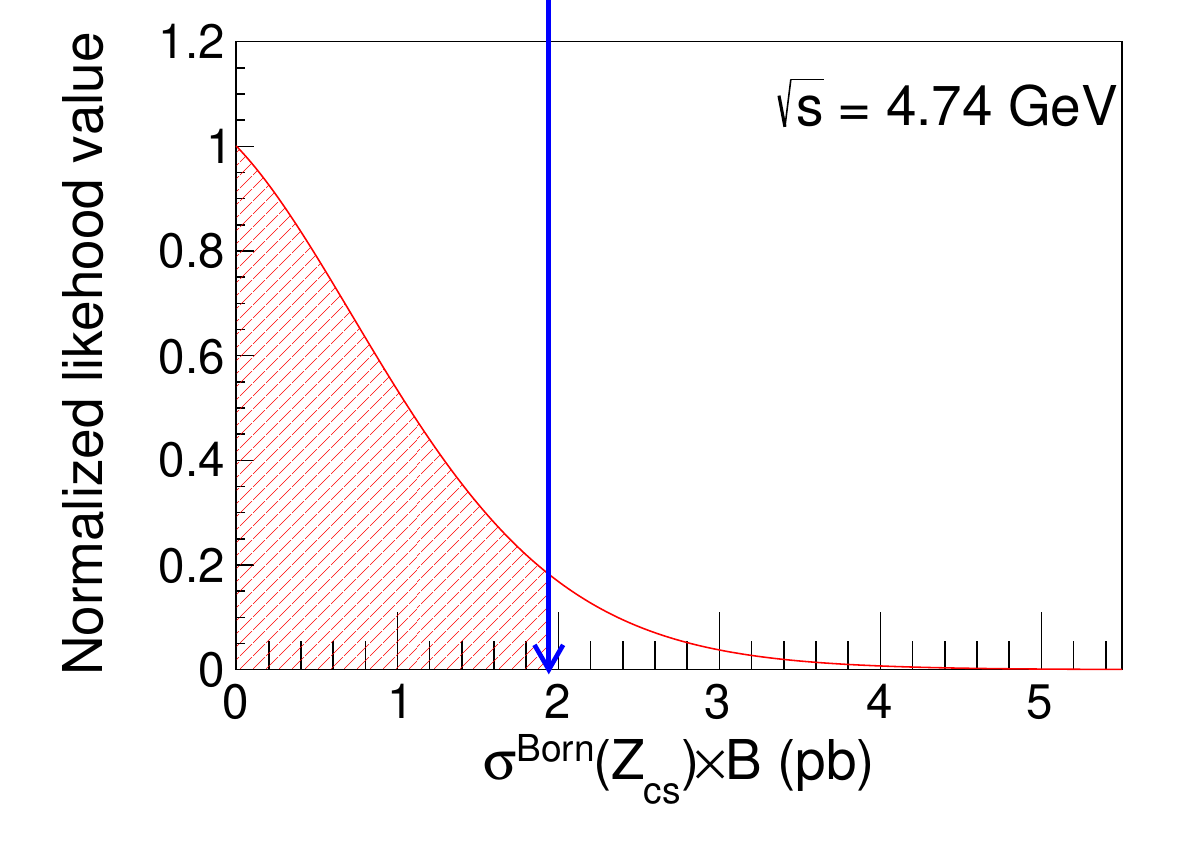}
		\includegraphics[width=0.3\textwidth]{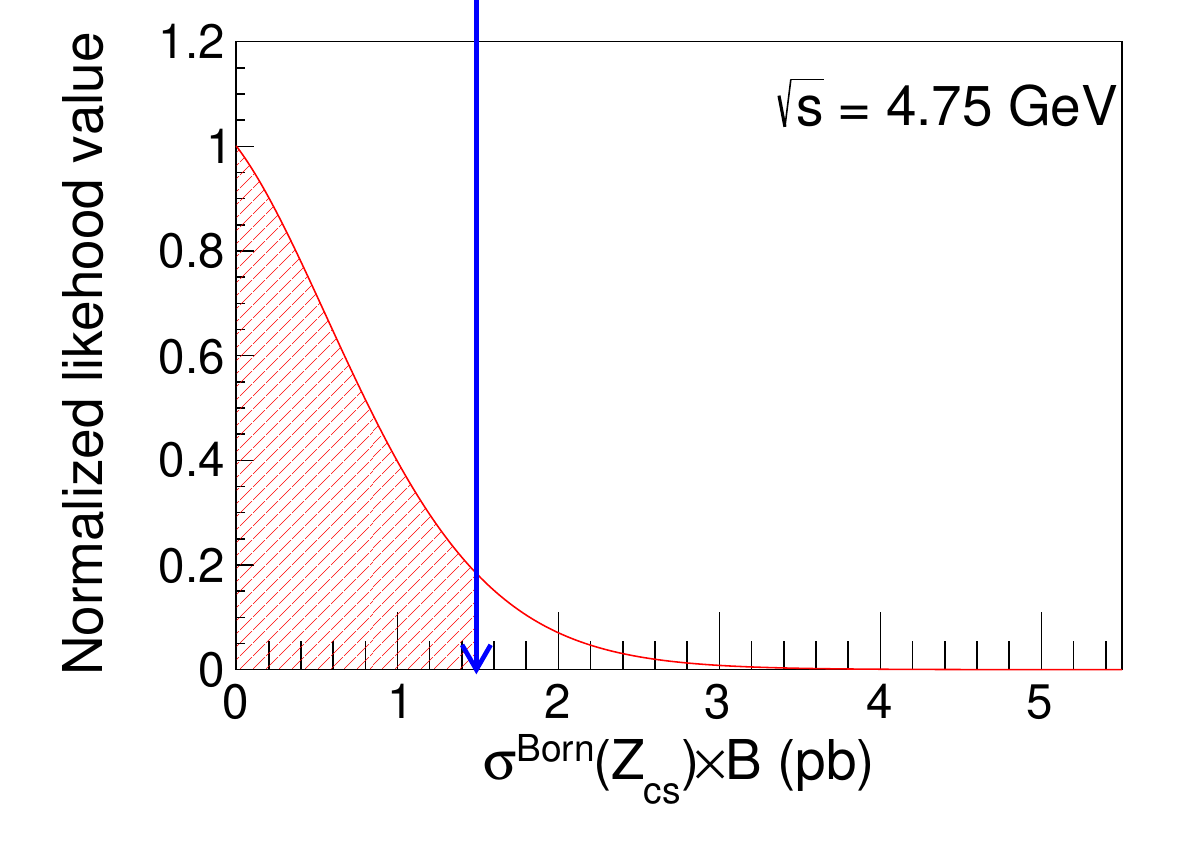}
		\includegraphics[width=0.3\textwidth]{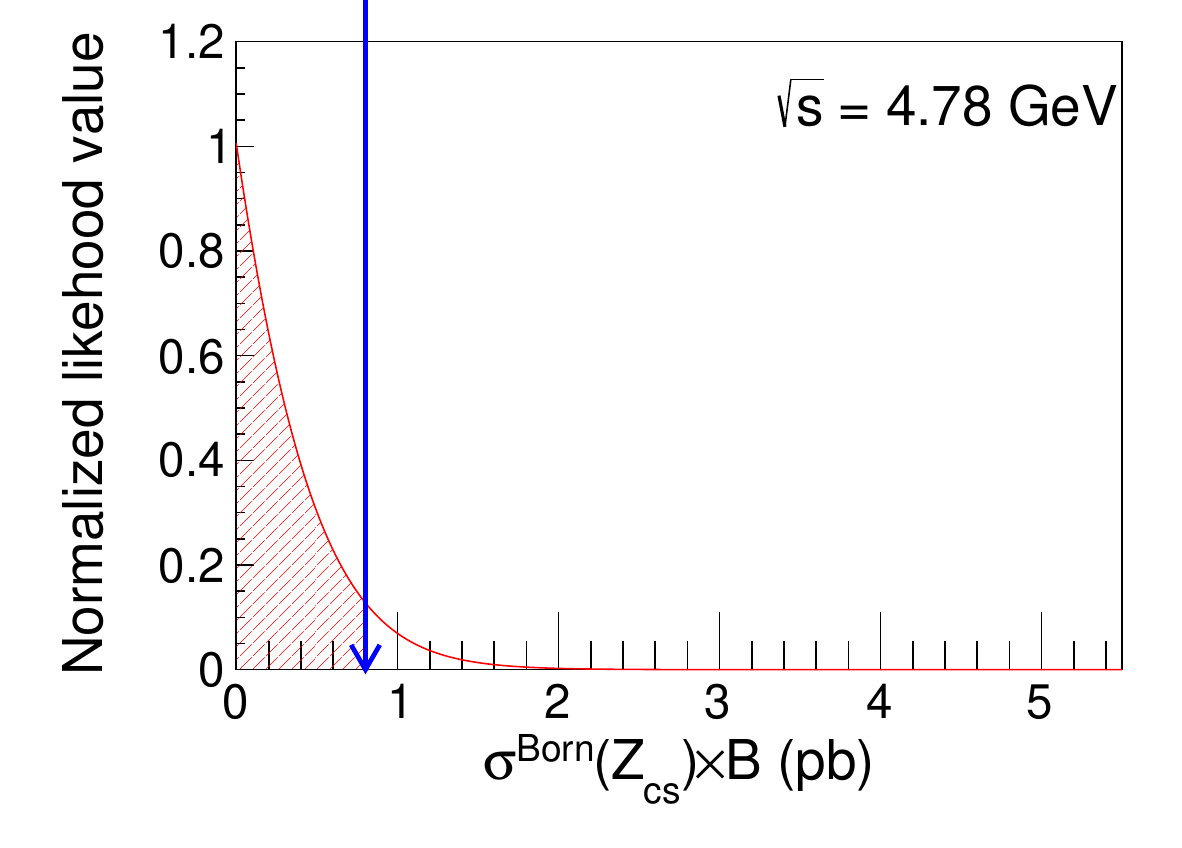}
		\includegraphics[width=0.3\textwidth]{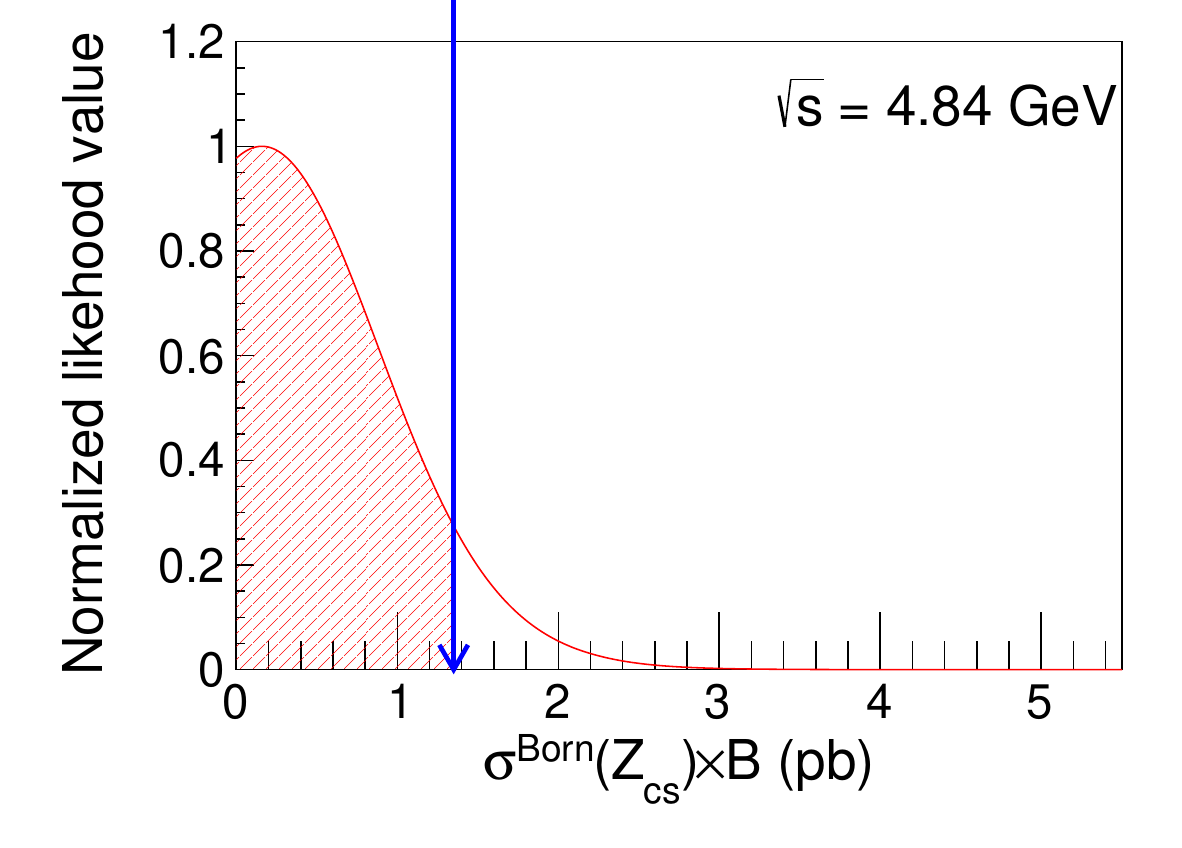}
		\includegraphics[width=0.3\textwidth]{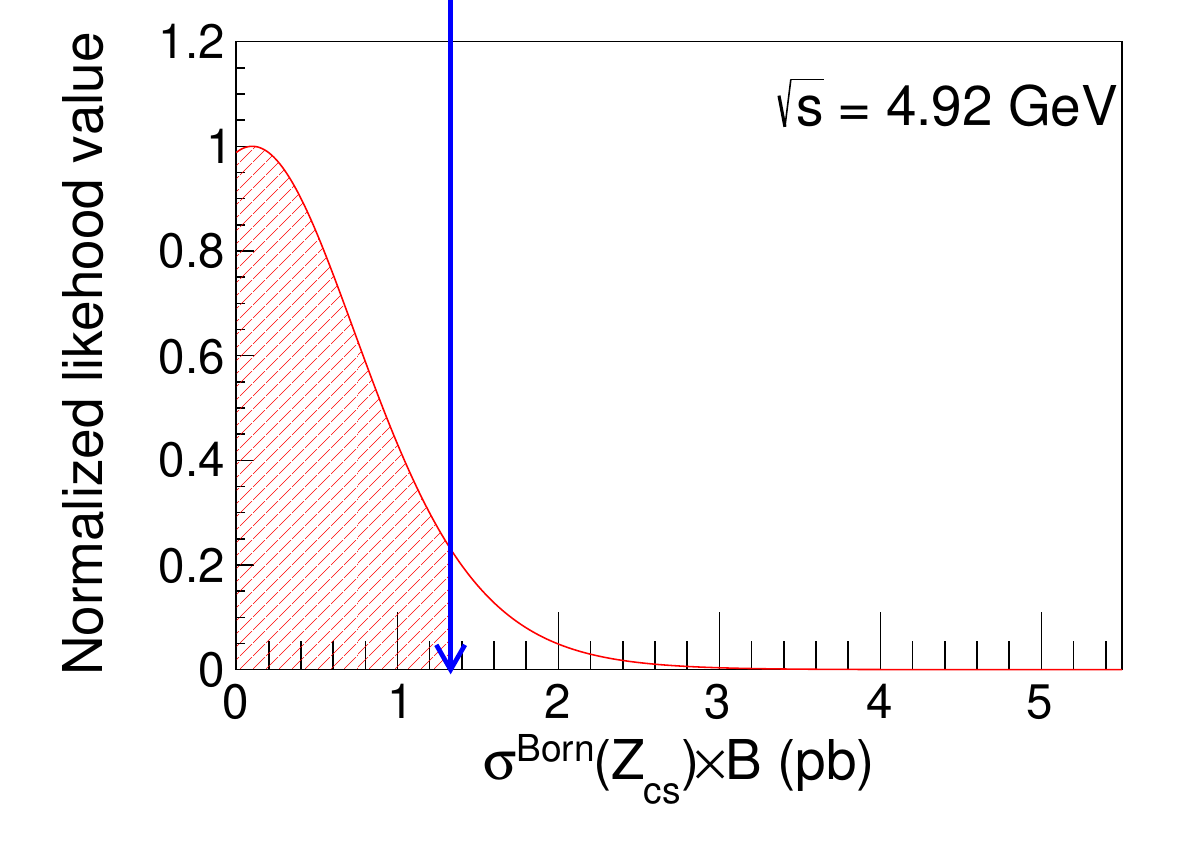}
		\caption{The scans for the upper limits of Born cross sections of $\ee\to K^- Z_{cs}(4000)^{+} + c.c.$. The blue arrows indicate the upper limits at the 90\% confidence level by integrating the red regions consisting of smeared likelihood values with systematic uncertainties considered.}
		\label{fig:UL Zcs4000}
	\end{figure}

	The convolved $-\ln L$ distributions are shown in red curves in Figs.~\ref{fig:UL Zcs3985} and \ref{fig:UL Zcs4000}, which are used to determine the upper limits at the 90\% confidence level. The upper limits are shown in Table~\ref{tab:UL}.
	
	\begin{table}[htpb]
		\centering
		\caption{The detection efficiencies $\epsilon$, and upper limits $\sigma^{\rm Born}(e^+e^-\to K^- Z_{cs}^{+} + c.c.)\cdot\mathcal{B}(Z_{cs}^{+}\to K^+ J/\psi)$ of Born cross sections for $Z_{cs}$ states at the 90\% confidence level, where the systematic uncertainties are incorporated. The VP and ISR factors are taken from Table~\ref{tab:bxs}.\label{tab:UL}}
		\label{tab:Zcs effi}
		\begin{tabular}{ccccc}
			\hline
   \hline
			$\sqrt{s}$ (GeV) & $\epsilon(Z_{cs}(3985))$ & $\epsilon(Z_{cs}(4000))$ & $\sigma^{\rm Born}\cdot\mathcal{B}[Z_{cs}(3985)]^{\rm UL}$ (pb) & $\sigma^{\rm Born}\cdot\mathcal{B}[Z_{cs}(4000)]^{\rm UL}$ (pb)\\
			\hline
			4.63   & $0.408$ & $0.409$    & 0.2 & 0.9\\
			4.64   & $0.402$ & $0.399$    & 0.2 & 0.7\\ 
			4.66   & $0.404$ & $0.396$    & 0.2 & 0.7\\ 
			4.68   & $0.415$ & $0.414$    & 0.1 & 0.8\\ 
			4.70  & $0.428$ & $0.429$    & 0.2 & 3.3\\ 
			4.74   & $0.456$ & $0.449$    & 0.6 & 1.9\\ 
			4.75   & $0.444$ & $0.445$    & 0.3 & 1.5\\ 
			4.78   & $0.426$ & $0.428$    & 0.3 & 0.8\\ 
			4.84   & $0.397$ & $0.401$    & 0.3 & 1.4\\ 
			4.92   & $0.376$ & $0.375$    & 0.6 & 1.3\\ 
			\hline
   \hline
		\end{tabular}
		
	\end{table}
	
	\clearpage
	\section{Partial Wave Analysis}
	
\begin{figure}[H]
		\centering
		\mbox{
			\begin{overpic}[width=1.0\linewidth,height=0.35\linewidth,angle=0]{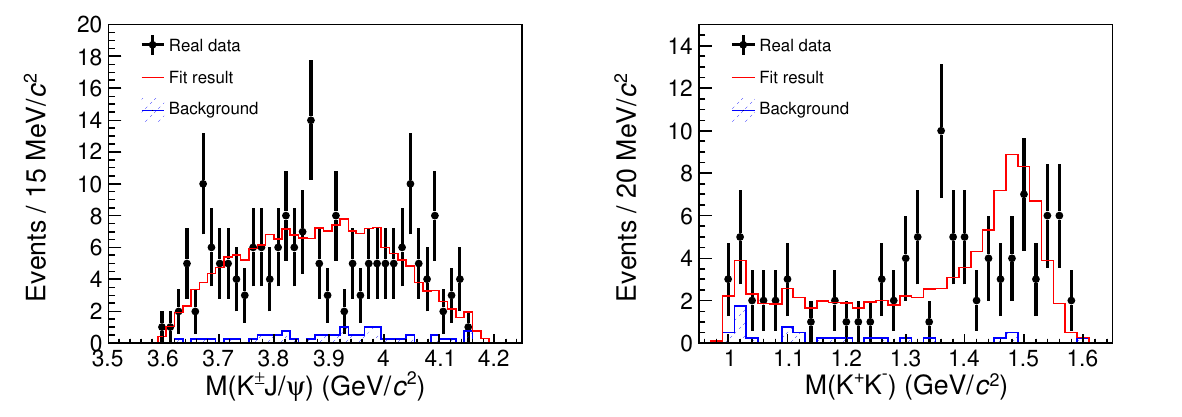}
				\put(38,29){(a)}
				\put(88,29){(b)}
			\end{overpic}
		}
		\mbox{
			\begin{overpic}[width=1.0\linewidth,height=0.35\linewidth,angle=0]{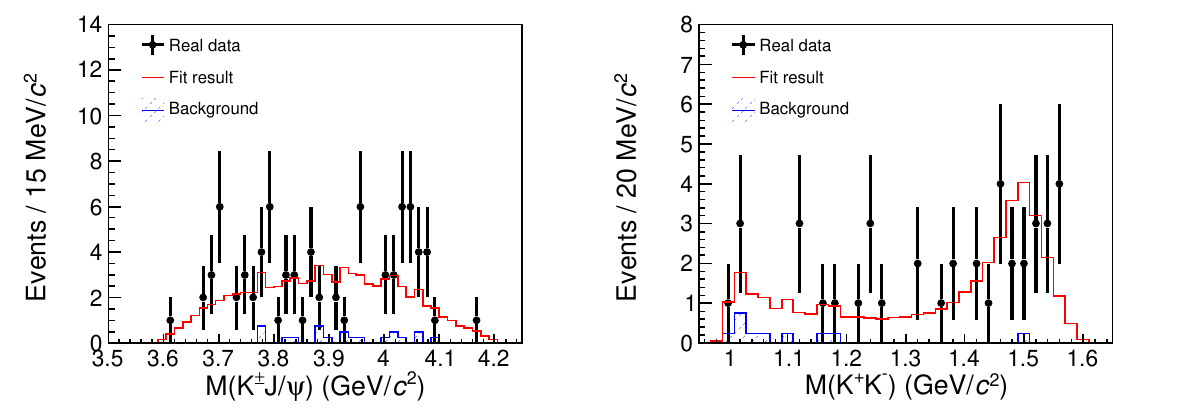}
				\put(38,29){(c)}
				\put(88,29){(d)}
			\end{overpic}
		}
		\caption{ The comparisons of $M(K^{\pm}J/\psi)$ and $M(K^+K^-)$ between real data and the PWA result. The black dots with error bars are  real data, the red line is the sum of the fit result and background from $J/\psi$ sideband.  
			The PWA results at (a, b) $\sqrt{s}=4.68$ GeV and (c, d) $\sqrt{s}=4.70$ GeV  are based on the $f_0(980)+f_0(1500)$ assumption.
		}
		\label{PWAprojection1}
	\end{figure}

\begin{figure}[H]
		\centering
		\mbox{
			\begin{overpic}[width=1.0\linewidth,height=0.35\linewidth,angle=0]{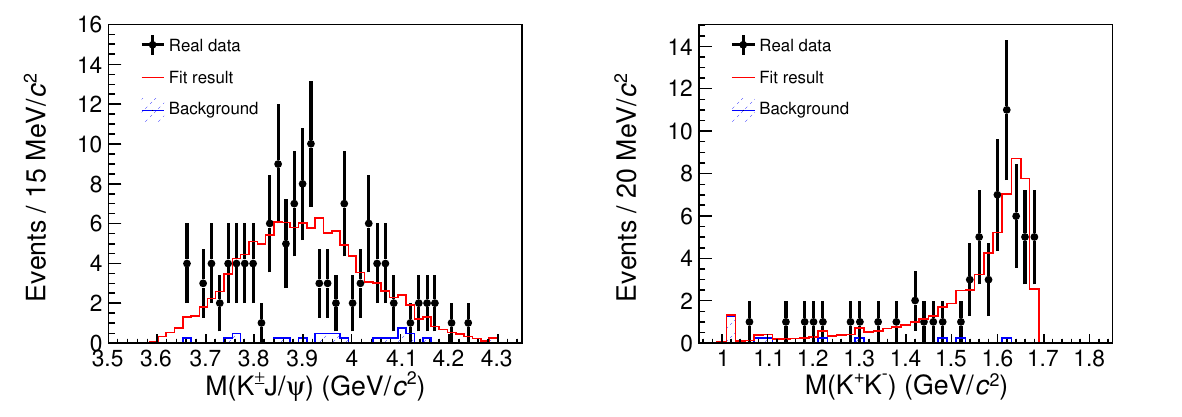}
				\put(38,29){(a)}
				\put(88,29){(b)}
			\end{overpic}
		}
		\mbox{
			\begin{overpic}[width=1.0\linewidth,height=0.35\linewidth,angle=0]{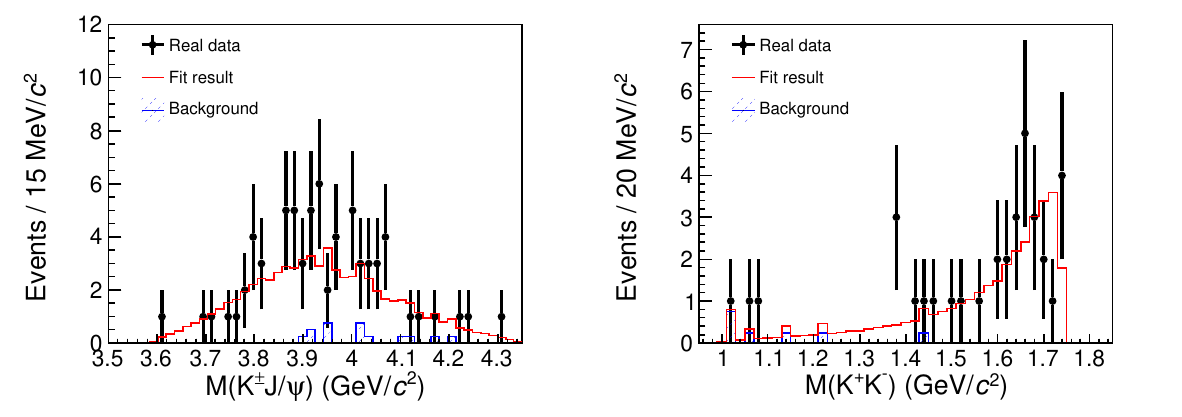}
				\put(38,29){(c)}
				\put(88,29){(d)}
			\end{overpic}
		}
		\caption{ The comparisons of $M(K^{\pm}J/\psi)$ and $M(K^+K^-)$ between real data and the PWA result. The black dots with error bars are  real data, the red line is the sum of the fit result and background from $J/\psi$ sideband.   
			The PWA results at (a, b) $\sqrt{s}=4.78$ GeV and (c, d) $\sqrt{s}=4.84$ GeV are based on a single $f_0(x)$  with mass and width free in the fit.
		}
		\label{PWAprojection2}
	\end{figure}

	Partial wave analysis (PWA) using helicity formalism is performed on each c.m.~energy to study the intermediate states. However, no significant $Z_{cs}$ signal is detected because of the limited statistics. The $f_0(0^{++})$ and $f_2(2^{++})$ components can be not well distinguished  either.  The PWA results with different 
	$f_{0,2}$ combinations are used to generate alternative signal MC samples to do efficiency uncertainty study for $e^+e^-\to K^+K^-J/\psi$ cross section measurement.  Figures~\ref{PWAprojection1} and \ref{PWAprojection2} show the comparison of distributions between data and one of the PWA results at four c.m.~energies with higher statistics ($\sqrt{s}=4.68, 4.70, 4.78, 4.84$ GeV). For the data sets with $\sqrt{s}\leq4.70$ GeV, the PWA results are based on the $f_0(980)+f_0(1500)$ assumption.  For the data sets with $\sqrt{s}>4.70$ GeV, the PWA results are based on a single $f_0(x)$ with mass and width free in the fit.

	\bibliographystyle{apsrev4-2}

\end{document}